\documentclass[aps,prl,reprint,amsmath,amssymb,superscriptaddress,floatfix,nofootinbib]{revtex4-2}

\usepackage{graphicx}
\usepackage{bm}
\usepackage{booktabs}
\usepackage{array}
\usepackage{listings}
\usepackage{xcolor}
\usepackage[colorlinks=true,linkcolor=blue,citecolor=blue,urlcolor=blue,filecolor=magenta, urlcolor=blue, runcolor=cyan]{hyperref}

\graphicspath{{figure/}}
\newcommand{\ket}[1]{\left|#1\right\rangle}

\newcommand{\addrA}{Beijing Key Laboratory of Fault-Tolerant Quantum Computing,\\Beijing Academy of Quantum Information Sciences, Beijing 100193, China}
\newcommand{\addrBH}{School of Physics, Beihang University, Beijing 100191, China}
\newcommand{\addrB}{Beijing National Laboratory for Condensed Matter Physics, Institute of Physics, Chinese Academy of Sciences, Beijing 100190, China}
\newcommand{\addrC}{School of Physical Sciences, University of Chinese Academy of Sciences, Beijing 100049, China}
\newcommand{\addrD}{Hefei National Laboratory, Hefei 230088, China}
\newcommand{\addrE}{Songshan Lake Materials Laboratory, Dongguan 523808, Guangdong, China}
\newcommand{\addrZJM}{Zhejiang Institute of Modern Physics, Zhejiang University, Hangzhou 310058, China}
\newcommand{\addrZJQ}{Zhejiang Key Laboratory of Micro-Nano Quantum Chips and Quantum Control,
School of Physics, Zhejiang University, Hangzhou 310027, China}

\begin{document}

\title{Hidden Coherence and Controlled Quasihole Motion in a Fermionic Laughlin State on a Superconducting Quantum Processor}

\author{Kui Zhao}
        \altaffiliation{These authors contributed equally to this work.}
	\affiliation{\addrA}

\author{Hao Wu}
        \altaffiliation{These authors contributed equally to this work.}
	\affiliation{\addrBH}	
	
\author{Lei-Yi-Nan Liu}
        \altaffiliation{These authors contributed equally to this work.}
	\affiliation{\addrBH}	
	
\author{Gui-Han Liang}
    \thanks{These authors contributed equally to this work.}
	\affiliation{\addrB}

\author{Cheng-Lin Deng}
	\affiliation{\addrA}

\author{Cai-Ping Fang}
	\affiliation{\addrB}
	\affiliation{\addrC}

\author{Hao-Tian Liu}
	\affiliation{\addrB}
	\affiliation{\addrC}

\author{Wei-Ping Yuan}
	\affiliation{\addrA}
	\affiliation{\addrB}
	
\author{Qingyuan Wu}
	\affiliation{\addrA}
	\affiliation{\addrB}	

\author{Ziting Wang}
	\affiliation{\addrA}

\author{Wei-Guo Ma}
	\affiliation{\addrB}

\author{Hao Li}
	\affiliation{\addrA}
	
\author{Jia-Chi Zhang}
	\affiliation{\addrB}
	\affiliation{\addrC}
    
\author{Tian-Ming Li}
	\affiliation{\addrB}
	\affiliation{\addrC}

\author{Jia-Cheng Song}
	\affiliation{\addrB}
	\affiliation{\addrC}

\author{Zhen-Ting Bao}
	\affiliation{\addrB}
	\affiliation{\addrC}

\author{Zheng-He Liu}
	\affiliation{\addrB}
	\affiliation{\addrC}	

\author{Yang He}
	\affiliation{\addrB}
	\affiliation{\addrC}

\author{Yong-Xi Xiao}
	\affiliation{\addrB}
	\affiliation{\addrC}
	
\author{Yu Liu}
	\affiliation{\addrB}
	
\author{Zhongcheng Xiang}
	\affiliation{\addrB}
	\affiliation{\addrC}
	\affiliation{\addrD}

\author{Dongning Zheng}
	\affiliation{\addrB}
	\affiliation{\addrC}
	\affiliation{\addrD}
	\affiliation{\addrE}

\author{Guangming Xue}
    \affiliation{\addrA}
    
\author{Haifeng Yu}
    \address{\addrA}
    \address{\addrD}

\author{Kai Xu}
	\affiliation{\addrA}
	\affiliation{\addrB}
	\affiliation{\addrC}
	\affiliation{\addrD}

\author{Zhao Liu}
	\affiliation{\addrZJM}
	\affiliation{\addrZJQ}

\author{Kaixuan Huang}
    \email{huangkx@baqis.ac.cn}
	\affiliation{\addrA}
	
\author{Jian Cui}
    \email{jiancui@buaa.edu.cn}
	\affiliation{\addrBH}	

\author{Heng Fan}
    \email{hfan@iphy.ac.cn}
	\affiliation{\addrA}
	\affiliation{\addrB}
	\affiliation{\addrC}
	\affiliation{\addrD}



\begin{abstract}

Fractional quantum Hall (FQH) states support fractionalized
excitations and geometric phases, yet experimentally
resolving the many-body coherence encoded in their wave
functions remains challenging.
Here, we prepare a three-electron $\nu=1/3$ Laughlin state
on the Haldane sphere using seven superconducting qubits
and reconstruct its many-body density matrix.
Through Read's composite-boson construction, we reveal
vortex-dressed coherence that remains appreciable over
a larger separation range than electron coherence.
The extended profile is suppressed in a Fock-diagonal reference
constructed from the measured populations, demonstrating its
dependence on off-diagonal many-body coherence.
We further prepare a one-flux-added quasihole and characterize
its fractional charge response against finite-size benchmarks.
Controlled quasihole motion yields an effective geometric-phase signature
from the reconstructed dominant-eigenstate trajectory.
These complementary measurements provide microscopic
access to the coherent structure and quasihole response
of a finite fermionic Laughlin state, demonstrating an
experimental route to probing strongly correlated FQH
wave functions on programmable quantum processors.

\end{abstract}

\maketitle

\textit{Introduction}---
Fractional quantum Hall (FQH) states provide a paradigmatic
example of emergent quantum matter, where intrinsic topological order,
fractionalized excitations, and geometric response arise
from strongly correlated many-body wave
functions~\cite{Stormer1999RMPFQH,
Wen1995AdvPhysTopologicalOrder,Haldane2011PRLGeometry,
Arovas1984PRLFractionalStatistics,Nayak2008RMPNonAbelianAnyons,Lapa2018PRBHallViscositygeometricresponse}.
The Laughlin state, the simplest microscopic realization
of an Abelian FQH phase, supports fractionally charged
quasiparticles and nonlocal many-body
correlations~\cite{Laughlin1983PRLFractionalCharge,
Haldane1983PRLHierarchy}.
The entangled many-body wave function carries a coherent structure that is not captured by conventional single-particle correlation functions.
Read's composite-boson construction interprets this structure physically, i.e., each electron, bound to the $m=1/\nu$ vortices of the Laughlin state, behaves as a boson, and the state is the Bose condensate of these composite objects~\cite{Read1989PRLCompositeBoson}. This composite-boson construction predicts non-trivial consequences such as off-diagonal long-range order (ODLRO) of the composite-boson field.
Experimentally accessing this vortex-dressed coherence is challenging because the corresponding correlator involves highly nonlocal many-body operators.  

Recent advances in quantum simulation have made finite
FQH states accessible in ultracold
atoms~\cite{Leonard2023NatureUltracoldFQH},
photonic systems~\cite{Clark2020NaturePhotonicLaughlin},
and superconducting
circuits~\cite{Wang2024SciencePhotonicFQH}.
Superconducting processors have also become versatile
platforms for exploring equilibrium and nonequilibrium
topology~\cite{Xiang2023Chern,Jin2025ZeroModes,
Qian2025HigherOrder,Zhou2025Qutrit,Koh2024HigherOrder,
Will2025Floquet,Koh2026Edge,Han2024Exceptional},
entanglement and quantum
criticality~\cite{Wang2025PRXBellCorrelation,Tan2026Critical,Zheng2023Cat,
Han2024Spectroscopy,Li2026NonEquilibriumSensing,Yan2026Projected},
geometric control~\cite{Xu2026Geometric},
and anyonic
phenomena~\cite{Andersen2023Braiding,Xu2024Fibonacci,
Niu2024Braiding}.
For Laughlin-type states, theoretical and experimental
efforts have explored state
preparation~\cite{Nakamura2012PPLFermionChain,
Rahmani2020PRXLinearDepthLaughlin,
Shen2026NatCommunLaughlin,Exposito2026arXivVQEFQH,
Xu2026arXivFQHFactory},
geometric response~\cite{Kirmani2022PRLGeometricExcitations,
Kirmani2026PRBHallViscosity},
and quasihole
braiding~\cite{Kirmani2023PRBQuasiholeBraiding}. Recent work has also prepared bosonic Laughlin-like fractional
Chern insulator and quasihole states on a superconducting
quantum processor~\cite{Dutkiewicz2026LaughlinFCI}.
These advances have expanded the preparation and control
of finite FQH states, while direct experimental access
to their many-body coherence remains challenging.
Vortex-dressed correlation functions provide a natural
probe of the coherent structure hidden in Laughlin
wave functions~\cite{Pauw2024PRRHiddenOrder}.

\textit{Model}---Here, we reveal vortex-dressed Read coherence in a finite
fermionic \(\nu=1/3\) Laughlin state on the Haldane sphere
using superconducting qubits.
The Read correlator remains appreciable over a larger
separation range than the fermionic single-particle coherence.
We further prepare a one-flux-added quasihole and
characterize its fractional charge response against
finite-size benchmarks.
Controlled quasihole motion provides an effective geometric-phase diagnostic
of the reconstructed dominant-eigenstate trajectory.
Together, these measurements offer microscopic access
to the coherent structure and quasihole response of a
finite Laughlin state, motivating studies of their
evolution with system size and quasihole configuration.


\begin{figure*}[!htbp]
\centering
\includegraphics[width=0.85\textwidth]
{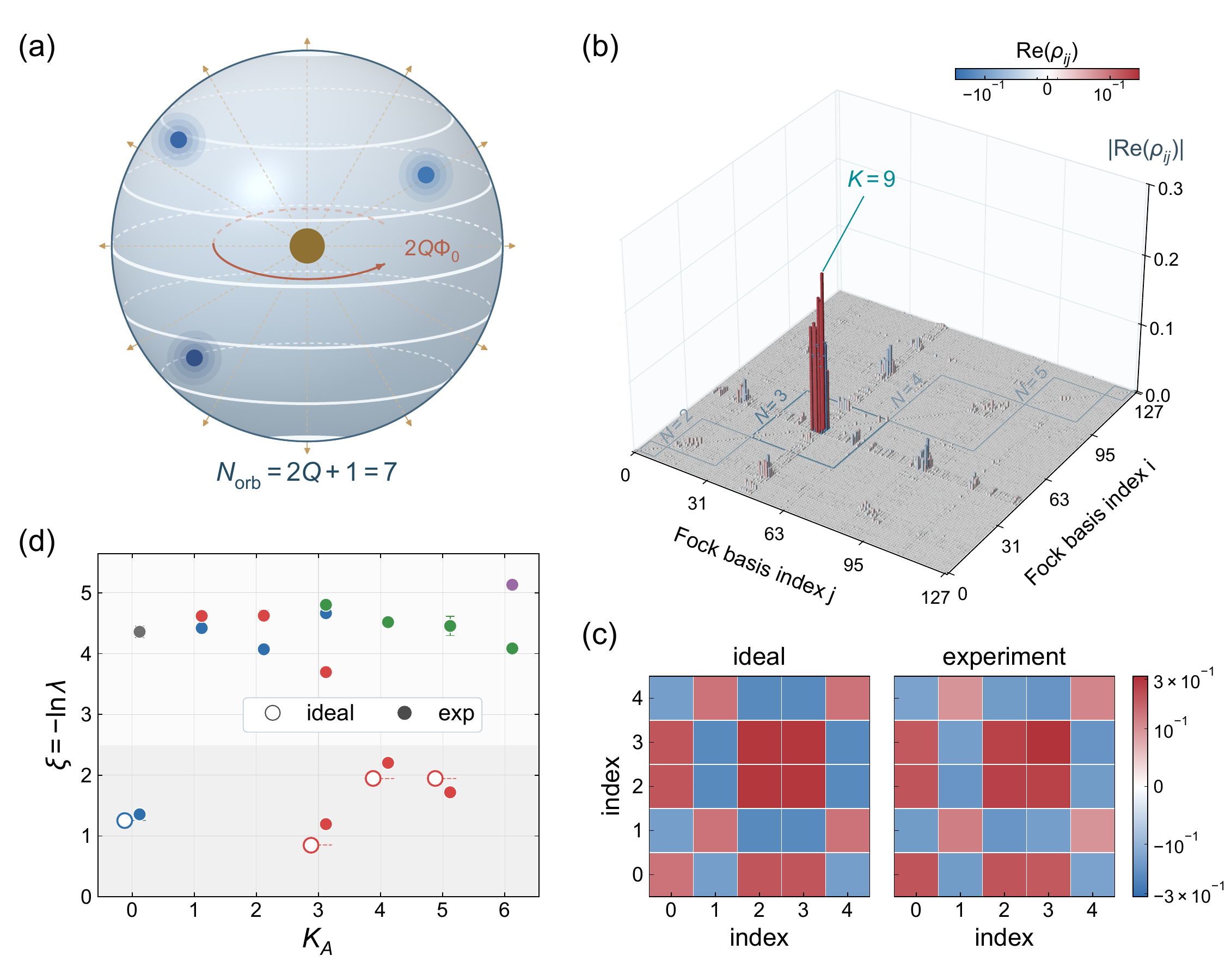}
\caption[Experimental reconstruction and entanglement structure of the Laughlin state.]{
Experimental reconstruction and entanglement structure of the Laughlin state.
(a) Haldane-sphere schematic with $N_\phi=2Q$ magnetic flux quanta. 
(b) Experimental density matrix reconstructed by full quantum state tomography
(FQST) in the Fock basis, reordered by particle number $N$ and orbital-index sum
$K=\sum_j j n_j$. Colored bars show the experimental matrix, with heights
$|\operatorname{Re}\rho_{ij}|$ and colors indicating the signed real parts.
Thin outlines mark the particle-number sectors; the cyan outline marks
the target Laughlin sector $(N_e,K)=(3,9)$.
(c) Ideal and experimental density matrices in the target Laughlin sector
$(N_e,K)=(3,9)$. The experimental submatrix from (b) is divided by its sector
weight to obtain a unit-trace conditional density matrix, resolving populations
and off-diagonal matrix elements. The measured state yields a full-state fidelity $F=0.662$, a
target-sector weight $w_{N_e,K}=0.747$, and a conditional fidelity $F_{N_e,K}=0.886$.
(d) 
Block-resolved reduced-density spectrum obtained from the
$(N_A,K_A)$ diagonal blocks of $\rho_A=\operatorname{Tr}_B(\rho_{\mathrm{exp}})$.
 Open markers indicate the ideal Schmidt levels, while filled markers denote the experimental levels, with colors distinguishing  $N_A$ sectors. 
The four lowest experimental levels occur in the same sectors as the ideal finite-size Laughlin state.
}
\label{fig:laughlin-full-tomo-entanglement}
\end{figure*}

Haldane's spherical geometry provides a natural finite-size setting for studying Laughlin correlations without physical boundaries, while retaining the coupling between magnetic flux and particle number. 
For the $\nu=1/3$ Laughlin state, the spherical flux--particle relation fixes the finite-size manifold as $N_\phi=3(N_e-1)$, where $N_\phi$ is the number of magnetic flux quanta.
A magnetic monopole of strength $Q=N_\phi/2$ at the center of the sphere generates $N_\phi+1$ degenerate lowest-Landau-level (LLL) orbitals, as illustrated in  Fig.~\ref{fig:laughlin-full-tomo-entanglement}(a). 
In spherical spinor coordinates $\chi=(u,v)^T$, the fermionic Laughlin wave function is $\Psi_{\rm L} = \prod_{i<j}(u_i v_j-v_i u_j)^3$.
For the orbital representation used in our experiment, let $c_j^\dagger$ denote the fermionic creation operator associated with the $j$-th LLL orbital. The Laughlin state can  be expanded in the fermionic Fock basis as
\begin{equation}
|\Psi_{\rm L}\rangle =
\sum_{\{n_j\}}
C_{n_0...n_{N_\phi}}
\prod_{j=0}^{N_\phi}
(c_j^\dagger)^{n_j}|{\rm vac}\rangle ,
\label{eq:fock-expansion}
\end{equation}
with $n_j\in\{0,1\}$ and total electron number $\sum_j n_j=N_e$.
For the experimental system with $N_e=3$ and $N_\phi=6$, 
the seven LLL orbital occupations are encoded into seven superconducting qubits, so that the computational basis represents the fermionic Fock basis.
The occupation basis provides direct access to the rotational symmetry sectors of the spherical geometry, with $N$ labeling the particle-number sectors of the full Fock space.
Using the shifted orbital index $j=m+Q$, we define $K=\sum_j jn_j$, which is related to the angular-momentum component in a particle-number sector $N$ by $L_z=K-QN$. Here $N=0,\ldots,7$ is a general particle-number-sector label in the seven-orbital Fock space, while $N_e=3$ specifies the target particle number.
The rotationally invariant Laughlin ground state with $L=0$ therefore lies in the target symmetry sector $(N_e,K)=(3,9)$. Within this sector, the state has support on a five-dimensional squeezing subspace generated from the root configuration $\ket{1001001}$, with the Laughlin correlations encoded in the
coherent amplitudes among these configurations~\cite{Bernevig2008PRLJackPolynomials}.

\textit{Results}---Experimentally, we first realize this seven-orbital encoding on seven selected qubits from an eight-qubit active subset of a
$3\times3$ array of nine frequency-tunable transmon qubits coupled through
tunable couplers~\cite{Wang2025arXivLGT}. The five amplitudes defining the Laughlin state are prepared by a binary-tree circuit and then resolved into the corresponding occupation-number configurations~\cite{wu2026,
Zhang2022PRLOptimalStatePreparation}. The logical circuit is mapped onto the selected physical-qubit topology before compilation into calibrated
single-qubit rotations and CZ gates, with average XEB Pauli errors of
$0.90\times10^{-3}$ and $4.9\times10^{-3}$ for gate durations of
$20\,\mathrm{ns}$ and $60\,\mathrm{ns}$, respectively (see the SM~\cite{SM}).
 We then reconstruct the seven-qubit density matrix $\rho_{\mathrm{exp}}$ by full quantum state tomography (FQST), yielding a full-state raw fidelity $F=\langle\Psi_L|\rho_{\mathrm{exp}}|\Psi_L\rangle=0.662$ [Fig.~\ref{fig:laughlin-full-tomo-entanglement}(b)].
To reveal the symmetry-sector structure of the prepared state, we reorder the Fock basis by particle number $N$ and, within each $N$ sector, by $K$. 
For a sector $\lambda=(N,K)$, denoting the corresponding projector by $P_\lambda$, the sector weight and normalized conditional state are defined as~\cite{Huggins2021npjQuantumMeasurement,Gupta2024nature,Zhang2026nano}
\begin{equation}
w_\lambda=\mathrm{Tr}(P_\lambda\rho_{\rm exp}),
\quad
\rho_\lambda=
\frac{P_\lambda\rho_{\rm exp}P_\lambda}{w_\lambda}.
\end{equation}
The conditional fidelity $F_\lambda=\langle\Psi_{\rm L}|\rho_\lambda|\Psi_{\rm L}\rangle$
quantifies the fidelity with the ideal Laughlin state after conditioning on the selected symmetry sector.
Figure~\ref{fig:laughlin-full-tomo-entanglement}(b) shows the FQST density matrix in the Fock basis reordered by $(N,K)$. The enlarged $(N_e,K)=(3,9)$ block in Fig.~\ref{fig:laughlin-full-tomo-entanglement}(c) reproduces the dominant features of the Laughlin superposition. Conditioning the reconstructed density matrix on the $N_e=3$ and $(N_e,K)=(3,9)$ sectors yields conditional fidelities $(F_{N_e},F_{N_e,K})=(0.824,0.886)$, with corresponding sector weights $(w_{N_e},w_{N_e,K})=(0.803,0.747)$,
respectively. For the selected sectors containing the ideal Laughlin state,
$P_\lambda|\Psi_L\rangle=|\Psi_L\rangle$, and hence
$F=w_\lambda F_\lambda$.
The measured sector weights and conditional fidelities show
that a substantial fraction of the full-state fidelity loss
is associated with population outside the selected sector.
The corresponding many-body structure is further examined through the orbital entanglement spectrum (OES)~\cite{LiHaldane2008PRLEntanglementSpectrum},
which provides a sector-resolved view of the entanglement structure and reveals the finite-size counting pattern of the Laughlin state.
For the orbital partition $A=\{0,1,2,3\}$, we construct
$\rho_A=\operatorname{Tr}_B(\rho_{\mathrm{exp}})$ and diagonalize its $(N_A,K_A)$ diagonal blocks separately, defining the entanglement levels as $\xi_\alpha=-\ln\lambda_\alpha$, where $\lambda_\alpha$ is the $\alpha\text{-}th$ eigenvalue.
The ideal state has four finite levels in the sectors
$(N_A,K_A)=(1,0),(2,3),(2,4),(2,5)$.
Experimental imperfections lift otherwise vanishing eigenvalues, producing additional levels at larger $\xi$ with correspondingly smaller reduced-density weights.
As shown in Fig.~\ref{fig:laughlin-full-tomo-entanglement}(d), the four lowest experimental levels retain the ideal sector pattern, providing a finite-size consistency check of the Laughlin counting structure.

\begin{figure*}[!htbp]
\centering
\includegraphics[width=0.96\textwidth]
{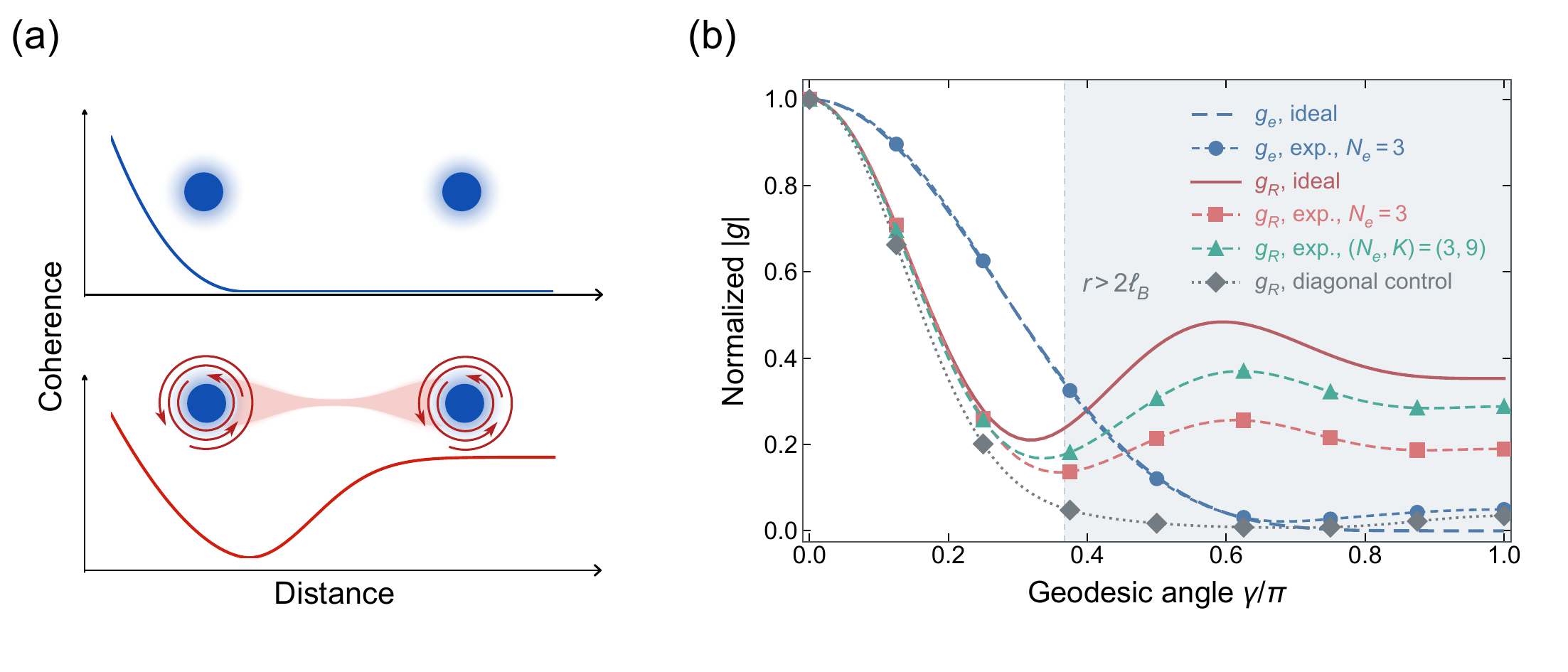}
\caption[Vortex-dressed coherence of the reconstructed Laughlin state.]{Vortex-dressed coherence of the reconstructed Laughlin state.
(a) Conceptual schematic comparison between the fermionic one-body coherence $g_e$ and the vortex-dressed Read coherence $g_R$. The Read coherence is obtained by dressing the fermionic field with three Laughlin correlation vortices.
(b) Coherence versus geodesic separation. The shaded region marks $r>2\ell_B$. Here $\ell_B=1/\sqrt3$ is the magnetic length of the unit-radius $Q=3$ sphere.
The vortex-dressed Read correlator $g_R$ reveals the nonlocal coherence encoded in the reconstructed Laughlin state. The diagonal control $\rho_{\rm diag}^{(N_e)}=\mathrm{diag}(\rho^{(N_e)})$ removes off-diagonal Fock-space coherences while preserving populations, leading to suppression of $g_R$. Projection onto the target $(N_e,K)=(3,9)$ sector shifts the experimental profile toward the ideal profile.
All experimental coherence profiles are calculated from the full-tomography density matrix after projection onto the $N_e=3$ sector.
}
\label{fig:read-correlator}
\end{figure*}


We next examine the many-body coherence of the reconstructed Laughlin state.
Its fermionic single-particle coherence does not exhibit ODLRO associated with a conventional condensate. Read's construction binds an electron to the $m=1/\nu$ vortices of the Laughlin wave function, defining a composite-boson field whose first-order coherence can exhibit long-range behavior in the thermodynamic limit~\cite{Read1989PRLCompositeBoson,Pauw2026arXivCompositeBoson}. Here, we use the corresponding vortex-dressed correlator as a finite-size diagnostic of the many-body coherence encoded in the reconstructed Laughlin state.
In the spherical LLL basis, the fermionic field operator is expanded as
\begin{equation}
\hat\psi(\Omega) = \sum_{j=0}^{N_\phi}
\phi_j(\Omega)c_j ,
\label{eq:lll-field-operator}
\end{equation}
where $\phi_j(\Omega)$ is the $j\text{-}th$ LLL orbital.
For two points $\Omega_a$ and $\Omega_b$ on the sphere, we 
compare fermionic single-particle coherence $g_e=\left\langle\tilde{\psi}_a^\dagger\tilde{\psi}_b\right\rangle$ with the vortex-dressed Read coherence $g_R(\Omega_a,\Omega_b)=\left\langle \tilde{\mathcal B}_a^\dagger \tilde{\mathcal B}_b \right\rangle$.
In our case $\nu=1/3$, the Read operator is constructed as $\hat{\mathcal B}^\dagger(\Omega)=\hat\psi^\dagger(\Omega)\hat U^3(\Omega)$, where each $\hat U(\Omega)$ attaches one Laughlin vortex at
$\Omega$, so that $\hat U^3$ implements the three-vortex dressing appropriate to $\nu=1/3$. The tildes denote local normalization (see the SM~\cite{SM}). We plot the magnitudes \(|g_e|\) and $|g_R|$ as functions of geodesic separation.
The vortex attachment defines an operator dressing of the Laughlin state, distinct from the one-flux-added quasihole excitation considered below. Experimental correlators are evaluated from the FQST density matrix projected and renormalized within the $N_e=3$ sector.

For the ideal $N_e=3$, $N_\phi=6$ Laughlin state, rotational symmetry gives $\langle c_m^\dagger c_n\rangle=(3/7)\delta_{mn}$, and therefore $|g_e(\gamma)| = \left| \cos\frac{\gamma}{2} \right|^6 $,  where $\gamma$ is the angular separation on the sphere. Thus the electronic coherence is intrinsically suppressed at large separation even for the ideal Laughlin state,  as illustrated schematically in Fig.~\ref{fig:read-correlator}(a).
Vortex dressing reveals the coherence structure of the Laughlin state, leading to a qualitatively different spatial behavior. Whereas the fermionic one-body coherence rapidly decays with separation, the Read coherence $g_R$ remains
appreciable over a substantially larger distance range [Fig.~\ref{fig:read-correlator}(b)]. We use the region $r>2\ell_B$ as a finite-size long-distance window, without associating it with a thermodynamic limit. 

The origin of this extended coherence is revealed by the diagonal control $\rho_{\rm diag}^{(N_e)}=\mathrm{diag}(\rho^{(N_e)})$, which retains the measured Fock-state populations while removing all off-diagonal coherences~\cite{Pauw2024PRRHiddenOrder}. 
The suppression of $g_R$ in this diagonal control shows that the extended Read-coherence profile of the reconstructed state depends on off-diagonal elements of density matrix in the Fock basis, rather than on the occupation probabilities alone.
Conditioning on the target \((N_e,K)=(3,9)\) sector removes contributions from other angular-momentum sectors and shifts the experimental Read-coherence profile toward the ideal result.
These results show that the vortex-dressed correlator probes coherence between correlated Fock configurations in the reconstructed Laughlin state.


\begin{figure}[!htbp]
\centering
\includegraphics[width=0.45\textwidth]
{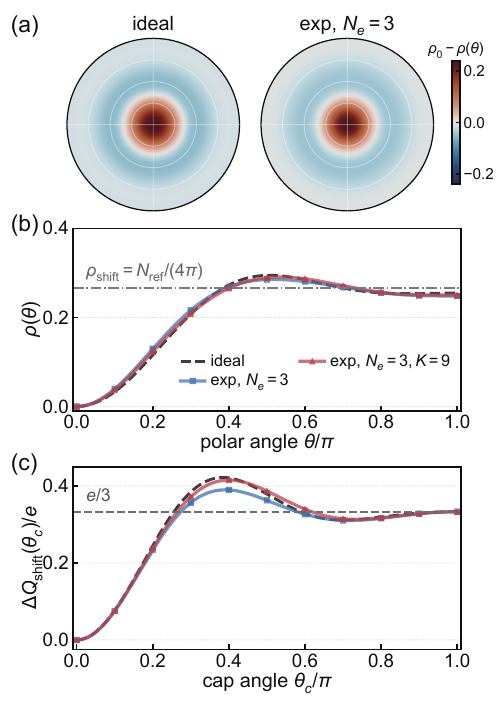}
\caption[Polar quasihole density and integrated deficit.]{
Polar quasihole density and integrated deficit.
(a) Ideal and experimental real-space density deficits, with the experimental profile obtained from the measured orbital occupations in the $N_e=3$ sector.
(b) Azimuthally averaged density of the ideal state and the prepared states after postselection. The dash-dotted line marks the sphere-shift reference density $\rho_{\rm shift}=10/(12\pi)$.
(c) Integrated density deficit relative to $\rho_{\rm shift}$ within a polar cap. The dashed line marks $e/3$. The experimental and ideal quasihole profiles show the polar-cap dependence of the density response. Data are averaged over three independent measurements.
}
\label{fig:polar-hole-density-charge}
\end{figure}


We next complement the coherence characterization with the density response of the corresponding one-flux-added Laughlin quasihole.
In the Laughlin picture, adding one flux quantum at fixed particle number creates a quasihole excitation carrying a fractionalized charge response. On the sphere, a quasihole located at $\Omega_h=(\theta_h,\phi_h)$ is described by
\begin{equation}
\Psi_{\rm qh}(\Omega_h)=\mathcal N_{\rm qh}
\prod_{i=1}^{N_e}(u_i v_h-v_i u_h)\Psi_{\rm L},
\label{eq:qh-wavefunction}
\end{equation}
where $\chi_h=(u_h,v_h)^T$ denotes the quasihole spinor and $\mathcal N_{\rm qh}$ ensures normalization. The quasihole factor increases the LLL polynomial degree by one, corresponding to the addition of one flux quantum. Its dependence on $\chi_h$ generates a spin-$N_e/2$ SU(2) manifold, providing a natural description of quasihole motion on the sphere. Experimentally, we prepare this quasihole state directly
in the eight-orbital Hilbert space using the sparse state-preparation circuit described in the SM~\cite{SM}.

For $N_e=3$ and $N_\phi=7$, the one-quasihole manifold forms a four-dimensional spin-$3/2$ multiplet. We first consider the quasihole at the north pole. The real-space density is
$\rho(\Omega)=\langle\hat{\psi}^{\dagger}(\Omega)
\hat{\psi}(\Omega)\rangle$.
For the ideal polar quasihole, axial symmetry makes the
one-body density matrix diagonal in the $L_z$ orbital basis.
The measured orbital occupations determine the azimuthally
averaged density, which we denote by $\rho(\theta)$:
\begin{equation}
\rho(\theta)
\equiv \frac{1}{2\pi}\int_0^{2\pi}
\mathrm{d}\phi\,\rho(\theta,\phi)
=\sum_{j=0}^{7}
\langle n_j\rangle|\phi_j(\theta)|^2.
\label{eq:polar-qh-density}
\end{equation}

Figure~\ref{fig:polar-hole-density-charge}(a) shows the depletion $\rho_0-\rho(\theta)$ relative to the uniform three-fermion density $\rho_0=3/(4\pi)$. For the integrated deficit, we use the uniform sphere-shift reference $\rho_{\rm shift}=N_{\rm ref}/(4\pi)$. The spherical $\nu=1/3$ Laughlin state has topological shift $\mathcal S=3$, so the no-quasihole relation is $N_\phi=3N_{\rm ref}-3$~\cite{WenZee1992PRLShift}. 
At $N_\phi=7$, this gives $N_{\rm ref}=10/3$ for the reference background, while the quasihole state contains three fermions.
The accumulated density deficit is obtained by integrating over a polar cap~\cite{Johri2014PRBQuasiholeSize,Liu2015PRBQuasiholeFCI},
\begin{equation}
\frac{\Delta Q(\theta_c)}{e}=
2\pi\int_0^{\theta_c}
d\theta\,\sin\theta\,[\rho_{\rm shift}-\rho(\theta)].
\end{equation}
The polar-cap integral follows the density deficit outward from the quasihole pole, with $\Delta Q(\pi)=e/3$ over the full sphere.
In larger Laughlin systems, a localized quasihole produces an intermediate-distance plateau of the integrated deficit near $e/3$. This plateau lies beyond the quasihole core and extends over a wider range as the system grows~\cite{Johri2014PRBQuasiholeSize}.
Figure~\ref{fig:polar-hole-density-charge} summarizes the reconstructed polar quasihole response.
The measured $N_e=3$ density rises from the depleted north pole and follows the ideal quasihole profile in Fig.~\ref{fig:polar-hole-density-charge}(b). Projection onto the $(N_e,K)=(3,9)$ symmetry sector brings both the radial density and integrated polar-cap profile closer to the ideal results in Fig.~\ref{fig:polar-hole-density-charge}(b,c). The polar depletion and surrounding density excess form the spatial signature of the prepared quasihole.

\begin{figure*}[!htbp]
\centering
\includegraphics[width=0.96\textwidth]{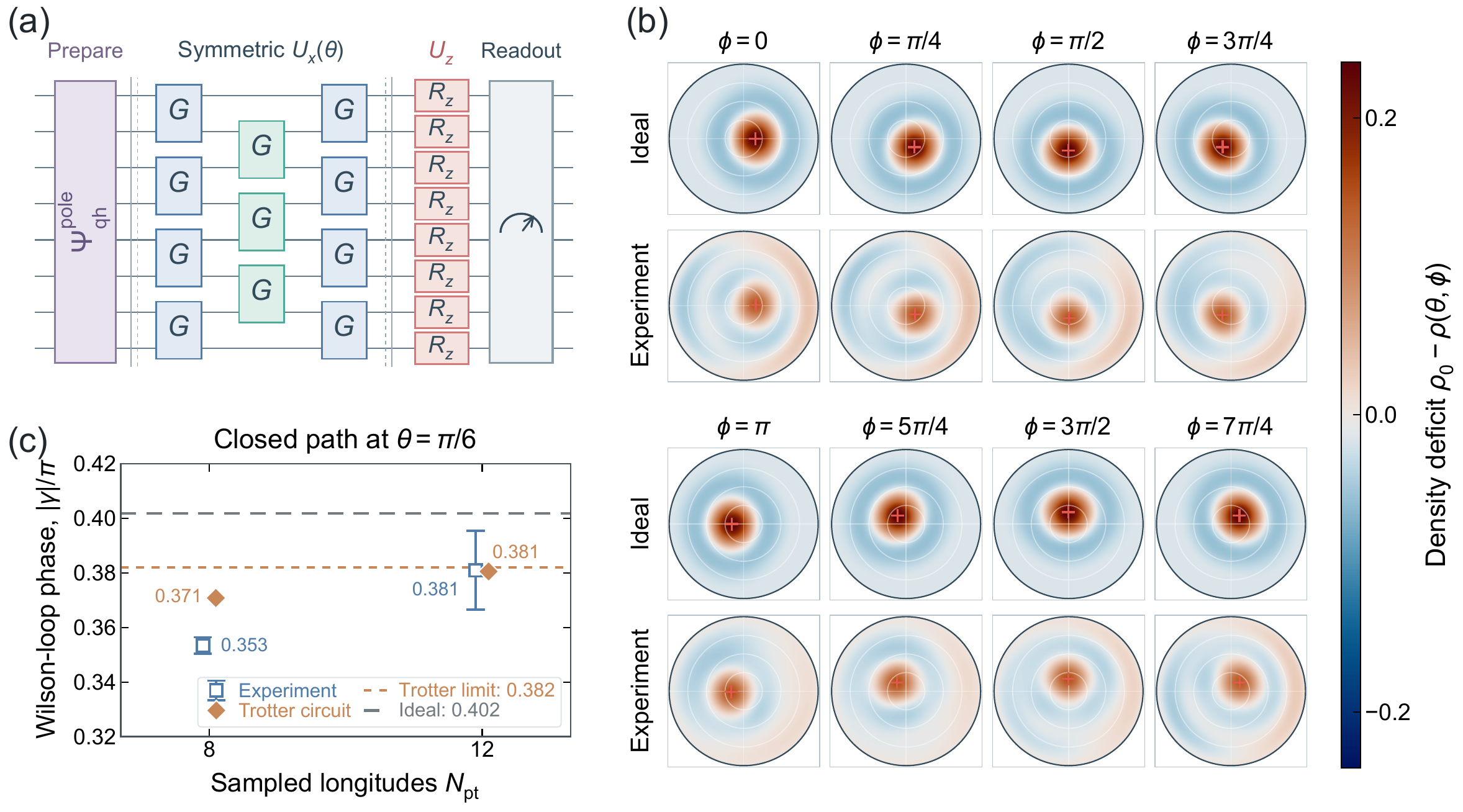}
\caption[Controlled quasihole motion and geometric diagnostics.]{
Controlled quasihole motion and geometric diagnostics.
(a) Digital implementation of quasihole displacement from the north pole. The polar quasihole state $\ket{\Psi_{\rm qh}^{\rm pole}}$ is prepared on eight orbitals, followed by the orbital rotation $U(\theta,\phi)=e^{-i(\phi+\pi/2)L_z}e^{-i\theta L_x}$.
The $L_x$ rotation is compiled into symmetric even- and odd-bond exchange layers, followed by a parallel $U_z(\phi+\pi/2)$ layer and measurement.
(b) Real-space density deficits for an eight-point longitude scan at fixed polar angle $\theta=\pi/6$. Ideal and reconstructed density deficits are shown at the sampled longitudes, demonstrating that the density minimum follows the programmed circular path.
(c) Wilson-loop phase extracted from the principal-eigenvector trajectory within fixed-$N_e=3$ tomography, providing an effective geometric-phase diagnostic. Blue open squares
show experimental values averaged over independent scans; muted-orange diamonds show noiseless simulations of the implemented Trotter circuits at the same sampled longitudes. The orange dashed line denotes the dense-sampling Trotter limit, $|\gamma_{\mathrm{Trotter}}^{\mathrm{dense}}|/\pi=0.382$, while the gray dashed line denotes the ideal spin-$3/2$ coherent-state value, $|\gamma_{\mathrm{id}}|/\pi=0.402$.
Error bars denote the standard error of the mean across independent scans.}
\label{fig:moved-qh-berry}
\end{figure*}


A quasihole is not merely a pointlike topological defect, but a spatially extended many-body excitation whose density profile and geometric evolution provide complementary probes of its internal structure~\cite{Liu2023PRBSpinStatistics}. 
Here we examine an effective geometric-phase signature associated with a closed trajectory of the reconstructed quasihole states.
For $N_e=3$ and $N_\phi=7$, the one-flux-added zero-mode sector forms a four-dimensional $L=3/2$ multiplet. A quasihole localized at $\Omega_h=(\theta,\phi)$ is a coherent state within this manifold, obtained from the polar quasihole state by a collective rotation,
\begin{equation}
|\Psi_{\rm qh}(\theta,\phi)\rangle =
e^{-i\phi L_z}e^{-i\theta L_y}
|\Psi_{\rm qh}^{\rm pole}\rangle.
\label{eq:qh-rotation}
\end{equation}
At fixed polar angle $\theta$, varying the longitude through $0\leq\phi<2\pi$ traces a closed path enclosing the solid angle $\Omega(\theta)=2\pi(1-\cos\theta)$. 
For the experimental path at $\theta=\pi/6$, the ideal spin-coherent-state phase magnitude is $\lvert\gamma_{\mathrm{id}}\rvert/\pi
=\frac{N_e\Omega(\theta)}{2\pi}
=N_e(1-\cos\theta)\simeq0.402$.

We implement this collective rotation digitally using $ U_{\rm move}(\theta,\phi) = e^{-i(\phi+\pi/2)L_z}e^{-i\theta L_x}$.
In the orbital basis, $L_x$ generates weighted nearest-neighbor fermionic exchanges, which map onto number-conserving $XX+YY$ operations between adjacent qubits. The $L_x$ evolution is compiled into a symmetric second-order Suzuki--Trotter sequence of even- and odd-bond exchange layers, followed by parallel virtual-$Z$ rotations implementing the azimuthal angle [Fig.~\ref{fig:moved-qh-berry}(a)]. Because $[U_{\rm move},\hat N]=0$, the ideal evolution acts independently within each particle-number sector.

At each programmed longitude $\phi_k$, 91 informationally complete product-Pauli settings and a constrained maximum-likelihood reconstruction yield a positive, unit-trace density matrix $\widetilde\rho^{(N_e)}(\phi_k)$ in the 56-dimensional $N_e=3$ manifold.
Independent all-$Z$ measurements give the sector weight $w^{(N_e)}=0.461\pm0.003$. 
Although the reconstructed states are mixed, the dominant component follows a well-defined trajectory through the programmed loop. The leading eigenvalue remains nondegenerate and spectrally well separated throughout the path (see the SM~\cite{SM}), allowing a continuous principal-eigenstate branch $|\psi_1(\phi)\rangle$ to be tracked. 
For the closed sequence $\phi_k=2\pi k/N_{\rm pt}$, we evaluate the gauge-invariant discrete Wilson product of the principal-eigenvector branch as an effective geometric-phase diagnostic of the reconstructed trajectory,
\begin{equation}
W=\prod_{k=0}^{N_{\rm pt}-1}
\langle\psi_1(\phi_k)|\psi_1(\phi_{k+1})\rangle,
\qquad
\phi_{N_{\rm pt}}\equiv\phi_0,
\label{eq:qh-wilson-loop}
\end{equation}
and report the phase magnitude $|\gamma_{\rm eff}|=|\arg W|$ as an effective geometric-phase diagnostic of the reconstructed principal-eigenvector trajectory. The reconstructed values are $|\gamma_{\rm eff}|/\pi=0.353\pm0.003$ for the eight-point path and $0.381\pm0.014$ for the twelve-point path [Fig.~\ref{fig:moved-qh-berry}(c)]. Noiseless simulations of the implemented Trotter circuits at the same sampled longitudes give $|\gamma_{\rm Trotter}|/\pi=0.371$ and $0.381$, respectively. Increasing the path-sampling density converges to the Trotter-circuit limit $|\gamma_{\rm Trotter}^{\rm dense}|/\pi=0.382$, while the exact spin-$3/2$ coherent-state loop gives $|\gamma_{\rm id}|/\pi=0.402$. The same-point Trotter results are therefore the direct circuit-level benchmarks for experiment; their convergence with increasing sampling separates path-discretization effects, while the residual difference between the dense Trotter limit and the exact value reflects the finite Trotter approximation. The twelve-point result agrees with the noiseless circuit benchmark at the same sampled longitudes. The eight-point result is lower by $0.018\pi$, a deviation larger than the reported statistical uncertainty and not explained by path discretization alone. The reconstructed density trajectory resolves the programmed quasihole motion, while the principal-eigenvector Wilson product provides an effective geometric-phase diagnostic of the reconstructed path.


\textit{Conclusion}---We have experimentally characterized the
many-body coherence and quasihole response of a finite
fermionic $\nu=1/3$ Laughlin state using superconducting
qubits.
Read's composite-boson construction reveals vortex-dressed
coherence that remains appreciable over a larger separation
range than fermionic one-body coherence.
The corresponding one-flux-added quasihole exhibits a
fractional charge response consistent with finite-size
benchmarks, while controlled motion provides an effective geometric-phase diagnostic
of the reconstructed dominant-eigenstate trajectory.
Together, these results provide complementary probes of
the coherent structure and fractionalized response of
a finite Laughlin state.
Extending these measurements to larger systems and multiple
quasiholes would enable studies of the finite-size scaling
of vortex-dressed coherence and its dependence on quasihole
configuration.


\begin{acknowledgments}
This work was supported by Ministry of Science and Technology project (Grant No. 2025YFE0217600), the National Natural Science
Foundation of China (NSFC) (Grant Nos.~12404578, 92265207, T2121001, 12122504, T2322030, 92365301, 92565301, 12447184, and 12375007), the National Key Research and Development Program of China (Grant No.~2021YFA1402001),
the Innovation Program for Quantum Science and Technology
(Grant No.~2021ZD0301800), the Beijing Nova Program
(Grant Nos.~20220484121 and 2022000216), the Young Elite Scientists Sponsorship Program of the Beijing High Innovation Plan (Grant No.~20250945), the Beijing Natural Science Foundation (Grant No.~1262048), the Beijing National Laboratory for Condensed Matter Physics (Grant No.~2024BNLCMPKF022), and the China Postdoctoral Science Foundation (Grant Nos.~GZB20240815 and GZC20252227).
\end{acknowledgments}



\begin{thebibliography}{54}%
\makeatletter
\providecommand \@ifxundefined [1]{%
 \@ifx{#1\undefined}
}%
\providecommand \@ifnum [1]{%
 \ifnum #1\expandafter \@firstoftwo
 \else \expandafter \@secondoftwo
 \fi
}%
\providecommand \@ifx [1]{%
 \ifx #1\expandafter \@firstoftwo
 \else \expandafter \@secondoftwo
 \fi
}%
\providecommand \natexlab [1]{#1}%
\providecommand \enquote  [1]{``#1''}%
\providecommand \bibnamefont  [1]{#1}%
\providecommand \bibfnamefont [1]{#1}%
\providecommand \citenamefont [1]{#1}%
\providecommand \href@noop [0]{\@secondoftwo}%
\providecommand \href [0]{\begingroup \@sanitize@url \@href}%
\providecommand \@href[1]{\@@startlink{#1}\@@href}%
\providecommand \@@href[1]{\endgroup#1\@@endlink}%
\providecommand \@sanitize@url [0]{\catcode `\\12\catcode `\$12\catcode
  `\&12\catcode `\#12\catcode `\^12\catcode `\_12\catcode `\%12\relax}%
\providecommand \@@startlink[1]{}%
\providecommand \@@endlink[0]{}%
\providecommand \url  [0]{\begingroup\@sanitize@url \@url }%
\providecommand \@url [1]{\endgroup\@href {#1}{\urlprefix }}%
\providecommand \urlprefix  [0]{URL }%
\providecommand \Eprint [0]{\href }%
\providecommand \doibase [0]{https://doi.org/}%
\providecommand \selectlanguage [0]{\@gobble}%
\providecommand \bibinfo  [0]{\@secondoftwo}%
\providecommand \bibfield  [0]{\@secondoftwo}%
\providecommand \translation [1]{[#1]}%
\providecommand \BibitemOpen [0]{}%
\providecommand \bibitemStop [0]{}%
\providecommand \bibitemNoStop [0]{.\EOS\space}%
\providecommand \EOS [0]{\spacefactor3000\relax}%
\providecommand \BibitemShut  [1]{\csname bibitem#1\endcsname}%
\let\auto@bib@innerbib\@empty
\bibitem [{\citenamefont {Stormer}\ \emph {et~al.}(1999)\citenamefont
  {Stormer}, \citenamefont {Tsui},\ and\ \citenamefont
  {Gossard}}]{Stormer1999RMPFQH}%
  \BibitemOpen
  \bibfield  {author} {\bibinfo {author} {\bibfnamefont {H.~L.}\ \bibnamefont
  {Stormer}}, \bibinfo {author} {\bibfnamefont {D.~C.}\ \bibnamefont {Tsui}},\
  and\ \bibinfo {author} {\bibfnamefont {A.~C.}\ \bibnamefont {Gossard}},\
  }\bibfield  {title} {\bibinfo {title} {The fractional quantum {Hall}
  effect},\ }\href {https://doi.org/10.1103/RevModPhys.71.S298} {\bibfield
  {journal} {\bibinfo  {journal} {Rev. Mod. Phys.}\ }\textbf {\bibinfo {volume}
  {71}},\ \bibinfo {pages} {S298} (\bibinfo {year} {1999})}\BibitemShut
  {NoStop}%
\bibitem [{\citenamefont {Wen}(1995)}]{Wen1995AdvPhysTopologicalOrder}%
  \BibitemOpen
  \bibfield  {author} {\bibinfo {author} {\bibfnamefont {X.-G.}\ \bibnamefont
  {Wen}},\ }\bibfield  {title} {\bibinfo {title} {Topological orders and edge
  excitations in fractional quantum {Hall} states},\ }\href
  {https://doi.org/10.1080/00018739500101566} {\bibfield  {journal} {\bibinfo
  {journal} {Adv. Phys.}\ }\textbf {\bibinfo {volume} {44}},\ \bibinfo {pages}
  {405} (\bibinfo {year} {1995})}\BibitemShut {NoStop}%
\bibitem [{\citenamefont {Haldane}(2011)}]{Haldane2011PRLGeometry}%
  \BibitemOpen
  \bibfield  {author} {\bibinfo {author} {\bibfnamefont {F.~D.~M.}\
  \bibnamefont {Haldane}},\ }\bibfield  {title} {\bibinfo {title} {Geometrical
  description of the fractional quantum {Hall} effect},\ }\href
  {https://doi.org/10.1103/PhysRevLett.107.116801} {\bibfield  {journal}
  {\bibinfo  {journal} {Phys. Rev. Lett.}\ }\textbf {\bibinfo {volume} {107}},\
  \bibinfo {pages} {116801} (\bibinfo {year} {2011})}\BibitemShut {NoStop}%
\bibitem [{\citenamefont {Arovas}\ \emph {et~al.}(1984)\citenamefont {Arovas},
  \citenamefont {Schrieffer},\ and\ \citenamefont
  {Wilczek}}]{Arovas1984PRLFractionalStatistics}%
  \BibitemOpen
  \bibfield  {author} {\bibinfo {author} {\bibfnamefont {D.}~\bibnamefont
  {Arovas}}, \bibinfo {author} {\bibfnamefont {J.~R.}\ \bibnamefont
  {Schrieffer}},\ and\ \bibinfo {author} {\bibfnamefont {F.}~\bibnamefont
  {Wilczek}},\ }\bibfield  {title} {\bibinfo {title} {Fractional statistics and
  the quantum {Hall} effect},\ }\href
  {https://doi.org/10.1103/PhysRevLett.53.722} {\bibfield  {journal} {\bibinfo
  {journal} {Phys. Rev. Lett.}\ }\textbf {\bibinfo {volume} {53}},\ \bibinfo
  {pages} {722} (\bibinfo {year} {1984})}\BibitemShut {NoStop}%
\bibitem [{\citenamefont {Nayak}\ \emph {et~al.}(2008)\citenamefont {Nayak},
  \citenamefont {Simon}, \citenamefont {Stern}, \citenamefont {Freedman},\ and\
  \citenamefont {Das~Sarma}}]{Nayak2008RMPNonAbelianAnyons}%
  \BibitemOpen
  \bibfield  {author} {\bibinfo {author} {\bibfnamefont {C.}~\bibnamefont
  {Nayak}}, \bibinfo {author} {\bibfnamefont {S.~H.}\ \bibnamefont {Simon}},
  \bibinfo {author} {\bibfnamefont {A.}~\bibnamefont {Stern}}, \bibinfo
  {author} {\bibfnamefont {M.}~\bibnamefont {Freedman}},\ and\ \bibinfo
  {author} {\bibfnamefont {S.}~\bibnamefont {Das~Sarma}},\ }\bibfield  {title}
  {\bibinfo {title} {{Non-Abelian} anyons and topological quantum
  computation},\ }\href {https://doi.org/10.1103/RevModPhys.80.1083} {\bibfield
   {journal} {\bibinfo  {journal} {Rev. Mod. Phys.}\ }\textbf {\bibinfo
  {volume} {80}},\ \bibinfo {pages} {1083} (\bibinfo {year}
  {2008})}\BibitemShut {NoStop}%
\bibitem [{\citenamefont {Lapa}\ and\ \citenamefont
  {Hughes}(2018)}]{Lapa2018PRBHallViscositygeometricresponse}%
  \BibitemOpen
  \bibfield  {author} {\bibinfo {author} {\bibfnamefont {M.~F.}\ \bibnamefont
  {Lapa}}\ and\ \bibinfo {author} {\bibfnamefont {T.~L.}\ \bibnamefont
  {Hughes}},\ }\bibfield  {title} {\bibinfo {title} {Hall viscosity and
  geometric response in the {Chern-Simons} matrix model of the {Laughlin}
  states},\ }\href {https://doi.org/10.1103/PhysRevB.97.205122} {\bibfield
  {journal} {\bibinfo  {journal} {Phys. Rev. B}\ }\textbf {\bibinfo {volume}
  {97}},\ \bibinfo {pages} {205122} (\bibinfo {year} {2018})}\BibitemShut
  {NoStop}%
\bibitem [{\citenamefont {Laughlin}(1983)}]{Laughlin1983PRLFractionalCharge}%
  \BibitemOpen
  \bibfield  {author} {\bibinfo {author} {\bibfnamefont {R.~B.}\ \bibnamefont
  {Laughlin}},\ }\bibfield  {title} {\bibinfo {title} {Anomalous quantum {Hall}
  effect: An incompressible quantum fluid with fractionally charged
  excitations},\ }\href {https://doi.org/10.1103/PhysRevLett.50.1395}
  {\bibfield  {journal} {\bibinfo  {journal} {Phys. Rev. Lett.}\ }\textbf
  {\bibinfo {volume} {50}},\ \bibinfo {pages} {1395} (\bibinfo {year}
  {1983})}\BibitemShut {NoStop}%
\bibitem [{\citenamefont {Haldane}(1983)}]{Haldane1983PRLHierarchy}%
  \BibitemOpen
  \bibfield  {author} {\bibinfo {author} {\bibfnamefont {F.~D.~M.}\
  \bibnamefont {Haldane}},\ }\bibfield  {title} {\bibinfo {title} {Fractional
  quantization of the {Hall} effect: A hierarchy of incompressible quantum
  fluid states},\ }\href {https://doi.org/10.1103/PhysRevLett.51.605}
  {\bibfield  {journal} {\bibinfo  {journal} {Phys. Rev. Lett.}\ }\textbf
  {\bibinfo {volume} {51}},\ \bibinfo {pages} {605} (\bibinfo {year}
  {1983})}\BibitemShut {NoStop}%
\bibitem [{\citenamefont {Read}(1989)}]{Read1989PRLCompositeBoson}%
  \BibitemOpen
  \bibfield  {author} {\bibinfo {author} {\bibfnamefont {N.}~\bibnamefont
  {Read}},\ }\bibfield  {title} {\bibinfo {title} {Order parameter and
  {Ginzburg-Landau} theory for the fractional quantum {Hall} effect},\ }\href
  {https://doi.org/10.1103/PhysRevLett.62.86} {\bibfield  {journal} {\bibinfo
  {journal} {Phys. Rev. Lett.}\ }\textbf {\bibinfo {volume} {62}},\ \bibinfo
  {pages} {86} (\bibinfo {year} {1989})}\BibitemShut {NoStop}%
\bibitem [{\citenamefont {L{\'e}onard}\ \emph {et~al.}(2023)\citenamefont
  {L{\'e}onard}, \citenamefont {Kim}, \citenamefont {Kwan}, \citenamefont
  {Segura}, \citenamefont {Grusdt}, \citenamefont {Repellin}, \citenamefont
  {Goldman},\ and\ \citenamefont {Greiner}}]{Leonard2023NatureUltracoldFQH}%
  \BibitemOpen
  \bibfield  {author} {\bibinfo {author} {\bibfnamefont {J.}~\bibnamefont
  {L{\'e}onard}}, \bibinfo {author} {\bibfnamefont {S.}~\bibnamefont {Kim}},
  \bibinfo {author} {\bibfnamefont {J.}~\bibnamefont {Kwan}}, \bibinfo {author}
  {\bibfnamefont {P.}~\bibnamefont {Segura}}, \bibinfo {author} {\bibfnamefont
  {F.}~\bibnamefont {Grusdt}}, \bibinfo {author} {\bibfnamefont
  {C.}~\bibnamefont {Repellin}}, \bibinfo {author} {\bibfnamefont
  {N.}~\bibnamefont {Goldman}},\ and\ \bibinfo {author} {\bibfnamefont
  {M.}~\bibnamefont {Greiner}},\ }\bibfield  {title} {\bibinfo {title}
  {Realization of a fractional quantum {Hall} state with ultracold atoms},\
  }\href {https://doi.org/10.1038/s41586-023-06122-4} {\bibfield  {journal}
  {\bibinfo  {journal} {Nature}\ }\textbf {\bibinfo {volume} {619}},\ \bibinfo
  {pages} {495} (\bibinfo {year} {2023})}\BibitemShut {NoStop}%
\bibitem [{\citenamefont {Clark}\ \emph {et~al.}(2020)\citenamefont {Clark},
  \citenamefont {Schine}, \citenamefont {Baum}, \citenamefont {Jia},\ and\
  \citenamefont {Simon}}]{Clark2020NaturePhotonicLaughlin}%
  \BibitemOpen
  \bibfield  {author} {\bibinfo {author} {\bibfnamefont {L.~W.}\ \bibnamefont
  {Clark}}, \bibinfo {author} {\bibfnamefont {N.}~\bibnamefont {Schine}},
  \bibinfo {author} {\bibfnamefont {C.}~\bibnamefont {Baum}}, \bibinfo {author}
  {\bibfnamefont {N.}~\bibnamefont {Jia}},\ and\ \bibinfo {author}
  {\bibfnamefont {J.}~\bibnamefont {Simon}},\ }\bibfield  {title} {\bibinfo
  {title} {Observation of {Laughlin} states made of light},\ }\href
  {https://doi.org/10.1038/s41586-020-2318-5} {\bibfield  {journal} {\bibinfo
  {journal} {Nature}\ }\textbf {\bibinfo {volume} {582}},\ \bibinfo {pages}
  {41} (\bibinfo {year} {2020})}\BibitemShut {NoStop}%
\bibitem [{\citenamefont {Wang}\ \emph {et~al.}(2024)\citenamefont {Wang},
  \citenamefont {Liu}, \citenamefont {Chen}, \citenamefont {Chen},
  \citenamefont {Zhao}, \citenamefont {Ying}, \citenamefont {Shang},
  \citenamefont {Wang}, \citenamefont {Huo}, \citenamefont {Peng},
  \citenamefont {Zhu}, \citenamefont {Lu},\ and\ \citenamefont
  {Pan}}]{Wang2024SciencePhotonicFQH}%
  \BibitemOpen
  \bibfield  {author} {\bibinfo {author} {\bibfnamefont {C.}~\bibnamefont
  {Wang}}, \bibinfo {author} {\bibfnamefont {F.-M.}\ \bibnamefont {Liu}},
  \bibinfo {author} {\bibfnamefont {M.-C.}\ \bibnamefont {Chen}}, \bibinfo
  {author} {\bibfnamefont {H.}~\bibnamefont {Chen}}, \bibinfo {author}
  {\bibfnamefont {X.-H.}\ \bibnamefont {Zhao}}, \bibinfo {author}
  {\bibfnamefont {C.}~\bibnamefont {Ying}}, \bibinfo {author} {\bibfnamefont
  {Z.-X.}\ \bibnamefont {Shang}}, \bibinfo {author} {\bibfnamefont {J.-W.}\
  \bibnamefont {Wang}}, \bibinfo {author} {\bibfnamefont {Y.-H.}\ \bibnamefont
  {Huo}}, \bibinfo {author} {\bibfnamefont {C.-Z.}\ \bibnamefont {Peng}},
  \bibinfo {author} {\bibfnamefont {X.}~\bibnamefont {Zhu}}, \bibinfo {author}
  {\bibfnamefont {C.-Y.}\ \bibnamefont {Lu}},\ and\ \bibinfo {author}
  {\bibfnamefont {J.-W.}\ \bibnamefont {Pan}},\ }\bibfield  {title} {\bibinfo
  {title} {Realization of fractional quantum {Hall} state with interacting
  photons},\ }\href {https://doi.org/10.1126/science.ado3912} {\bibfield
  {journal} {\bibinfo  {journal} {Science}\ }\textbf {\bibinfo {volume}
  {384}},\ \bibinfo {pages} {579} (\bibinfo {year} {2024})}\BibitemShut
  {NoStop}%
\bibitem [{\citenamefont {Xiang}\ \emph {et~al.}(2023)\citenamefont {Xiang},
  \citenamefont {Huang}, \citenamefont {Zhang} \emph
  {et~al.}}]{Xiang2023Chern}%
  \BibitemOpen
  \bibfield  {author} {\bibinfo {author} {\bibfnamefont {Z.-C.}\ \bibnamefont
  {Xiang}}, \bibinfo {author} {\bibfnamefont {K.}~\bibnamefont {Huang}},
  \bibinfo {author} {\bibfnamefont {Y.-R.}\ \bibnamefont {Zhang}}, \emph
  {et~al.},\ }\bibfield  {title} {\bibinfo {title} {{Simulating Chern
  insulators on a superconducting quantum processor}},\ }\href
  {https://doi.org/10.1038/s41467-023-41230-9} {\bibfield  {journal} {\bibinfo
  {journal} {Nat. Commun.}\ }\textbf {\bibinfo {volume} {14}},\ \bibinfo
  {pages} {5433} (\bibinfo {year} {2023})}\BibitemShut {NoStop}%
\bibitem [{\citenamefont {Jin}\ \emph {et~al.}(2025)\citenamefont {Jin} \emph
  {et~al.}}]{Jin2025ZeroModes}%
  \BibitemOpen
  \bibfield  {author} {\bibinfo {author} {\bibfnamefont {F.}~\bibnamefont
  {Jin}} \emph {et~al.},\ }\bibfield  {title} {\bibinfo {title} {{Topological
  prethermal strong zero modes on superconducting processors}},\ }\href
  {https://doi.org/10.1038/s41586-025-09476-z} {\bibfield  {journal} {\bibinfo
  {journal} {Nature}\ }\textbf {\bibinfo {volume} {645}},\ \bibinfo {pages}
  {626} (\bibinfo {year} {2025})}\BibitemShut {NoStop}%
\bibitem [{\citenamefont {Qian}\ \emph {et~al.}(2025)\citenamefont {Qian},
  \citenamefont {Gong}, \citenamefont {Zhang} \emph
  {et~al.}}]{Qian2025HigherOrder}%
  \BibitemOpen
  \bibfield  {author} {\bibinfo {author} {\bibfnamefont {H.}~\bibnamefont
  {Qian}}, \bibinfo {author} {\bibfnamefont {M.}~\bibnamefont {Gong}}, \bibinfo
  {author} {\bibfnamefont {J.}~\bibnamefont {Zhang}}, \emph {et~al.},\
  }\bibfield  {title} {\bibinfo {title} {{Programmable higher-order
  nonequilibrium topological phases on a superconducting quantum processor}},\
  }\href {https://doi.org/10.1126/science.adp6802} {\bibfield  {journal}
  {\bibinfo  {journal} {Science}\ }\textbf {\bibinfo {volume} {390}},\ \bibinfo
  {pages} {930} (\bibinfo {year} {2025})}\BibitemShut {NoStop}%
\bibitem [{\citenamefont {Zhou}\ \emph {et~al.}(2025)\citenamefont {Zhou},
  \citenamefont {Xu}, \citenamefont {Su} \emph {et~al.}}]{Zhou2025Qutrit}%
  \BibitemOpen
  \bibfield  {author} {\bibinfo {author} {\bibfnamefont {K.}~\bibnamefont
  {Zhou}}, \bibinfo {author} {\bibfnamefont {J.-W.}\ \bibnamefont {Xu}},
  \bibinfo {author} {\bibfnamefont {Q.-P.}\ \bibnamefont {Su}}, \emph
  {et~al.},\ }\bibfield  {title} {\bibinfo {title} {{Demonstration of
  Discrete-Time Quantum Walks and Observation of Topological Edge States in a
  Superconducting Qutrit Chain}},\ }\href {https://doi.org/10.1103/5lm1-2kpk}
  {\bibfield  {journal} {\bibinfo  {journal} {Phys. Rev. Lett.}\ }\textbf
  {\bibinfo {volume} {135}},\ \bibinfo {pages} {250601} (\bibinfo {year}
  {2025})}\BibitemShut {NoStop}%
\bibitem [{\citenamefont {Koh}\ \emph {et~al.}(2024)\citenamefont {Koh},
  \citenamefont {Tai},\ and\ \citenamefont {Lee}}]{Koh2024HigherOrder}%
  \BibitemOpen
  \bibfield  {author} {\bibinfo {author} {\bibfnamefont {J.~M.}\ \bibnamefont
  {Koh}}, \bibinfo {author} {\bibfnamefont {T.}~\bibnamefont {Tai}},\ and\
  \bibinfo {author} {\bibfnamefont {C.~H.}\ \bibnamefont {Lee}},\ }\bibfield
  {title} {\bibinfo {title} {{Realization of higher-order topological lattices
  on a quantum computer}},\ }\href {https://doi.org/10.1038/s41467-024-49648-5}
  {\bibfield  {journal} {\bibinfo  {journal} {Nat. Commun.}\ }\textbf {\bibinfo
  {volume} {15}},\ \bibinfo {pages} {5807} (\bibinfo {year}
  {2024})}\BibitemShut {NoStop}%
\bibitem [{\citenamefont {Will}\ \emph {et~al.}(2025)\citenamefont {Will},
  \citenamefont {Cochran} \emph {et~al.}}]{Will2025Floquet}%
  \BibitemOpen
  \bibfield  {author} {\bibinfo {author} {\bibfnamefont {M.}~\bibnamefont
  {Will}}, \bibinfo {author} {\bibfnamefont {T.~A.}\ \bibnamefont {Cochran}},
  \emph {et~al.},\ }\bibfield  {title} {\bibinfo {title} {{Probing
  non-equilibrium topological order on a quantum processor}},\ }\href
  {https://doi.org/10.1038/s41586-025-09456-3} {\bibfield  {journal} {\bibinfo
  {journal} {Nature}\ }\textbf {\bibinfo {volume} {645}},\ \bibinfo {pages}
  {348} (\bibinfo {year} {2025})}\BibitemShut {NoStop}%
\bibitem [{\citenamefont {Koh}\ \emph {et~al.}(2026)\citenamefont {Koh},
  \citenamefont {Xue}, \citenamefont {Tai}, \citenamefont {Koh},\ and\
  \citenamefont {Lee}}]{Koh2026Edge}%
  \BibitemOpen
  \bibfield  {author} {\bibinfo {author} {\bibfnamefont {J.~M.}\ \bibnamefont
  {Koh}}, \bibinfo {author} {\bibfnamefont {W.-T.}\ \bibnamefont {Xue}},
  \bibinfo {author} {\bibfnamefont {T.}~\bibnamefont {Tai}}, \bibinfo {author}
  {\bibfnamefont {D.~E.}\ \bibnamefont {Koh}},\ and\ \bibinfo {author}
  {\bibfnamefont {C.~H.}\ \bibnamefont {Lee}},\ }\bibfield  {title} {\bibinfo
  {title} {{Interacting non-Hermitian edge and cluster bursts on a digital
  quantum processor}},\ }\href {https://doi.org/10.1038/s41467-026-75904-x}
  {\bibfield  {journal} {\bibinfo  {journal} {Nat. Commun.}\ }\textbf {\bibinfo
  {volume} {17}},\ \bibinfo {pages} {9120} (\bibinfo {year}
  {2026})}\BibitemShut {NoStop}%
\bibitem [{\citenamefont {Han}\ \emph {et~al.}(2024{\natexlab{a}})\citenamefont
  {Han}, \citenamefont {Ning}, \citenamefont {Huang} \emph
  {et~al.}}]{Han2024Exceptional}%
  \BibitemOpen
  \bibfield  {author} {\bibinfo {author} {\bibfnamefont {P.-R.}\ \bibnamefont
  {Han}}, \bibinfo {author} {\bibfnamefont {W.}~\bibnamefont {Ning}}, \bibinfo
  {author} {\bibfnamefont {X.-J.}\ \bibnamefont {Huang}}, \emph {et~al.},\
  }\bibfield  {title} {\bibinfo {title} {{Measuring topological invariants for
  higher-order exceptional points in quantum three-mode systems}},\ }\href
  {https://doi.org/10.1038/s41467-024-54662-8} {\bibfield  {journal} {\bibinfo
  {journal} {Nat. Commun.}\ }\textbf {\bibinfo {volume} {15}},\ \bibinfo
  {pages} {10293} (\bibinfo {year} {2024}{\natexlab{a}})}\BibitemShut {NoStop}%
\bibitem [{\citenamefont {Wang}\ \emph
  {et~al.}(2025{\natexlab{a}})\citenamefont {Wang}, \citenamefont {Li},
  \citenamefont {Xu} \emph {et~al.}}]{Wang2025PRXBellCorrelation}%
  \BibitemOpen
  \bibfield  {author} {\bibinfo {author} {\bibfnamefont {K.}~\bibnamefont
  {Wang}}, \bibinfo {author} {\bibfnamefont {W.}~\bibnamefont {Li}}, \bibinfo
  {author} {\bibfnamefont {S.}~\bibnamefont {Xu}}, \emph {et~al.},\ }\bibfield
  {title} {\bibinfo {title} {Probing many-body {Bell} correlation depth with
  superconducting qubits},\ }\href {https://doi.org/10.1103/PhysRevX.15.021024}
  {\bibfield  {journal} {\bibinfo  {journal} {Phys. Rev. X}\ }\textbf {\bibinfo
  {volume} {15}},\ \bibinfo {pages} {021024} (\bibinfo {year}
  {2025}{\natexlab{a}})}\BibitemShut {NoStop}%
\bibitem [{\citenamefont {Tan}\ \emph {et~al.}(2026)\citenamefont {Tan},
  \citenamefont {Wang}, \citenamefont {Yang} \emph {et~al.}}]{Tan2026Critical}%
  \BibitemOpen
  \bibfield  {author} {\bibinfo {author} {\bibfnamefont {Z.}~\bibnamefont
  {Tan}}, \bibinfo {author} {\bibfnamefont {K.}~\bibnamefont {Wang}}, \bibinfo
  {author} {\bibfnamefont {S.}~\bibnamefont {Yang}}, \emph {et~al.},\
  }\bibfield  {title} {\bibinfo {title} {{Exploring nontrivial topology at
  quantum criticality on a superconducting processor}},\ }\href
  {https://doi.org/10.1038/s42005-026-02569-9} {\bibfield  {journal} {\bibinfo
  {journal} {Commun. Phys.}\ }\textbf {\bibinfo {volume} {9}},\ \bibinfo
  {pages} {136} (\bibinfo {year} {2026})}\BibitemShut {NoStop}%
\bibitem [{\citenamefont {Zheng}\ \emph {et~al.}(2023)\citenamefont {Zheng},
  \citenamefont {Ning}, \citenamefont {Chen} \emph {et~al.}}]{Zheng2023Cat}%
  \BibitemOpen
  \bibfield  {author} {\bibinfo {author} {\bibfnamefont {R.-H.}\ \bibnamefont
  {Zheng}}, \bibinfo {author} {\bibfnamefont {W.}~\bibnamefont {Ning}},
  \bibinfo {author} {\bibfnamefont {Y.-H.}\ \bibnamefont {Chen}}, \emph
  {et~al.},\ }\bibfield  {title} {\bibinfo {title} {{Observation of a
  Superradiant Phase Transition with Emergent Cat States}},\ }\href
  {https://doi.org/10.1103/PhysRevLett.131.113601} {\bibfield  {journal}
  {\bibinfo  {journal} {Phys. Rev. Lett.}\ }\textbf {\bibinfo {volume} {131}},\
  \bibinfo {pages} {113601} (\bibinfo {year} {2023})}\BibitemShut {NoStop}%
\bibitem [{\citenamefont {Han}\ \emph {et~al.}(2024{\natexlab{b}})\citenamefont
  {Han}, \citenamefont {Lyu}, \citenamefont {Zhou} \emph
  {et~al.}}]{Han2024Spectroscopy}%
  \BibitemOpen
  \bibfield  {author} {\bibinfo {author} {\bibfnamefont {Z.}~\bibnamefont
  {Han}}, \bibinfo {author} {\bibfnamefont {C.}~\bibnamefont {Lyu}}, \bibinfo
  {author} {\bibfnamefont {Y.}~\bibnamefont {Zhou}}, \emph {et~al.},\
  }\bibfield  {title} {\bibinfo {title} {{Multilevel variational spectroscopy
  using a programmable quantum simulator}},\ }\href
  {https://doi.org/10.1103/PhysRevResearch.6.013015} {\bibfield  {journal}
  {\bibinfo  {journal} {Phys. Rev. Research}\ }\textbf {\bibinfo {volume}
  {6}},\ \bibinfo {pages} {013015} (\bibinfo {year}
  {2024}{\natexlab{b}})}\BibitemShut {NoStop}%
\bibitem [{\citenamefont {Li}\ \emph {et~al.}(2026)\citenamefont {Li},
  \citenamefont {Yang}, \citenamefont {Shi} \emph
  {et~al.}}]{Li2026NonEquilibriumSensing}%
  \BibitemOpen
  \bibfield  {author} {\bibinfo {author} {\bibfnamefont {H.}~\bibnamefont
  {Li}}, \bibinfo {author} {\bibfnamefont {Y.}~\bibnamefont {Yang}}, \bibinfo
  {author} {\bibfnamefont {Y.-H.}\ \bibnamefont {Shi}}, \emph {et~al.},\
  }\bibfield  {title} {\bibinfo {title} {{Non-equilibrium criticality-enhanced
  quantum sensing with superconducting qubits}},\ }\href
  {https://doi.org/10.1016/j.scib.2026.05.023} {\bibfield  {journal} {\bibinfo
  {journal} {Sci. Bull.}\ }\textbf {\bibinfo {volume} {71}},\ \bibinfo {pages}
  {2996} (\bibinfo {year} {2026})}\BibitemShut {NoStop}%
\bibitem [{\citenamefont {Yan}\ \emph {et~al.}(2026)\citenamefont {Yan},
  \citenamefont {Ge}, \citenamefont {Li}, \citenamefont {Zhang}, \citenamefont
  {Nori},\ and\ \citenamefont {Nakamura}}]{Yan2026Projected}%
  \BibitemOpen
  \bibfield  {author} {\bibinfo {author} {\bibfnamefont {Z.}~\bibnamefont
  {Yan}}, \bibinfo {author} {\bibfnamefont {Z.-Y.}\ \bibnamefont {Ge}},
  \bibinfo {author} {\bibfnamefont {R.}~\bibnamefont {Li}}, \bibinfo {author}
  {\bibfnamefont {Y.-R.}\ \bibnamefont {Zhang}}, \bibinfo {author}
  {\bibfnamefont {F.}~\bibnamefont {Nori}},\ and\ \bibinfo {author}
  {\bibfnamefont {Y.}~\bibnamefont {Nakamura}},\ }\bibfield  {title} {\bibinfo
  {title} {{Characterizing many-body dynamics with projected ensembles on a
  superconducting quantum processor}},\ }\href
  {https://doi.org/10.1126/sciadv.aeb8213} {\bibfield  {journal} {\bibinfo
  {journal} {Sci. Adv.}\ }\textbf {\bibinfo {volume} {12}},\ \bibinfo {pages}
  {eaeb8213} (\bibinfo {year} {2026})}\BibitemShut {NoStop}%
\bibitem [{\citenamefont {Xu}\ \emph {et~al.}(2026{\natexlab{a}})\citenamefont
  {Xu}, \citenamefont {Zhou}, \citenamefont {Sun} \emph
  {et~al.}}]{Xu2026Geometric}%
  \BibitemOpen
  \bibfield  {author} {\bibinfo {author} {\bibfnamefont {Y.}~\bibnamefont
  {Xu}}, \bibinfo {author} {\bibfnamefont {Y.}~\bibnamefont {Zhou}}, \bibinfo
  {author} {\bibfnamefont {L.}~\bibnamefont {Sun}}, \emph {et~al.},\ }\bibfield
   {title} {\bibinfo {title} {{High-Fidelity Controlled-Phase Gate for Binomial
  Codes via Geometric Phase Engineering}},\ }\href
  {https://doi.org/10.1103/yv9x-1sqp} {\bibfield  {journal} {\bibinfo
  {journal} {Phys. Rev. Lett.}\ }\textbf {\bibinfo {volume} {136}},\ \bibinfo
  {pages} {210602} (\bibinfo {year} {2026}{\natexlab{a}})}\BibitemShut
  {NoStop}%
\bibitem [{\citenamefont {{Google Quantum AI and
  Collaborators}}(2023)}]{Andersen2023Braiding}%
  \BibitemOpen
  \bibfield  {author} {\bibinfo {author} {\bibnamefont {{Google Quantum AI and
  Collaborators}}},\ }\bibfield  {title} {\bibinfo {title} {{Non-Abelian
  braiding of graph vertices in a superconducting processor}},\ }\href
  {https://doi.org/10.1038/s41586-023-05954-4} {\bibfield  {journal} {\bibinfo
  {journal} {Nature}\ }\textbf {\bibinfo {volume} {618}},\ \bibinfo {pages}
  {264} (\bibinfo {year} {2023})}\BibitemShut {NoStop}%
\bibitem [{\citenamefont {Xu}\ \emph {et~al.}(2024)\citenamefont {Xu},
  \citenamefont {Sun}, \citenamefont {Wang} \emph {et~al.}}]{Xu2024Fibonacci}%
  \BibitemOpen
  \bibfield  {author} {\bibinfo {author} {\bibfnamefont {S.}~\bibnamefont
  {Xu}}, \bibinfo {author} {\bibfnamefont {Z.-Z.}\ \bibnamefont {Sun}},
  \bibinfo {author} {\bibfnamefont {K.}~\bibnamefont {Wang}}, \emph {et~al.},\
  }\bibfield  {title} {\bibinfo {title} {{Non-Abelian braiding of Fibonacci
  anyons with a superconducting processor}},\ }\href
  {https://doi.org/10.1038/s41567-024-02529-6} {\bibfield  {journal} {\bibinfo
  {journal} {Nat. Phys.}\ }\textbf {\bibinfo {volume} {20}},\ \bibinfo {pages}
  {1469} (\bibinfo {year} {2024})}\BibitemShut {NoStop}%
\bibitem [{\citenamefont {Niu}\ \emph {et~al.}(2024)\citenamefont {Niu},
  \citenamefont {Li}, \citenamefont {Zhang} \emph {et~al.}}]{Niu2024Braiding}%
  \BibitemOpen
  \bibfield  {author} {\bibinfo {author} {\bibfnamefont {J.}~\bibnamefont
  {Niu}}, \bibinfo {author} {\bibfnamefont {Y.}~\bibnamefont {Li}}, \bibinfo
  {author} {\bibfnamefont {L.}~\bibnamefont {Zhang}}, \emph {et~al.},\
  }\bibfield  {title} {\bibinfo {title} {{Demonstrating Path-Independent
  Anyonic Braiding on a Modular Superconducting Quantum Processor}},\ }\href
  {https://doi.org/10.1103/PhysRevLett.132.020601} {\bibfield  {journal}
  {\bibinfo  {journal} {Phys. Rev. Lett.}\ }\textbf {\bibinfo {volume} {132}},\
  \bibinfo {pages} {020601} (\bibinfo {year} {2024})}\BibitemShut {NoStop}%
\bibitem [{\citenamefont {Nakamura}\ \emph {et~al.}(2012)\citenamefont
  {Nakamura}, \citenamefont {Wang},\ and\ \citenamefont
  {Bergholtz}}]{Nakamura2012PPLFermionChain}%
  \BibitemOpen
  \bibfield  {author} {\bibinfo {author} {\bibfnamefont {M.}~\bibnamefont
  {Nakamura}}, \bibinfo {author} {\bibfnamefont {Z.-Y.}\ \bibnamefont {Wang}},\
  and\ \bibinfo {author} {\bibfnamefont {E.~J.}\ \bibnamefont {Bergholtz}},\
  }\bibfield  {title} {\bibinfo {title} {Exactly solvable fermion chain
  describing a $\ensuremath{\nu}=1/3$ fractional quantum {Hall} state},\ }\href
  {https://doi.org/10.1103/PhysRevLett.109.016401} {\bibfield  {journal}
  {\bibinfo  {journal} {Phys. Rev. Lett.}\ }\textbf {\bibinfo {volume} {109}},\
  \bibinfo {pages} {016401} (\bibinfo {year} {2012})}\BibitemShut {NoStop}%
\bibitem [{\citenamefont {Rahmani}\ \emph {et~al.}(2020)\citenamefont
  {Rahmani}, \citenamefont {Sung}, \citenamefont {Putterman}, \citenamefont
  {Roushan}, \citenamefont {Ghaemi},\ and\ \citenamefont
  {Jiang}}]{Rahmani2020PRXLinearDepthLaughlin}%
  \BibitemOpen
  \bibfield  {author} {\bibinfo {author} {\bibfnamefont {A.}~\bibnamefont
  {Rahmani}}, \bibinfo {author} {\bibfnamefont {K.~J.}\ \bibnamefont {Sung}},
  \bibinfo {author} {\bibfnamefont {H.}~\bibnamefont {Putterman}}, \bibinfo
  {author} {\bibfnamefont {P.}~\bibnamefont {Roushan}}, \bibinfo {author}
  {\bibfnamefont {P.}~\bibnamefont {Ghaemi}},\ and\ \bibinfo {author}
  {\bibfnamefont {Z.}~\bibnamefont {Jiang}},\ }\bibfield  {title} {\bibinfo
  {title} {Creating and manipulating a {Laughlin}-type $\ensuremath{\nu}=1/3$
  fractional quantum {Hall} state on a quantum computer with linear depth
  circuits},\ }\href {https://doi.org/10.1103/PRXQuantum.1.020309} {\bibfield
  {journal} {\bibinfo  {journal} {PRX Quantum}\ }\textbf {\bibinfo {volume}
  {1}},\ \bibinfo {pages} {020309} (\bibinfo {year} {2020})}\BibitemShut
  {NoStop}%
\bibitem [{\citenamefont {Shen}\ \emph {et~al.}(2026)\citenamefont {Shen},
  \citenamefont {Lin}, \citenamefont {Lin}, \citenamefont {Xiao},\ and\
  \citenamefont {Cao}}]{Shen2026NatCommunLaughlin}%
  \BibitemOpen
  \bibfield  {author} {\bibinfo {author} {\bibfnamefont {L.}~\bibnamefont
  {Shen}}, \bibinfo {author} {\bibfnamefont {M.}~\bibnamefont {Lin}}, \bibinfo
  {author} {\bibfnamefont {C.~Y.-Y.}\ \bibnamefont {Lin}}, \bibinfo {author}
  {\bibfnamefont {D.}~\bibnamefont {Xiao}},\ and\ \bibinfo {author}
  {\bibfnamefont {T.}~\bibnamefont {Cao}},\ }\bibfield  {title} {\bibinfo
  {title} {Realization of fermionic {Laughlin} state on a quantum processor},\
  }\href {https://doi.org/10.1038/s41467-026-72769-y} {\bibfield  {journal}
  {\bibinfo  {journal} {Nat. Commun.}\ }\textbf {\bibinfo {volume} {17}},\
  \bibinfo {pages} {4919} (\bibinfo {year} {2026})}\BibitemShut {NoStop}%
\bibitem [{\citenamefont {Exp\'osito}\ \emph {et~al.}(2026)\citenamefont
  {Exp\'osito}, \citenamefont {Aseginolaza}, \citenamefont {Guerrero-Avil\'es},
  \citenamefont {Jornet-Somoza}, \citenamefont {Guinea},\ and\ \citenamefont
  {Borge}}]{Exposito2026arXivVQEFQH}%
  \BibitemOpen
  \bibfield  {author} {\bibinfo {author} {\bibfnamefont {S.~F.}\ \bibnamefont
  {Exp\'osito}}, \bibinfo {author} {\bibfnamefont {U.}~\bibnamefont
  {Aseginolaza}}, \bibinfo {author} {\bibfnamefont {R.}~\bibnamefont
  {Guerrero-Avil\'es}}, \bibinfo {author} {\bibfnamefont {J.}~\bibnamefont
  {Jornet-Somoza}}, \bibinfo {author} {\bibfnamefont {F.}~\bibnamefont
  {Guinea}},\ and\ \bibinfo {author} {\bibfnamefont {J.}~\bibnamefont
  {Borge}},\ }\href@noop {} {\bibinfo {title} {Toward quantum utility in
  correlated topological matter: Variational preparation of fractional quantum
  {Hall} manifolds}} (\bibinfo {year} {2026}),\ \Eprint
  {https://arxiv.org/abs/2607.11380} {arXiv:2607.11380} \BibitemShut {NoStop}%
\bibitem [{\citenamefont {Xu}\ \emph {et~al.}(2026{\natexlab{b}})\citenamefont
  {Xu}, \citenamefont {Lee}, \citenamefont {Tu},\ and\ \citenamefont
  {Zhang}}]{Xu2026arXivFQHFactory}%
  \BibitemOpen
  \bibfield  {author} {\bibinfo {author} {\bibfnamefont {C.}~\bibnamefont
  {Xu}}, \bibinfo {author} {\bibfnamefont {C.~H.}\ \bibnamefont {Lee}},
  \bibinfo {author} {\bibfnamefont {H.-H.}\ \bibnamefont {Tu}},\ and\ \bibinfo
  {author} {\bibfnamefont {Y.}~\bibnamefont {Zhang}},\ }\href@noop {} {\bibinfo
  {title} {A fractional quantum {Hall} factory on quantum processors:
  constant-depth preparation of clustered {Non-Abelian} states}} (\bibinfo
  {year} {2026}{\natexlab{b}}),\ \Eprint {https://arxiv.org/abs/2608.05140}
  {arXiv:2608.05140} \BibitemShut {NoStop}%
\bibitem [{\citenamefont {Kirmani}\ \emph {et~al.}(2022)\citenamefont
  {Kirmani}, \citenamefont {Bull}, \citenamefont {Hou}, \citenamefont
  {Saravanan}, \citenamefont {Saeed}, \citenamefont
  {Papi\ifmmode~\acute{c}\else \'{c}\fi{}}, \citenamefont {Rahmani},\ and\
  \citenamefont {Ghaemi}}]{Kirmani2022PRLGeometricExcitations}%
  \BibitemOpen
  \bibfield  {author} {\bibinfo {author} {\bibfnamefont {A.}~\bibnamefont
  {Kirmani}}, \bibinfo {author} {\bibfnamefont {K.}~\bibnamefont {Bull}},
  \bibinfo {author} {\bibfnamefont {C.-Y.}\ \bibnamefont {Hou}}, \bibinfo
  {author} {\bibfnamefont {V.}~\bibnamefont {Saravanan}}, \bibinfo {author}
  {\bibfnamefont {S.~M.}\ \bibnamefont {Saeed}}, \bibinfo {author}
  {\bibfnamefont {Z.}~\bibnamefont {Papi\ifmmode~\acute{c}\else \'{c}\fi{}}},
  \bibinfo {author} {\bibfnamefont {A.}~\bibnamefont {Rahmani}},\ and\ \bibinfo
  {author} {\bibfnamefont {P.}~\bibnamefont {Ghaemi}},\ }\bibfield  {title}
  {\bibinfo {title} {Probing geometric excitations of fractional quantum {Hall}
  states on quantum computers},\ }\href
  {https://doi.org/10.1103/PhysRevLett.129.056801} {\bibfield  {journal}
  {\bibinfo  {journal} {Phys. Rev. Lett.}\ }\textbf {\bibinfo {volume} {129}},\
  \bibinfo {pages} {056801} (\bibinfo {year} {2022})}\BibitemShut {NoStop}%
\bibitem [{\citenamefont {Kirmani}\ \emph {et~al.}(2026)\citenamefont
  {Kirmani}, \citenamefont {Allocca}, \citenamefont {Zhu}, \citenamefont
  {Rahmani}, \citenamefont {Ganeshan},\ and\ \citenamefont
  {Ghaemi}}]{Kirmani2026PRBHallViscosity}%
  \BibitemOpen
  \bibfield  {author} {\bibinfo {author} {\bibfnamefont {A.}~\bibnamefont
  {Kirmani}}, \bibinfo {author} {\bibfnamefont {A.~A.}\ \bibnamefont
  {Allocca}}, \bibinfo {author} {\bibfnamefont {J.-X.}\ \bibnamefont {Zhu}},
  \bibinfo {author} {\bibfnamefont {A.}~\bibnamefont {Rahmani}}, \bibinfo
  {author} {\bibfnamefont {S.}~\bibnamefont {Ganeshan}},\ and\ \bibinfo
  {author} {\bibfnamefont {P.}~\bibnamefont {Ghaemi}},\ }\bibfield  {title}
  {\bibinfo {title} {{Hall} viscosity of the {Laughlin} state on noisy quantum
  computers},\ }\href {https://doi.org/10.1103/2g6v-fqbr} {\bibfield  {journal}
  {\bibinfo  {journal} {Phys. Rev. B}\ }\textbf {\bibinfo {volume} {114}},\
  \bibinfo {pages} {185121} (\bibinfo {year} {2026})}\BibitemShut {NoStop}%
\bibitem [{\citenamefont {Kirmani}\ \emph {et~al.}(2023)\citenamefont
  {Kirmani}, \citenamefont {Wang}, \citenamefont {Ghaemi},\ and\ \citenamefont
  {Rahmani}}]{Kirmani2023PRBQuasiholeBraiding}%
  \BibitemOpen
  \bibfield  {author} {\bibinfo {author} {\bibfnamefont {A.}~\bibnamefont
  {Kirmani}}, \bibinfo {author} {\bibfnamefont {D.~S.}\ \bibnamefont {Wang}},
  \bibinfo {author} {\bibfnamefont {P.}~\bibnamefont {Ghaemi}},\ and\ \bibinfo
  {author} {\bibfnamefont {A.}~\bibnamefont {Rahmani}},\ }\bibfield  {title}
  {\bibinfo {title} {Braiding fractional quantum {Hall} quasiholes on a
  superconducting quantum processor},\ }\href
  {https://doi.org/10.1103/PhysRevB.108.064303} {\bibfield  {journal} {\bibinfo
   {journal} {Phys. Rev. B}\ }\textbf {\bibinfo {volume} {108}},\ \bibinfo
  {pages} {064303} (\bibinfo {year} {2023})}\BibitemShut {NoStop}%
\bibitem [{\citenamefont {Dutkiewicz}\ \emph {et~al.}(2026)\citenamefont
  {Dutkiewicz}, \citenamefont {Andrade}, \citenamefont {Giedke},\ and\
  \citenamefont {Grass}}]{Dutkiewicz2026LaughlinFCI}%
  \BibitemOpen
  \bibfield  {author} {\bibinfo {author} {\bibfnamefont {K.}~\bibnamefont
  {Dutkiewicz}}, \bibinfo {author} {\bibfnamefont {B.}~\bibnamefont {Andrade}},
  \bibinfo {author} {\bibfnamefont {G.}~\bibnamefont {Giedke}},\ and\ \bibinfo
  {author} {\bibfnamefont {T.}~\bibnamefont {Grass}},\ }\href@noop {} {\bibinfo
  {title} {Preparation of a {Laughlin}-like fractional {Chern} insulator on a
  superconducting quantum processor}} (\bibinfo {year} {2026}),\ \Eprint
  {https://arxiv.org/abs/2610.41017} {arXiv:2610.41017} \BibitemShut {NoStop}%
\bibitem [{\citenamefont {Pauw}\ \emph {et~al.}(2024)\citenamefont {Pauw},
  \citenamefont {Palm}, \citenamefont {Schollw\"ock}, \citenamefont {Bohrdt},
  \citenamefont {Paeckel},\ and\ \citenamefont
  {Grusdt}}]{Pauw2024PRRHiddenOrder}%
  \BibitemOpen
  \bibfield  {author} {\bibinfo {author} {\bibfnamefont {F.}~\bibnamefont
  {Pauw}}, \bibinfo {author} {\bibfnamefont {F.~A.}\ \bibnamefont {Palm}},
  \bibinfo {author} {\bibfnamefont {U.}~\bibnamefont {Schollw\"ock}}, \bibinfo
  {author} {\bibfnamefont {A.}~\bibnamefont {Bohrdt}}, \bibinfo {author}
  {\bibfnamefont {S.}~\bibnamefont {Paeckel}},\ and\ \bibinfo {author}
  {\bibfnamefont {F.}~\bibnamefont {Grusdt}},\ }\bibfield  {title} {\bibinfo
  {title} {Detecting hidden order in fractional {Chern} insulators},\ }\href
  {https://doi.org/10.1103/PhysRevResearch.6.023180} {\bibfield  {journal}
  {\bibinfo  {journal} {Phys. Rev. Res.}\ }\textbf {\bibinfo {volume} {6}},\
  \bibinfo {pages} {023180} (\bibinfo {year} {2024})}\BibitemShut {NoStop}%
\bibitem [{\citenamefont {Bernevig}\ and\ \citenamefont
  {Haldane}(2008)}]{Bernevig2008PRLJackPolynomials}%
  \BibitemOpen
  \bibfield  {author} {\bibinfo {author} {\bibfnamefont {B.~A.}\ \bibnamefont
  {Bernevig}}\ and\ \bibinfo {author} {\bibfnamefont {F.~D.~M.}\ \bibnamefont
  {Haldane}},\ }\bibfield  {title} {\bibinfo {title} {Model fractional quantum
  {Hall} states and {Jack} polynomials},\ }\href
  {https://doi.org/10.1103/PhysRevLett.100.246802} {\bibfield  {journal}
  {\bibinfo  {journal} {Phys. Rev. Lett.}\ }\textbf {\bibinfo {volume} {100}},\
  \bibinfo {pages} {246802} (\bibinfo {year} {2008})}\BibitemShut {NoStop}%
\bibitem [{\citenamefont {Wang}\ \emph
  {et~al.}(2025{\natexlab{b}})\citenamefont {Wang}, \citenamefont {Ge},
  \citenamefont {Shi}, \citenamefont {Wang}, \citenamefont {Zhou},
  \citenamefont {Li}, \citenamefont {Zhao}, \citenamefont {Xu}, \citenamefont
  {Ma}, \citenamefont {Liu}, \citenamefont {Fang}, \citenamefont {Song},
  \citenamefont {Li}, \citenamefont {Zhang}, \citenamefont {Liu}, \citenamefont
  {Deng}, \citenamefont {Xue}, \citenamefont {Yu}, \citenamefont {Xu},
  \citenamefont {Huang}, \citenamefont {Nori},\ and\ \citenamefont
  {Fan}}]{Wang2025arXivLGT}%
  \BibitemOpen
  \bibfield  {author} {\bibinfo {author} {\bibfnamefont {Z.}~\bibnamefont
  {Wang}}, \bibinfo {author} {\bibfnamefont {Z.-Y.}\ \bibnamefont {Ge}},
  \bibinfo {author} {\bibfnamefont {Y.-H.}\ \bibnamefont {Shi}}, \bibinfo
  {author} {\bibfnamefont {Z.-A.}\ \bibnamefont {Wang}}, \bibinfo {author}
  {\bibfnamefont {S.-Y.}\ \bibnamefont {Zhou}}, \bibinfo {author}
  {\bibfnamefont {H.}~\bibnamefont {Li}}, \bibinfo {author} {\bibfnamefont
  {K.}~\bibnamefont {Zhao}}, \bibinfo {author} {\bibfnamefont {Y.-S.}\
  \bibnamefont {Xu}}, \bibinfo {author} {\bibfnamefont {W.-G.}\ \bibnamefont
  {Ma}}, \bibinfo {author} {\bibfnamefont {H.-T.}\ \bibnamefont {Liu}},
  \bibinfo {author} {\bibfnamefont {C.-P.}\ \bibnamefont {Fang}}, \bibinfo
  {author} {\bibfnamefont {J.-C.}\ \bibnamefont {Song}}, \bibinfo {author}
  {\bibfnamefont {T.-M.}\ \bibnamefont {Li}}, \bibinfo {author} {\bibfnamefont
  {J.-C.}\ \bibnamefont {Zhang}}, \bibinfo {author} {\bibfnamefont
  {Y.}~\bibnamefont {Liu}}, \bibinfo {author} {\bibfnamefont {C.-L.}\
  \bibnamefont {Deng}}, \bibinfo {author} {\bibfnamefont {G.}~\bibnamefont
  {Xue}}, \bibinfo {author} {\bibfnamefont {H.}~\bibnamefont {Yu}}, \bibinfo
  {author} {\bibfnamefont {K.}~\bibnamefont {Xu}}, \bibinfo {author}
  {\bibfnamefont {K.}~\bibnamefont {Huang}}, \bibinfo {author} {\bibfnamefont
  {F.}~\bibnamefont {Nori}},\ and\ \bibinfo {author} {\bibfnamefont
  {H.}~\bibnamefont {Fan}},\ }\href@noop {} {\bibinfo {title} {Observation of
  inelastic meson scattering in a {Floquet} system using a digital quantum
  simulator}} (\bibinfo {year} {2025}{\natexlab{b}}),\ \Eprint
  {https://arxiv.org/abs/2508.20759} {arXiv:2508.20759} \BibitemShut {NoStop}%
\bibitem [{\citenamefont {Wu}\ \emph {et~al.}(2026)\citenamefont {Wu},
  \citenamefont {Liu}, \citenamefont {Pei}, \citenamefont {Zhai}, \citenamefont
  {Luo}, \citenamefont {Liu},\ and\ \citenamefont {Cui}}]{wu2026}%
  \BibitemOpen
  \bibfield  {author} {\bibinfo {author} {\bibfnamefont {H.}~\bibnamefont
  {Wu}}, \bibinfo {author} {\bibfnamefont {L.-Y.-N.}\ \bibnamefont {Liu}},
  \bibinfo {author} {\bibfnamefont {Z.-X.}\ \bibnamefont {Pei}}, \bibinfo
  {author} {\bibfnamefont {Y.-X.}\ \bibnamefont {Zhai}}, \bibinfo {author}
  {\bibfnamefont {Z.-X.}\ \bibnamefont {Luo}}, \bibinfo {author} {\bibfnamefont
  {Z.}~\bibnamefont {Liu}},\ and\ \bibinfo {author} {\bibfnamefont
  {J.}~\bibnamefont {Cui}},\ }\href@noop {} {\bibinfo {title} {Preparation of
  fractional quantum hall states on quantum computers}} (\bibinfo {year}
  {2026}),\ \Eprint {https://arxiv.org/abs/2606.16548} {arXiv:2606.16548}
  \BibitemShut {NoStop}%
\bibitem [{\citenamefont {Zhang}\ \emph {et~al.}(2022)\citenamefont {Zhang},
  \citenamefont {Li},\ and\ \citenamefont
  {Yuan}}]{Zhang2022PRLOptimalStatePreparation}%
  \BibitemOpen
  \bibfield  {author} {\bibinfo {author} {\bibfnamefont {X.-M.}\ \bibnamefont
  {Zhang}}, \bibinfo {author} {\bibfnamefont {T.}~\bibnamefont {Li}},\ and\
  \bibinfo {author} {\bibfnamefont {X.}~\bibnamefont {Yuan}},\ }\bibfield
  {title} {\bibinfo {title} {Quantum state preparation with optimal circuit
  depth: Implementations and applications},\ }\href
  {https://doi.org/10.1103/PhysRevLett.129.230504} {\bibfield  {journal}
  {\bibinfo  {journal} {Phys. Rev. Lett.}\ }\textbf {\bibinfo {volume} {129}},\
  \bibinfo {pages} {230504} (\bibinfo {year} {2022})}\BibitemShut {NoStop}%
\bibitem [{SM()}]{SM}%
  \BibitemOpen
  \href@noop {} {}\bibinfo {note} {See Supplemental Material.}\BibitemShut
  {Stop}%
\bibitem [{\citenamefont {Huggins}\ \emph {et~al.}(2021)\citenamefont
  {Huggins}, \citenamefont {McClean}, \citenamefont {Rubin}, \citenamefont
  {Jiang}, \citenamefont {Wiebe}, \citenamefont {Whaley},\ and\ \citenamefont
  {Babbush}}]{Huggins2021npjQuantumMeasurement}%
  \BibitemOpen
  \bibfield  {author} {\bibinfo {author} {\bibfnamefont {W.~J.}\ \bibnamefont
  {Huggins}}, \bibinfo {author} {\bibfnamefont {J.~R.}\ \bibnamefont
  {McClean}}, \bibinfo {author} {\bibfnamefont {N.~C.}\ \bibnamefont {Rubin}},
  \bibinfo {author} {\bibfnamefont {Z.}~\bibnamefont {Jiang}}, \bibinfo
  {author} {\bibfnamefont {N.}~\bibnamefont {Wiebe}}, \bibinfo {author}
  {\bibfnamefont {K.~B.}\ \bibnamefont {Whaley}},\ and\ \bibinfo {author}
  {\bibfnamefont {R.}~\bibnamefont {Babbush}},\ }\bibfield  {title} {\bibinfo
  {title} {Efficient and noise resilient measurements for quantum chemistry on
  near-term quantum computers},\ }\href
  {https://doi.org/10.1038/s41534-020-00341-7} {\bibfield  {journal} {\bibinfo
  {journal} {npj Quantum Inf.}\ }\textbf {\bibinfo {volume} {7}},\ \bibinfo
  {pages} {23} (\bibinfo {year} {2021})}\BibitemShut {NoStop}%
\bibitem [{\citenamefont {Gupta}\ \emph {et~al.}(2024)\citenamefont {Gupta},
  \citenamefont {Sundaresan}, \citenamefont {Alexander}, \citenamefont {Wood},
  \citenamefont {Merkel}, \citenamefont {Healy}, \citenamefont {Hillenbrand},
  \citenamefont {Jochym-O'Connor}, \citenamefont {Wootton}, \citenamefont
  {Yoder}, \citenamefont {Cross}, \citenamefont {Takita},\ and\ \citenamefont
  {Brown}}]{Gupta2024nature}%
  \BibitemOpen
  \bibfield  {author} {\bibinfo {author} {\bibfnamefont {R.~S.}\ \bibnamefont
  {Gupta}}, \bibinfo {author} {\bibfnamefont {N.}~\bibnamefont {Sundaresan}},
  \bibinfo {author} {\bibfnamefont {T.}~\bibnamefont {Alexander}}, \bibinfo
  {author} {\bibfnamefont {C.~J.}\ \bibnamefont {Wood}}, \bibinfo {author}
  {\bibfnamefont {S.~T.}\ \bibnamefont {Merkel}}, \bibinfo {author}
  {\bibfnamefont {M.~B.}\ \bibnamefont {Healy}}, \bibinfo {author}
  {\bibfnamefont {M.}~\bibnamefont {Hillenbrand}}, \bibinfo {author}
  {\bibfnamefont {T.}~\bibnamefont {Jochym-O'Connor}}, \bibinfo {author}
  {\bibfnamefont {J.~R.}\ \bibnamefont {Wootton}}, \bibinfo {author}
  {\bibfnamefont {T.~J.}\ \bibnamefont {Yoder}}, \bibinfo {author}
  {\bibfnamefont {A.~W.}\ \bibnamefont {Cross}}, \bibinfo {author}
  {\bibfnamefont {M.}~\bibnamefont {Takita}},\ and\ \bibinfo {author}
  {\bibfnamefont {B.~J.}\ \bibnamefont {Brown}},\ }\bibfield  {title} {\bibinfo
  {title} {Encoding a magic state with beyond break-even fidelity},\ }\href
  {https://doi.org/10.1038/s41586-023-06846-3} {\bibfield  {journal} {\bibinfo
  {journal} {Nature}\ }\textbf {\bibinfo {volume} {625}},\ \bibinfo {pages}
  {259} (\bibinfo {year} {2024})}\BibitemShut {NoStop}%
\bibitem [{\citenamefont {Zhang}\ \emph {et~al.}(2026)\citenamefont {Zhang},
  \citenamefont {Xu}, \citenamefont {Zhang}, \citenamefont {Duan},
  \citenamefont {Zhong}, \citenamefont {Bai}, \citenamefont {Wang},
  \citenamefont {Huang}, \citenamefont {Deng}, \citenamefont {Gao},
  \citenamefont {Zhang}, \citenamefont {Liu}, \citenamefont {Li}, \citenamefont
  {Jiang}, \citenamefont {Zhao}, \citenamefont {Shu}, \citenamefont {Wu},
  \citenamefont {Shi}, \citenamefont {Ding}, \citenamefont {Tian},
  \citenamefont {Wang}, \citenamefont {Yuan}, \citenamefont {Xin},
  \citenamefont {Hu}, \citenamefont {Liu}, \citenamefont {Pan}, \citenamefont
  {Huang}, \citenamefont {He},\ and\ \citenamefont {Yu}}]{Zhang2026nano}%
  \BibitemOpen
  \bibfield  {author} {\bibinfo {author} {\bibfnamefont {C.}~\bibnamefont
  {Zhang}}, \bibinfo {author} {\bibfnamefont {F.}~\bibnamefont {Xu}}, \bibinfo
  {author} {\bibfnamefont {S.}~\bibnamefont {Zhang}}, \bibinfo {author}
  {\bibfnamefont {M.}~\bibnamefont {Duan}}, \bibinfo {author} {\bibfnamefont
  {D.}~\bibnamefont {Zhong}}, \bibinfo {author} {\bibfnamefont
  {X.}~\bibnamefont {Bai}}, \bibinfo {author} {\bibfnamefont {H.}~\bibnamefont
  {Wang}}, \bibinfo {author} {\bibfnamefont {C.}~\bibnamefont {Huang}},
  \bibinfo {author} {\bibfnamefont {Y.}~\bibnamefont {Deng}}, \bibinfo {author}
  {\bibfnamefont {M.}~\bibnamefont {Gao}}, \bibinfo {author} {\bibfnamefont
  {Y.-N.}\ \bibnamefont {Zhang}}, \bibinfo {author} {\bibfnamefont
  {J.}~\bibnamefont {Liu}}, \bibinfo {author} {\bibfnamefont {C.}~\bibnamefont
  {Li}}, \bibinfo {author} {\bibfnamefont {Y.}~\bibnamefont {Jiang}}, \bibinfo
  {author} {\bibfnamefont {B.}~\bibnamefont {Zhao}}, \bibinfo {author}
  {\bibfnamefont {H.}~\bibnamefont {Shu}}, \bibinfo {author} {\bibfnamefont
  {K.}~\bibnamefont {Wu}}, \bibinfo {author} {\bibfnamefont {K.}~\bibnamefont
  {Shi}}, \bibinfo {author} {\bibfnamefont {Q.}~\bibnamefont {Ding}}, \bibinfo
  {author} {\bibfnamefont {Z.}~\bibnamefont {Tian}}, \bibinfo {author}
  {\bibfnamefont {G.}~\bibnamefont {Wang}}, \bibinfo {author} {\bibfnamefont
  {X.}~\bibnamefont {Yuan}}, \bibinfo {author} {\bibfnamefont {T.}~\bibnamefont
  {Xin}}, \bibinfo {author} {\bibfnamefont {G.}~\bibnamefont {Hu}}, \bibinfo
  {author} {\bibfnamefont {S.}~\bibnamefont {Liu}}, \bibinfo {author}
  {\bibfnamefont {T.}~\bibnamefont {Pan}}, \bibinfo {author} {\bibfnamefont
  {P.}~\bibnamefont {Huang}}, \bibinfo {author} {\bibfnamefont
  {Y.}~\bibnamefont {He}},\ and\ \bibinfo {author} {\bibfnamefont
  {D.}~\bibnamefont {Yu}},\ }\bibfield  {title} {\bibinfo {title} {Universal
  logical operations in a silicon quantum processor},\ }\href
  {https://doi.org/10.1038/s41565-026-02140-1} {\bibfield  {journal} {\bibinfo
  {journal} {Nat. Nanotechnol.}\ }\textbf {\bibinfo {volume} {21}},\ \bibinfo
  {pages} {635} (\bibinfo {year} {2026})}\BibitemShut {NoStop}%
\bibitem [{\citenamefont {Li}\ and\ \citenamefont
  {Haldane}(2008)}]{LiHaldane2008PRLEntanglementSpectrum}%
  \BibitemOpen
  \bibfield  {author} {\bibinfo {author} {\bibfnamefont {H.}~\bibnamefont
  {Li}}\ and\ \bibinfo {author} {\bibfnamefont {F.~D.~M.}\ \bibnamefont
  {Haldane}},\ }\bibfield  {title} {\bibinfo {title} {Entanglement spectrum as
  a generalization of entanglement entropy: Identification of topological order
  in {Non-Abelian} fractional quantum {Hall} effect states},\ }\href
  {https://doi.org/10.1103/PhysRevLett.101.010504} {\bibfield  {journal}
  {\bibinfo  {journal} {Phys. Rev. Lett.}\ }\textbf {\bibinfo {volume} {101}},\
  \bibinfo {pages} {010504} (\bibinfo {year} {2008})}\BibitemShut {NoStop}%
\bibitem [{\citenamefont {Pauw}\ \emph {et~al.}(2026)\citenamefont {Pauw},
  \citenamefont {Goldman},\ and\ \citenamefont
  {Palm}}]{Pauw2026arXivCompositeBoson}%
  \BibitemOpen
  \bibfield  {author} {\bibinfo {author} {\bibfnamefont {F.~J.}\ \bibnamefont
  {Pauw}}, \bibinfo {author} {\bibfnamefont {N.}~\bibnamefont {Goldman}},\ and\
  \bibinfo {author} {\bibfnamefont {F.~A.}\ \bibnamefont {Palm}},\ }\href@noop
  {} {\bibinfo {title} {Long-range order and composite boson condensation in
  lattice quantum {Hall} states}} (\bibinfo {year} {2026}),\ \Eprint
  {https://arxiv.org/abs/2608.24628} {arXiv:2608.24628} \BibitemShut {NoStop}%
\bibitem [{\citenamefont {Wen}\ and\ \citenamefont
  {Zee}(1992)}]{WenZee1992PRLShift}%
  \BibitemOpen
  \bibfield  {author} {\bibinfo {author} {\bibfnamefont {X.~G.}\ \bibnamefont
  {Wen}}\ and\ \bibinfo {author} {\bibfnamefont {A.}~\bibnamefont {Zee}},\
  }\bibfield  {title} {\bibinfo {title} {Shift and spin vector: New topological
  quantum numbers for the {Hall} fluids},\ }\href
  {https://doi.org/10.1103/PhysRevLett.69.953} {\bibfield  {journal} {\bibinfo
  {journal} {Phys. Rev. Lett.}\ }\textbf {\bibinfo {volume} {69}},\ \bibinfo
  {pages} {953} (\bibinfo {year} {1992})}\BibitemShut {NoStop}%
\bibitem [{\citenamefont {Johri}\ \emph {et~al.}(2014)\citenamefont {Johri},
  \citenamefont {Papi\ifmmode~\acute{c}\else \'{c}\fi{}}, \citenamefont
  {Bhatt},\ and\ \citenamefont {Schmitteckert}}]{Johri2014PRBQuasiholeSize}%
  \BibitemOpen
  \bibfield  {author} {\bibinfo {author} {\bibfnamefont {S.}~\bibnamefont
  {Johri}}, \bibinfo {author} {\bibfnamefont {Z.}~\bibnamefont
  {Papi\ifmmode~\acute{c}\else \'{c}\fi{}}}, \bibinfo {author} {\bibfnamefont
  {R.~N.}\ \bibnamefont {Bhatt}},\ and\ \bibinfo {author} {\bibfnamefont
  {P.}~\bibnamefont {Schmitteckert}},\ }\bibfield  {title} {\bibinfo {title}
  {Quasiholes of $\frac{1}{3}$ and $\frac{7}{3}$ quantum {Hall} states: Size
  estimates via exact diagonalization and density-matrix renormalization
  group},\ }\href {https://doi.org/10.1103/PhysRevB.89.115124} {\bibfield
  {journal} {\bibinfo  {journal} {Phys. Rev. B}\ }\textbf {\bibinfo {volume}
  {89}},\ \bibinfo {pages} {115124} (\bibinfo {year} {2014})}\BibitemShut
  {NoStop}%
\bibitem [{\citenamefont {Liu}\ \emph {et~al.}(2015)\citenamefont {Liu},
  \citenamefont {Bhatt},\ and\ \citenamefont
  {Regnault}}]{Liu2015PRBQuasiholeFCI}%
  \BibitemOpen
  \bibfield  {author} {\bibinfo {author} {\bibfnamefont {Z.}~\bibnamefont
  {Liu}}, \bibinfo {author} {\bibfnamefont {R.~N.}\ \bibnamefont {Bhatt}},\
  and\ \bibinfo {author} {\bibfnamefont {N.}~\bibnamefont {Regnault}},\
  }\bibfield  {title} {\bibinfo {title} {Characterization of quasiholes in
  fractional {Chern} insulators},\ }\href
  {https://doi.org/10.1103/PhysRevB.91.045126} {\bibfield  {journal} {\bibinfo
  {journal} {Phys. Rev. B}\ }\textbf {\bibinfo {volume} {91}},\ \bibinfo
  {pages} {045126} (\bibinfo {year} {2015})}\BibitemShut {NoStop}%
\bibitem [{\citenamefont {Liu}\ \emph {et~al.}(2023)\citenamefont {Liu},
  \citenamefont {Balram}, \citenamefont {Papi\'{c}},\ and\ \citenamefont
  {Gromov}}]{Liu2023PRBSpinStatistics}%
\BibitemOpen
  \bibfield {author} {
    \bibinfo {author} {\bibfnamefont {H.~Q.}\ \bibnamefont {Trung}},
    \bibinfo {author} {\bibfnamefont {Y.}\ \bibnamefont {Wang}},\ and
    \bibinfo {author} {\bibfnamefont {B.}\ \bibnamefont {Yang}},\
  }
  \bibfield {title} {
    \bibinfo {title} {Spin-statistics relation and {Abelian} braiding
    phase for anyons in the fractional quantum {Hall} effect},\
  }
  \href {https://doi.org/10.1103/PhysRevB.107.L201301} {
    \bibfield {journal} {\bibinfo {journal} {Phys. Rev. B}\ }
    \textbf {\bibinfo {volume} {107}},
    \bibinfo {pages} {L201301} (\bibinfo {year} {2023})
  }\BibitemShut {NoStop}%
\end{thebibliography}
%

\end{document}


\renewcommand{\theequation}{S\arabic{equation}}
\renewcommand{\thefigure}{S\arabic{figure}}
\renewcommand{\thetable}{S\arabic{table}}
\renewcommand{\thesection}{\Roman{section}}
\renewcommand{\thesubsection}{\Alph{subsection}}
\setcounter{secnumdepth}{2}
\setcounter{tocdepth}{2}

\title{Supplemental Material for ``Hidden Coherence and Controlled Quasihole Motion in a Fermionic Laughlin State on a Superconducting Quantum Processor''}


\maketitle

\tableofcontents

\section{Haldane-sphere representation and Fock-space encoding of Laughlin states}
\label{app:haldane-sphere}

The Haldane sphere provides a finite rotationally invariant geometry for studying fractional quantum Hall states. A Dirac monopole of strength $Q=N_\phi/2$ is placed at the center of the sphere, producing a magnetic flux
$N_\phi$ through the surface. The single-particle states are classified by the total angular momentum
$\ell=Q+n$, where $n$ denotes the Landau-level index. The lowest Landau level (LLL) corresponds to $n=0$ and therefore contains $N_{\rm orb}=2Q+1=N_\phi+1$
degenerate orbitals with magnetic quantum numbers $m=-Q,\ldots,Q$.
A point on the sphere $\Omega=(\theta,\varphi)$ is represented by the spinor coordinates
\begin{equation}
u=\cos\left(\frac{\theta}{2}\right)e^{i\varphi/2},
\quad
v=\sin\left(\frac{\theta}{2}\right)e^{-i\varphi/2}.
\end{equation}
Using the shifted orbital index
\begin{equation}
j=m+Q,\qquad j=0,\ldots,N_\phi ,
\end{equation}
the normalized LLL orbitals can be written as
\begin{equation}
\phi_j(u,v)
=
\left[
\frac{N_\phi+1}{4\pi}
\binom{N_\phi}{j}
\right]^{1/2}
u^jv^{N_\phi-j},
\label{eq:app-normalized-lll-orbital}
\end{equation}
where $m=j-Q$. These orbitals form the single-particle basis used for the orbital encoding in the experiment.

For the fermionic $\nu=1/3$ Laughlin state, the wave function in spherical spinor coordinates is
\begin{equation}
\Psi_{\rm L}=\prod_{i<j}
(u_i v_j-v_i u_j)^3 .
\label{eq:app-sphere-laughlin}
\end{equation}
Since all particles are restricted to the LLL, the many-body Hilbert space is spanned by Slater determinants constructed from the LLL orbitals $\{\phi_j\}$. 
Introducing $c_j^\dagger$ as the fermionic creation operator of the $j$th LLL orbital, the LLL-projected field operator is defined as
\begin{equation}
\hat\psi(\Omega) =
\sum_{j=0}^{N_\phi}
\phi_j(\Omega)c_j,
\label{eq:app-projected-field}
\end{equation}
where $c_j^\dagger$ creates an electron in orbital $j$, while
$\hat\psi^\dagger(\Omega)$ creates an electron localized at position
$\Omega$ within the LLL.
The Laughlin state in Eq.~\eqref{eq:app-sphere-laughlin} can therefore be expressed in the fermionic Fock basis as
\begin{equation}
|\Psi_{\rm L}\rangle =
\sum_{\{n_j\}}
C_{n_0\cdots n_{N_\phi}}
\prod_{j=0}^{N_\phi}
(c_j^\dagger)^{n_j}
|{\rm vac}\rangle ,
\label{eq:app-fock-expansion}
\end{equation}
where $n_j=0,1$ denotes the occupation of orbital $j$ and
$\sum_j n_j=N_e$. The coefficients
$C_{n_0\cdots n_{N_\phi}}$ encode the many-body correlations of the Laughlin wave function.
In the superconducting-qubit implementation, 
each LLL orbital occupation is encoded into one qubit, $|n_0 n_1\cdots n_{N_\phi}\rangle$, so that computational-basis measurements directly access the fermionic occupation configurations.
For the experimental Laughlin state with $N_e=3$ and $N_\phi=6$, the seven LLL orbitals are encoded into seven superconducting qubits. The orbital occupation basis provides direct access to the rotational symmetry sectors of the spherical geometry. Using the shifted orbital index, the conserved angular-momentum component is 
\begin{equation}
K=\sum_j j n_j,
\qquad {L_z=K-QN} ,
\end{equation}
where $Q=N_\phi/2$. The integer $N=\sum_j n_j$ labels a general Fock-space particle-number sector, ranging from $0$ to $7$ for seven orbitals. We reserve $N_e=3$ for the particle number of the target state.
The rotationally invariant Laughlin ground state with $L=0$ belongs to the symmetry sector $(N_e,K)=(3,9)$. The exact zero-energy state of the fermionic $V_1$ Haldane pseudopotential Hamiltonian in this sector gives the following normalized occupation-basis representation~\cite{Haldane1983PRLHierarchy}. For $N_e=3$ and $N_\phi=6$, the state is supported on five squeezed configurations generated from the root configuration $\ket{1001001}$:
\begin{equation}
\mathcal B_{\rm L} =\{
\ket{1001001},
\ket{1000110},
\ket{0110001},
\ket{0101010},
\ket{0011100}
\},
\end{equation}
with the corresponding normalized amplitudes $
\bm C_{\rm L}= 1/\sqrt{7}
\left(
1,-\sqrt{2},-\sqrt{2},1,-1
\right)$,
up to an overall phase. These configurations form the squeezing manifold $\mathcal B_{\rm L}$ obtained by inward pair squeezing from the root configuration $\ket{1001001}$
at fixed $N_e$ and $K$~\cite{Bernevig2008PRLJackPolynomials}. This sparse structure leads to the digital sparse-prefix preparation circuit used in the experiment, as described in Sec.~\ref{app:sparse}.

\begin{figure*}[!htbp]
\centering
\includegraphics[width=0.95\textwidth]{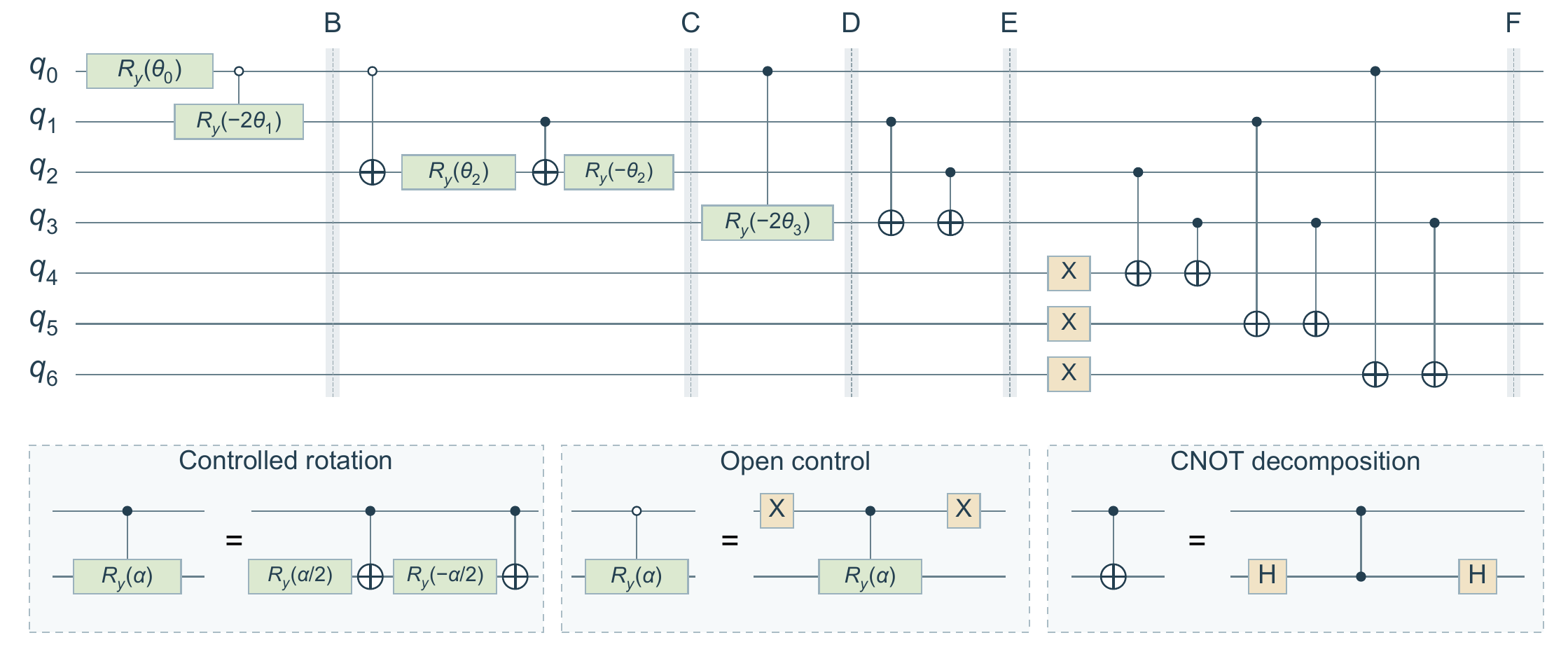}
\caption[Sparse preparation circuit for the seven-orbital Laughlin state]{
  Sparse preparation circuit for the seven-orbital sphere Laughlin state.
  The labels $q_j$ denote logical occupation qubits for orbital $j$, not fixed physical sites on the processor; the hardware assignment is introduced in Fig.~\ref{fig:app-laughlin-compiled-schedule}.
  The barriers B--F separate successive circuit stages. Before barrier B, the root and open-controlled rotations load three prefix branches. Between barriers B and C, two CNOTs split and relabel the occupied branches. The C--D stage creates the fifth branch, and D--E relabels the prefix. The E--F stage writes the remaining occupations $(q_4,q_5,q_6)$. The lower diagrams show the gate identities used in compilation.
  The complete logical circuit uses 14 CNOT gates and prepares the target Laughlin state exactly up to an irrelevant global phase.
  }
\label{fig:app-laughlin-binary-tree-circuit}
\end{figure*}

\section{Sparse preparation of the seven-orbital sphere Laughlin state}
\label{app:sparse}

For $N_e=3$ fermions in seven sphere orbitals, the normalized $\nu=1/3$ Laughlin state is
\begin{equation}
|\Psi_{\rm L}\rangle=\frac{1}{\sqrt7}\Big(
|1001001\rangle+|0101010\rangle-|0011100\rangle
-\sqrt2|1000110\rangle-\sqrt2|0110001\rangle
\Big).
\label{eq:app-seven-laughlin}
\end{equation}
At fixed particle number $N_e=3$, the orbital-index constraint $K=\sum_{j=0}^{6}j n_j=9$ leaves exactly these five Fock configurations.
The circuit first loads five signed prefix amplitudes, then unfolds them into the full seven-orbital occupation basis.
The five-state squeezed support identified above also makes the digital preparation compact. Instead of preparing a generic seven-qubit state, we first load the amplitudes of the allowed Laughlin configurations into a four-qubit prefix register and then reconstruct the remaining orbital occupations by reversible Boolean relations. The corresponding decomposition is
\begin{equation}
\begin{array}{c|c|c}
\text{prefix} & \text{suffix} & \text{amplitude}\\ \hline
1001 & 001 & 1/\sqrt7\\
0101 & 010 & 1/\sqrt7\\
0011 & 100 & -1/\sqrt7\\
1000 & 110 & -\sqrt{2/7}\\
0110 & 001 & -\sqrt{2/7}
\end{array}.
\label{tab:app-prefix}
\end{equation}
The preparation therefore separates into
\begin{equation}
|0000000\rangle
\xrightarrow{\,U_{\rm pre}\,}
|\psi_{\rm pre}\rangle|000\rangle
\xrightarrow{\,U_{\rm aff}\,}
|\Psi_{\rm L}\rangle ,
\label{eq:app-preparation}
\end{equation}
where
\begin{equation}
\begin{aligned}
|\psi_{\rm pre}\rangle=\frac{1}{\sqrt7}\Big(
|1001\rangle+|0101\rangle-|0011\rangle
-\sqrt2|1000\rangle-\sqrt2|0110\rangle
\Big).
\end{aligned}
\label{eq:app-prefix-state}
\end{equation}
The logical circuit implementing this two-stage construction is shown in Fig.~\ref{fig:app-laughlin-binary-tree-circuit}. The left block prepares the prefix amplitudes, while the right block unfolds the prefix state into the full seven-orbital occupation basis.
Here $q_j$ labels the logical qubit carrying orbital occupation $n_j$. These labels specify the state-preparation logic independently of the physical processor; mapping to connected hardware sites is described in Sec.~\ref{app:circuit-compilation}.
The prefix amplitudes are generated by a binary amplitude-splitting construction with
\begin{equation}
\theta_0=2\sin^{-1}\sqrt{\frac37},\qquad
\theta_1=\frac{\pi}{3},\qquad
\theta_2=\cos^{-1}\frac{1}{\sqrt3},\qquad
\theta_3=\sin^{-1}\frac{1}{\sqrt3}.
\label{eq:app-prefix-angles}
\end{equation}
The corresponding conditional rotations are decomposed into single-qubit rotations and CNOT gates.
On this support, the remaining occupations obey
\begin{equation}
q_4=1\oplus q_2\oplus q_3,\qquad
q_5=1\oplus q_1\oplus q_3,\qquad
q_6=1\oplus q_0\oplus q_3,
\label{eq:app-affine}
\end{equation}
where $\oplus$ denotes addition modulo two. The reconstruction $U_{\rm aff}$ is therefore implemented by $X$ gates on $q_4,q_5,q_6$ followed by
\begin{equation}
\mathrm{CX}(q_2,q_4), \mathrm{CX}(q_3,q_4), \quad
\mathrm{CX}(q_1,q_5), \mathrm{CX}(q_3,q_5), \quad
\mathrm{CX}(q_0,q_6), \mathrm{CX}(q_3,q_6).
\label{eq:app-affine-circuit}
\end{equation}
The affine reconstruction maps each populated prefix to its corresponding seven-orbital Fock configuration while preserving the amplitudes and relative phases.

\begin{figure*}[!htbp]
\centering
\includegraphics[width=\textwidth]{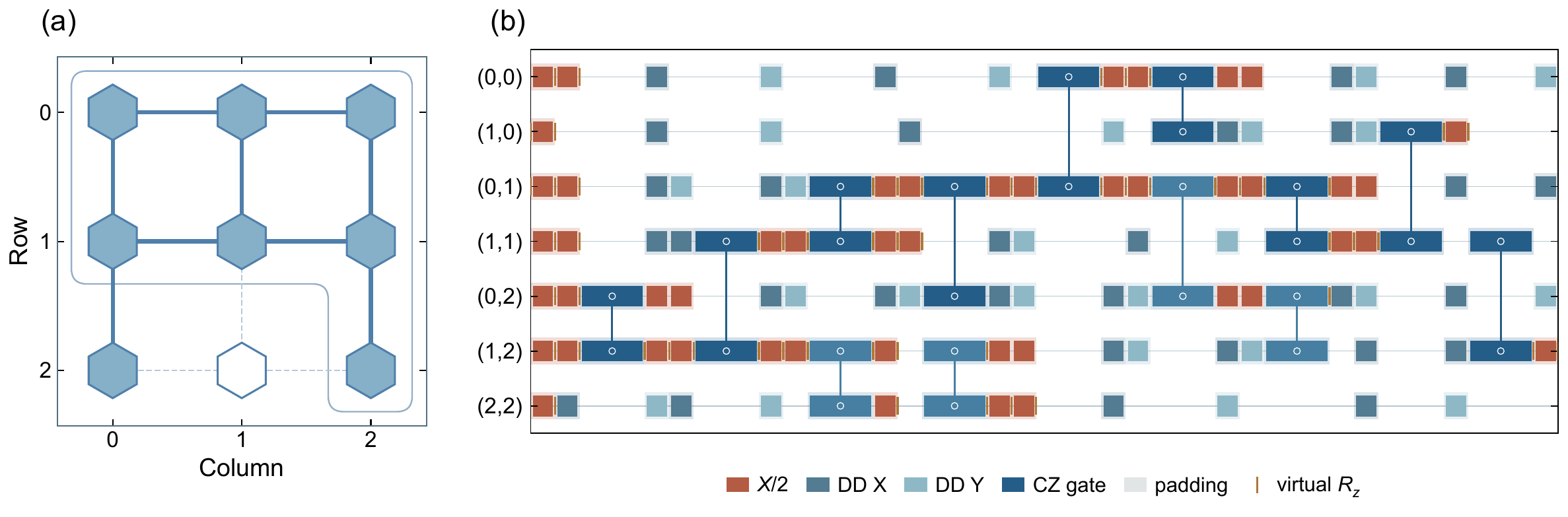}
\caption[Compiled Laughlin circuit]{(a) Eight experimental qubits in the $3\times3$ processor. The rounded outline marks the seven-qubit subset used for Laughlin-state preparation. (b) Final circuit compiled for hardware execution of the seven-orbital Laughlin-state preparation.}
\label{fig:app-laughlin-compiled-schedule}
\end{figure*}

\subsection{Circuit compilation}
\label{app:circuit-compilation}
The seven-orbital Laughlin-state preparation circuit is mapped onto the processor topology through connectivity-aware qubit mapping and routing. In the implementation shown, the logical qubit $q_j$ encoding orbital $j$ is assigned to physical qubit $\pi(j)$ as follows:
\begin{equation}
\begin{array}{c|ccccccc}
j & 0 & 1 & 2 & 3 & 4 & 5 & 6 \\
\pi(j) & {(0,2)} & {(2,2)} & {(0,0)} & {(1,1)} & {(1,0)} & {(0,1)} & {(1,2)}
\end{array}.
\label{eq:app-laughlin-logical-to-physical}
\end{equation}

For hardware execution, the logical circuit in Fig.~\ref{fig:app-laughlin-binary-tree-circuit} is mapped to the device connectivity and compiled into arbitrary single-qubit rotations and two-qubit CZ gates. Microwave-driven single-qubit gates and flux-driven CZ gates are scheduled in separate layers to reduce control crosstalk~\cite{WangKeprx2025bell}.
At the single-qubit level, an arbitrary $U$ gate can be expressed using microwave rotations and virtual $R_z$ gates. The latter update the phase of subsequent microwave drives and require no physical pulse time~\cite{McKay2017PRAVirtualZ}. A general $U$ gate can be synthesized with two calibrated $X_{\pi/2}$ microwave pulses and virtual $R_z$ rotations:
\begin{equation}
U(\theta,\phi,\lambda)
\doteq R_z(\phi)R_x(-\pi/2)R_z(\theta)R_x(\pi/2)R_z(\lambda).
\label{eq:app-compilation-u3}
\end{equation}
Here $R_\mu(\alpha)=\exp[-i\alpha\sigma_\mu/2]$, and $\doteq$ denotes equality up to a global phase.
For circuit-level optimization, we first use Qiskit~\cite{QiskitContributors2023} to map each theoretical circuit onto the connected device layout and compile it into the $\{U(\theta,\varphi,\lambda), \text{CZ}\}$ gate set, seeking candidates with fewer CZ gates~\cite{Xiang2024nc}. For the Laughlin-state preparation illustrated in Fig.~\ref{fig:app-laughlin-compiled-schedule}, 
we further optimize the CZ count by removing or relocating CZ gates while refitting all $U$-gate parameters to preserve the prepared state.
The resulting $U$ gates are then resolved into calibrated microwave pulses and virtual phase updates, and the native operations are scheduled with their gate durations and padding. Finally, dynamical decoupling (DD) pulse sequences are inserted into suitable idle intervals~\cite{Zhu2026Combinatorial,DDPRApplied2022,DDPRApplied2025,DD2025prxq,DDPRApplied2026}. Figure~\ref{fig:app-laughlin-compiled-schedule} shows the eight-site active layout, its selected seven-qubit subset, and the fully compiled Laughlin-state preparation circuit scheduled for execution on the processor.

\begin{figure*}[!htbp]
\centering
\includegraphics[width=0.96\textwidth]{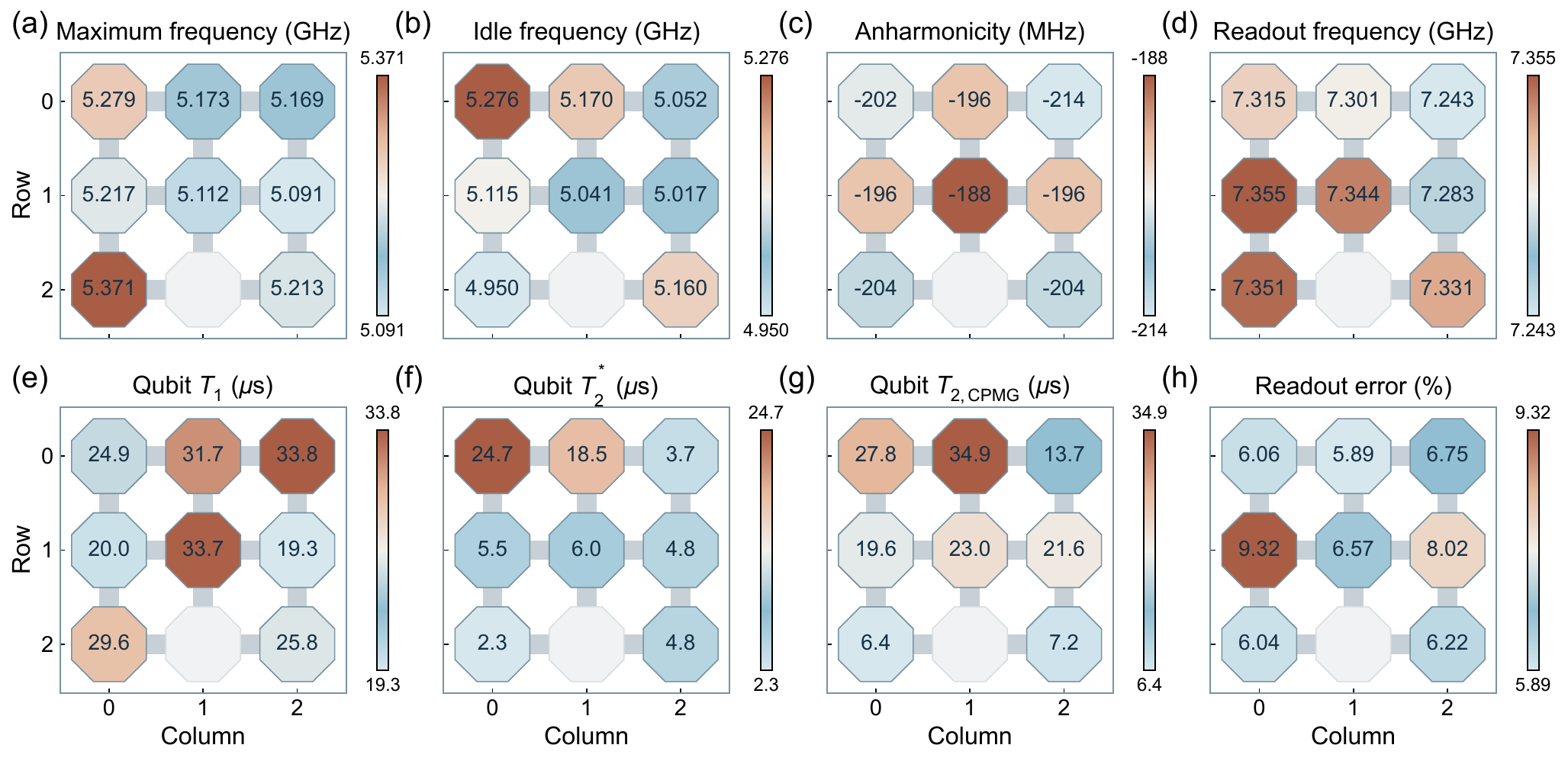}
\caption[Device parameters]{Single-qubit parameters.  (a) Maximum qubit frequency. (b) Idle qubit frequency. (c) Qubit anharmonicity. (d) Readout-resonator frequency. (e) Energy-relaxation time $T_1$. (f) Ramsey dephasing time $T_2^*$. (g) CPMG echo coherence time $T_{2,\mathrm{CPMG}}$. (h) Average readout error.}
\label{fig:app-chip-qubit-parameters}
\end{figure*}

\section{Device information}
\label{app:device-information}

The experiment is performed on a superconducting quantum processor comprising
nine frequency-tunable transmon qubits arranged in a $3\times3$ square lattice,
with 12 tunable couplers connecting nearest-neighbor qubits. The device is
fabricated using flip-chip integration with a tantalum thin-film base
circuit~\cite{Li2021APLFlipChip,Li2024PRRTopologyLocalization}. Each qubit is
independently controlled through microwave and flux lines and measured by
dispersive readout. The tunable couplers provide controllable
nearest-neighbor interactions for implementing two-qubit CZ gates. Connected
subsets of seven and eight qubits are selected for the experiments reported
here: seven qubits encode the seven orbitals of the lowest Landau level for
the $N_\phi=6$ Laughlin state, whereas eight qubits encode the one-flux-added
system with $N_\phi=7$.
In this encoding, the computational
basis state $\ket{n_0n_1\cdots n_{N_\phi}}$ represents a fermionic occupation
configuration of the spherical LLL orbitals.

\begin{figure*}[!htbp]
\centering
\includegraphics[width=0.85\linewidth]{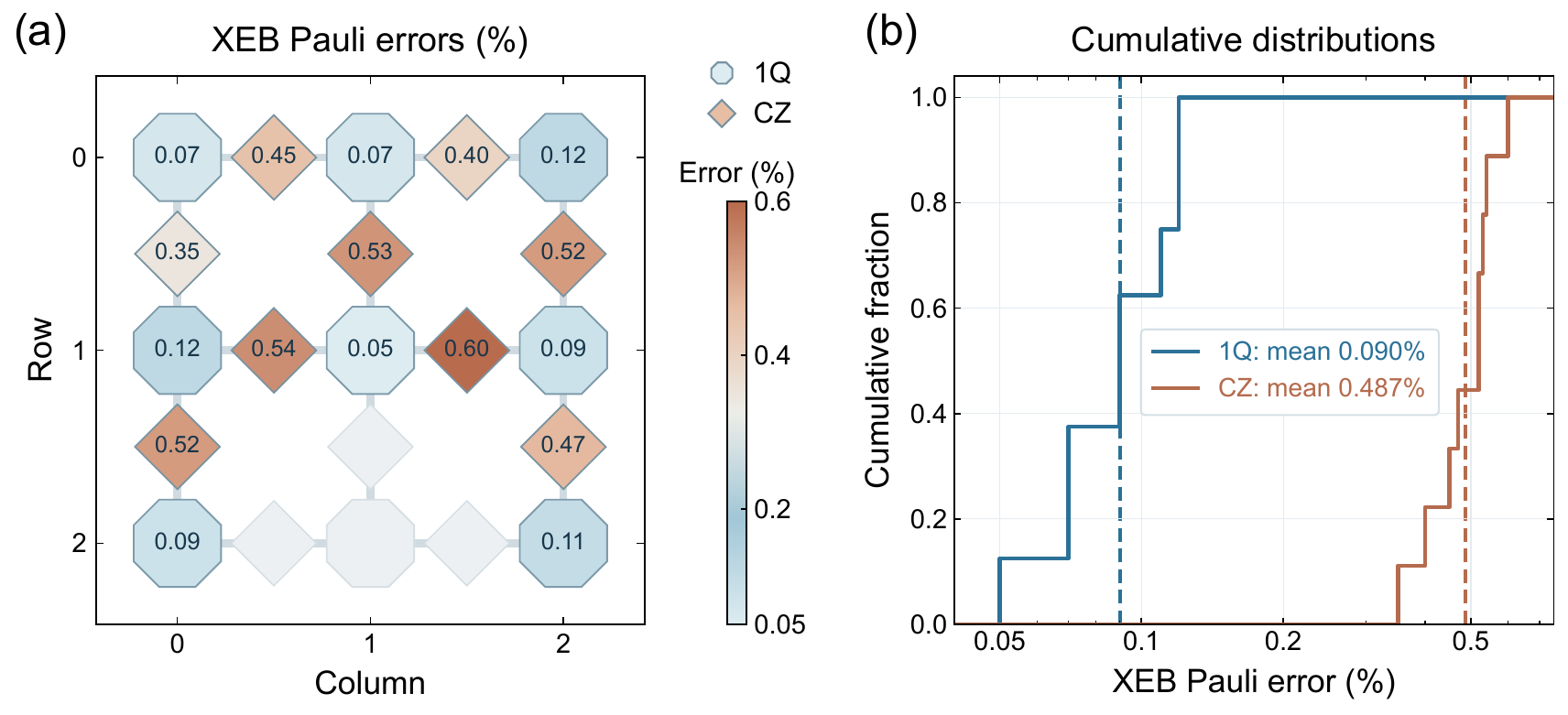}
\caption[XEB Pauli error maps and distributions]{XEB Pauli error maps and distributions. 
The mean Pauli errors are $0.90\times10^{-3}$ for $20\,\mathrm{ns}$ single-qubit gates and $4.9\times10^{-3}$ for $60\,\mathrm{ns}$ CZ gates.}
\label{fig:app-chip-xeb-errors}
\end{figure*}

\begin{figure*}[!htbp]
\centering
\includegraphics[width=0.85\linewidth]{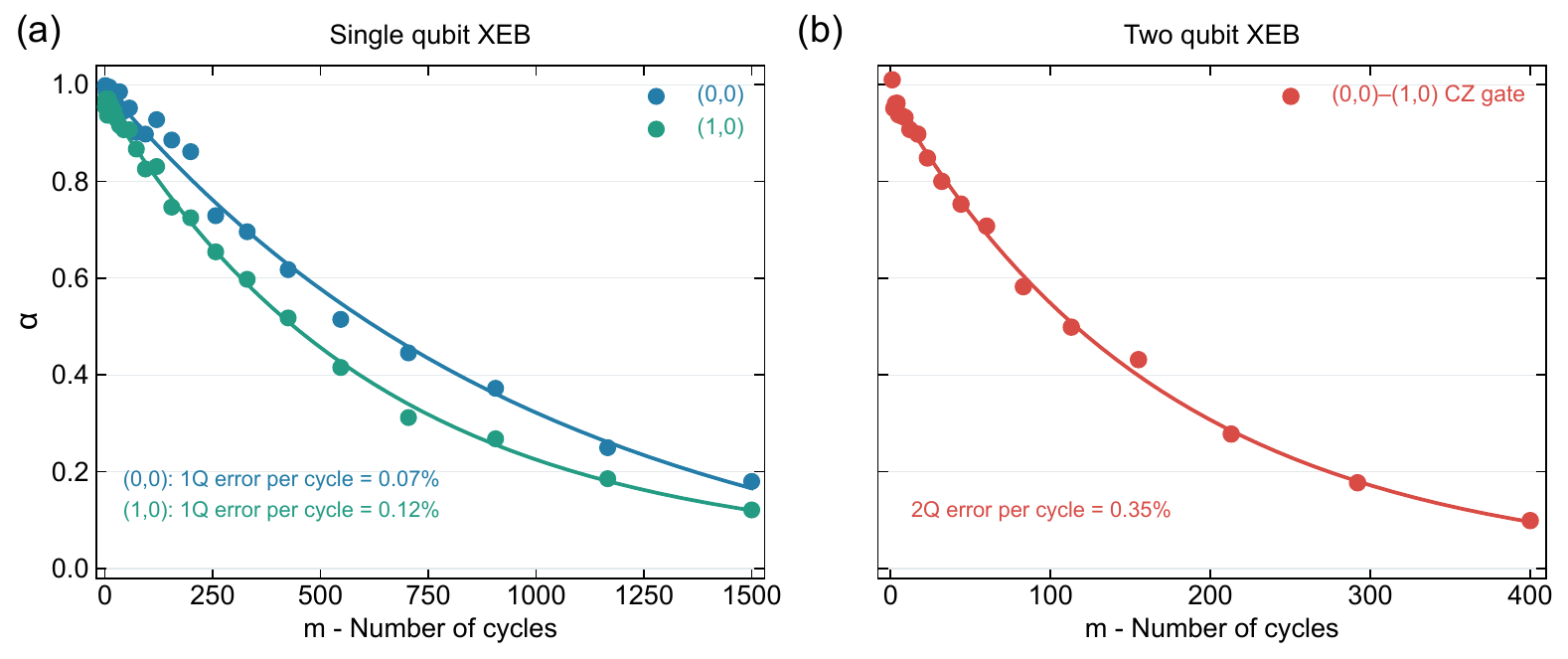}
\caption[Representative XEB decay data]{Cross-entropy benchmarking for (a) simultaneous $20\,\mathrm{ns}$ single-qubit gates at physical sites $(0,0)$ and $(1,0)$ and (b) the $60\,\mathrm{ns}$ CZ gate on the $(0,0)$--$(1,0)$ pair.}
\label{fig:app-q1q2-xeb-only}
\end{figure*}

Physical single-qubit gates are driven by microwave pulses
with a duration of $20\,\mathrm{ns}$. The $60\,\mathrm{ns}$ diabatic CZ gate
is realized by tuning the two-transmon levels $\ket{11}$ and $\ket{20}$
close to resonance while activating their interaction through the tunable
coupler.
Cross-entropy benchmarking (XEB) gives mean Pauli errors of $0.90\times10^{-3}$ and $4.9\times10^{-3}$ for the single-qubit and CZ gates,
respectively, as summarized in Fig.~\ref{fig:app-chip-xeb-errors}.
Representative XEB decay data for the physical $(0,0)$--$(1,0)$ pair are shown in
Fig.~\ref{fig:app-q1q2-xeb-only}.
The qubit frequencies, anharmonicities, coherence times, and readout errors
of the active subset are summarized in
Fig.~\ref{fig:app-chip-qubit-parameters} and
Table~\ref{tab:chip-parameter-summary}. For further details on the device
architecture, measurement setup, and calibration procedures, see
Refs.~\cite{Wang2026PRLBoseHubbard,Wang2024PRRMaxwellDemon}.

\begin{table*}[t]
\caption[Summary of device parameters]{Summary of device parameters for the active eight-qubit subset.}
\label{tab:chip-parameter-summary}
\begin{ruledtabular}
\begin{tabular}{lccc}
Parameter & Mean & Median & Stdev. \\
Qubit maximum frequency (GHz) & 5.203 & 5.193 & 0.085 \\
Qubit idle frequency (GHz) & 5.098 & 5.083 & 0.097 \\
Qubit anharmonicity (MHz) & -200.0 & -199.0 & 7.3 \\
Readout drive frequency (GHz) & 7.315 & 7.323 & 0.036 \\
$T_1$ ($\mu$s) & 27.3 & 27.7 & 5.4 \\
$T_2^*$ ($\mu$s) & 8.8 & 5.2 & 7.6 \\
$T_{2,\mathrm{CPMG}}$ ($\mu$s) & 19.3 & 20.6 & 9.2 \\
Readout $|0\rangle$ error (\%) & 4.81 & 4.62 & 1.52 \\
Readout $|1\rangle$ error (\%) & 8.91 & 8.73 & 1.13 \\
Readout average error (\%) & 6.86 & 6.39 & 1.13 \\
1Q XEB Pauli error ($10^{-3}$) & 0.90 & 0.90 & 0.24 \\
CZ XEB Pauli error (\%) & 0.49 & 0.52 & 0.07 \\
\end{tabular}
\end{ruledtabular}
\end{table*}

\section{Full quantum-state tomography}
\label{app:full-quantum-state-tomography}

Full quantum-state tomography reconstructs the density matrix in the complete
$N$-qubit Hilbert space from an informationally complete set of local
measurements. Since tensor-product Pauli operators form an orthogonal basis of
operator space, the density matrix can be expanded as
\begin{equation}
\rho=
\frac{1}{2^N}
\sum_{\boldsymbol{\alpha}\in\{I,X,Y,Z\}^N}
\left\langle P_{\boldsymbol{\alpha}}\right\rangle
P_{\boldsymbol{\alpha}},
\qquad
P_{\boldsymbol{\alpha}}
=
\bigotimes_{j=0}^{N-1}\sigma_{\alpha_j}^{(j)} .
\label{eq:app-pauli-expansion-full-tomo}
\end{equation}
The tomography problem therefore reduces to determining a complete set of
multi-qubit Pauli moments.

Experimentally, each qubit is measured in one of the three local Pauli bases.
The corresponding basis transformation is implemented by a single-qubit rotation followed by computational-basis readout. With
\begin{equation}
R_\mu(\theta)=\exp\left(-i\theta\sigma_\mu/2\right),
\end{equation}
our pulse convention corresponds to
\begin{equation}
I\rightarrow Z,\qquad
R_x(\pi/2)\rightarrow Y,\qquad
R_y(\pi/2)\rightarrow -X,
\label{eq:app-full-tomo-analysis-bases}
\end{equation}
where the final sign follows from
$R_y^\dagger(\pi/2)ZR_y(\pi/2)=-X$.
Taking all tensor products of these three local bases gives the complete set of
$3^N$ tomography settings.
For a local measurement axis $a\in\{Z,Y,X\}$ and binary outcome $b$, we define
\begin{equation}
\Pi_{a,b}
=
\frac{1}{2}\left[I+(-1)^b\tau_a\right],
\qquad
\tau_Z=Z,\quad
\tau_Y=Y,\quad
\tau_X=-X.
\label{eq:app-single-qubit-tomo-projector}
\end{equation}
For a setting $s=(s_0,\ldots,s_{N-1})$ and outcome string
$b=(b_0,\ldots,b_{N-1})$, the corresponding product measurement
operator and the ideal outcome probability are given by
\begin{equation}
M_{s,\bm b}
=
\bigotimes_{j=0}^{N-1}\Pi_{s_j,b_j},
\qquad
p_s(\bm b)
=
\operatorname{Tr}\!\left(\rho M_{s,\bm b}\right).
\label{eq:app-full-tomo-born-probability}
\end{equation}
The six single-qubit projectors associated with the three Pauli bases form an
overcomplete operator frame. 
We perform the linear reconstruction using the
corresponding canonical dual frame~\cite{Fischer2024PRADualFrame},
\begin{equation}
D_{a,b}
=
\frac{I}{6}
+
\frac{(-1)^b}{2}\tau_a,
\label{eq:app-single-qubit-dual-frame}
\end{equation}
which satisfies
\begin{equation}
O
=
\sum_{a\in\{Z,Y,X\}}
\sum_{b=0}^{1}
\operatorname{Tr}\!\left(O\Pi_{a,b}\right)
D_{a,b}
\label{eq:app-single-qubit-dual-completeness}
\end{equation}
for an arbitrary single-qubit operator $O$. 
The factor $I/6$ distributes the normalization information uniformly over the three redundant
measurement bases, while the second term extracts the Pauli component along the measured axis.
Because both the measurement operators and the dual frame factorize over qubits, the $N$-qubit dual operators are
\begin{equation}
D_{s,\bm b}
=
\bigotimes_{j=0}^{N-1}D_{s_j,b_j},
\end{equation}
and the linear tomographic density matrix becomes
\begin{equation}
\rho^{\mathrm{lin}}
=
\sum_{s\in\{Z,Y,X\}^N}
\sum_{\bm b\in\{0,1\}^N}
p_s(\bm b)\,
D_{s,\bm b}.
\label{eq:app-dual-tensor-tomography}
\end{equation}
We first Hermitize the linear estimate,
\(\rho_H=(\rho_{\mathrm{lin}}+
\rho_{\mathrm{lin}}^\dagger)/2\).
Writing \(\rho_H=\sum_\alpha\lambda_\alpha
|\alpha\rangle\langle\alpha|\), we obtain the physical
estimate by setting negative eigenvalues to zero and
renormalizing:
\[
\rho_{\mathrm{exp}}
=
\frac{\sum_\alpha\max(\lambda_\alpha,0)
|\alpha\rangle\langle\alpha|}
{\sum_\alpha\max(\lambda_\alpha,0)}.
\]

As the local Pauli measurements form an overcomplete operator frame,
a lower-order Pauli moment is generally sampled by several compatible
tomography settings. The canonical-dual reconstruction combines this
redundant information automatically.
In the experimental analysis, readout-assignment errors are corrected using the calibrated single-qubit matrices
$A_j=\bigl(\begin{smallmatrix}
F_0^{(j)} & 1-F_1^{(j)}\\
1-F_0^{(j)} & F_1^{(j)}
\end{smallmatrix}\bigr)$,
with $A=\bigotimes_j A_j$ and
$\bm p_s^{\,\mathrm{corr}}=A^{-1}\bm p_s^{\,\mathrm{meas}}$.
The corrected distributions are then used directly in the tensor-product reconstruction of Eq.~\eqref{eq:app-dual-tensor-tomography}.

\begin{figure*}[!htbp] 
\centering \includegraphics[width=0.75\textwidth] 
{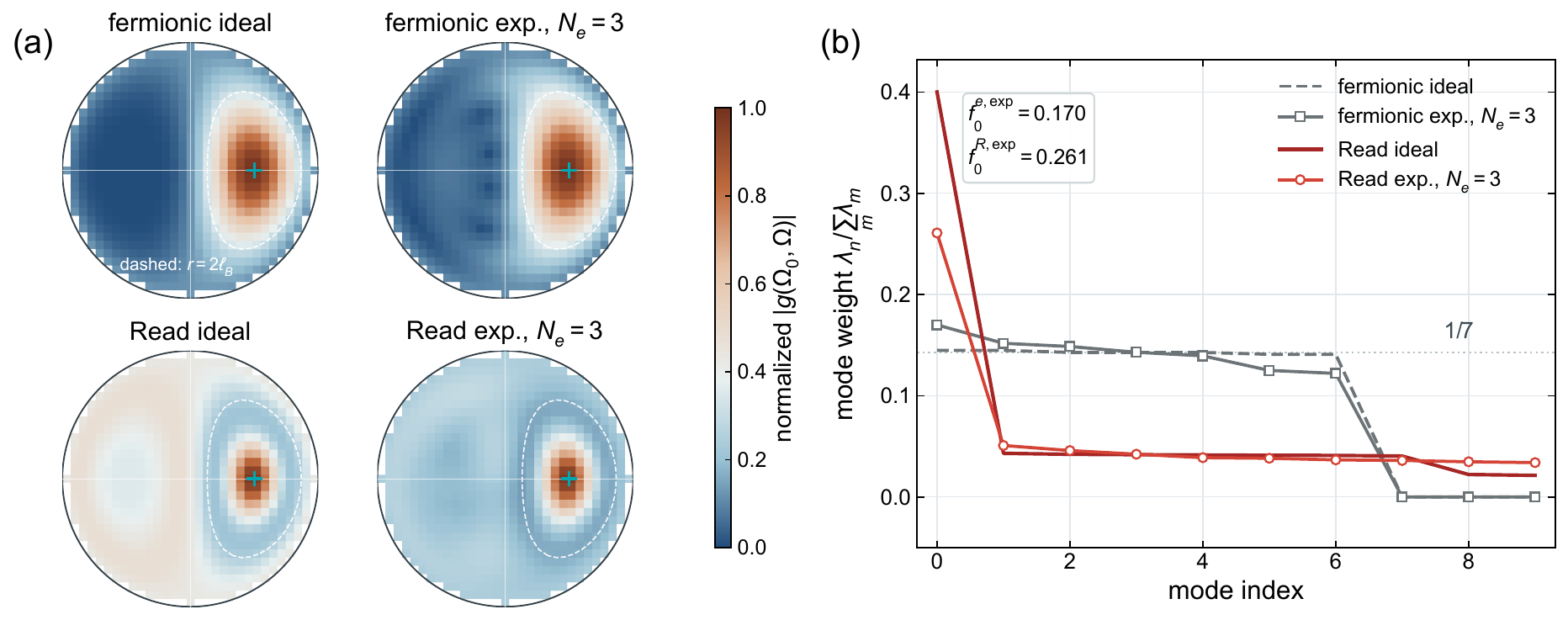}
\caption[Spatial and spectral characterization of the vortex-dressed Read coherence.]{
Spatial and spectral characterization of the vortex-dressed Read coherence.
(a) Spatial maps of the normalized fermionic and Read coherences,
$|g_e(\Omega_0,\Omega)|$ and $|g_R(\Omega_0,\Omega)|$, for the ideal and reconstructed $N_e=3$ Laughlin states.
The cross marks the reference point $\Omega_0$, and the dashed contour marks the finite-size long-distance boundary $r=2\ell_B$. 
Here $\ell_B=R/\sqrt{Q}=1/\sqrt{3}$ is the magnetic length for the unit-radius sphere with $Q=N_\phi/2=3$.
(b) Natural-mode spectra of the unnormalized fermionic and Read correlation kernels.
The mode weights are $f_n=\lambda_n/\sum_m\lambda_m$.
The ideal fermionic kernel exhibits the uniform seven-mode distribution $f_n=1/7$, whereas the Read kernel redistributes coherence weight toward the leading natural mode.
The experimental leading-mode fractions are $f_0^{e,\mathrm{exp}}=0.170$ and $f_0^{R,\mathrm{exp}}=0.261$.
}
\label{fig:supp-read-correlator} 
\end{figure*}


\section{Reconstruction of Read composite-boson coherence}
\label{sec:supp-read-reconstruction}

The reconstructed many-body density matrix allows direct evaluation of different coherence functions of the Laughlin state. We compare the conventional fermionic first-order coherence with the vortex-dressed coherence defined through Read's composite-boson construction.

The FQST density matrix $\rho_{\rm exp}$ is projected onto the fixed-particle-number sector relevant to the Laughlin state,
\begin{equation}
\rho^{(N_e)} = \frac{ P_{N_e}\rho_{\rm exp}P_{N_e} }{ \mathrm{Tr}(P_{N_e}\rho_{\rm exp}) },
\label{eq:supp-rhoN}
\end{equation}
with $N_e=3$.

The fermionic field operator $\hat\psi(\Omega)$ in the spherical LLL basis is defined in Eq.~\eqref{eq:app-projected-field}. The normalized electronic coherence between two points $\Omega_a$ and $\Omega_b$ on the sphere is
\begin{equation}
g_e(\Omega_a,\Omega_b) = \frac{ \langle \hat\psi^\dagger(\Omega_a) \hat\psi(\Omega_b) \rangle }{ \sqrt{ \langle \hat n(\Omega_a)\rangle \langle \hat n(\Omega_b)\rangle } },
\label{eq:electron-coherence}
\end{equation}
where $\hat n(\Omega)=\hat\psi^\dagger(\Omega)\hat\psi(\Omega)$.

To probe the hidden coherence structure of the Laughlin state, we employ Read's composite-boson construction. The electron operator is dressed by three correlation vortices associated with the fermionic $\nu=1/3$ Laughlin state~\cite{Girvin1987PRLODLRO,Rezayi1988PRLODLRO,Read1989PRLCompositeBoson},
\begin{equation}
\hat{\mathcal B}^{\dagger}(\Omega) = \hat\psi^\dagger(\Omega)\hat U^3(\Omega),
\label{eq:read-operator-definition}
\end{equation}
where $\hat U(\Omega)$ denotes the single-vortex attachment operator. The corresponding normalized Read coherence is
\begin{equation}
g_R(\Omega_a,\Omega_b) = \frac{ \langle \hat{\mathcal B}^{\dagger}(\Omega_a) \hat{\mathcal B}(\Omega_b) \rangle }{ \sqrt{ \langle \hat{\mathcal B}^{\dagger}(\Omega_a) \hat{\mathcal B}(\Omega_a) \rangle \langle \hat{\mathcal B}^{\dagger}(\Omega_b) \hat{\mathcal B}(\Omega_b) \rangle } }.
\label{eq:read-coherence}
\end{equation}
For numerical evaluation, we construct the corresponding correlation kernels,
\begin{equation}
G^e_{ab} = \mathrm{Tr}[ \rho^{(N_e)} \hat\psi^\dagger(\Omega_a) \hat\psi(\Omega_b) ],
\label{eq:supp-electron-kernel}
\end{equation}
and
\begin{equation}
G^R_{ab} = \mathrm{Tr}[ \rho^{(N_e)} \hat{\mathcal B}^{\dagger}(\Omega_a) \hat{\mathcal B}(\Omega_b) ].
\label{eq:supp-read-kernel}
\end{equation}
The normalized quantities are obtained as
\begin{equation}
g_\alpha(\Omega_a,\Omega_b) = \frac{ G^\alpha_{ab} }{ \sqrt{ G^\alpha_{aa}G^\alpha_{bb} } }, \qquad \alpha=e,R .
\label{eq:supp-normalized-coherence}
\end{equation}
Both coherences are therefore evaluated from the same reconstructed
$N_e=3$ density matrix, while probing different correlation structures.

On the finite Haldane sphere, the operator $\hat U^3(\Omega)$ is constructed through successive single-vortex attachment maps. A single correlation vortex is introduced by the factor
\begin{equation}
\prod_{i=1}^{N_e} (u_i v_h-v_i u_h),
\label{eq:supp-single-vortex}
\end{equation}
which increases the LLL flux by one quantum and maps an $N_\phi$-flux LLL state to the corresponding $(N_\phi+1)$-flux Hilbert space. The associated single-particle map is
\begin{equation}
T_{N_\phi}(\Omega): \mathbb C^{N_\phi+1} \rightarrow \mathbb C^{N_\phi+2}.
\end{equation}

In the normalized spherical-orbital convention of
Eq.~\eqref{eq:app-normalized-lll-orbital}, the only nonzero matrix elements are
\begin{equation}
[T_{N_\phi}(\Omega)]_{j,j}
=
-u_h\sqrt{\frac{N_\phi+1-j}{N_\phi+2}},
\qquad
[T_{N_\phi}(\Omega)]_{j+1,j}
=
v_h\sqrt{\frac{j+1}{N_\phi+2}},
\quad j=0,\ldots,N_\phi .
\end{equation}
For fermionic many-body states, the corresponding transformation is the antisymmetrized exterior lift,
\begin{equation}
[ \mathcal U_{N_\phi}(\Omega) ]_{\mathbf n',\mathbf n} = \det [ T_{N_\phi}(\Omega) ]_{\mathbf n',\mathbf n}.
\label{eq:supp-exterior-map}
\end{equation}
Here $\mathbf n$ and $\mathbf n'$ denote the occupied input and output orbitals in ascending order, and the determinant automatically enforces fermionic antisymmetry. 
The three-vortex dressing $\hat U^3(\Omega)$ is represented by three successive vortex-attachment maps,
\begin{equation}
\mathcal H_{3,6} \xrightarrow{\mathcal U} \mathcal H_{3,7} \xrightarrow{\mathcal U} \mathcal H_{3,8} \xrightarrow{\mathcal U} \mathcal H_{3,9}.
\label{eq:supp-read-auxiliary-spaces}
\end{equation}
Here $\mathcal H_{N_e,N_\phi}^{(D)}$ denotes the fixed-$N_e$ many-body Hilbert space in an $N_\phi$-flux LLL, with
$D=\binom{N_\phi+1}{N_e}$ its dimension.
These intermediate Hilbert spaces are auxiliary constructions required to represent the algebraic vortex attachment and do not correspond to physical quasihole creation or flux insertion. 
The vortex-attachment maps preserve $N_e=3$. In the composite product $\hat{B}^{\dagger}(\Omega_a)\hat{B}(\Omega_b)$, the intermediate particle-number and flux indices are contracted, yielding an operator that acts within the original $\mathcal{H}_{3,6}$ space.
For the spatial maps shown in Fig.~\ref{fig:supp-read-correlator}(a), one reference point $\Omega_0$ is fixed and the second coordinate is scanned over the sphere, yielding $|g_\alpha(\Omega_0,\Omega)|$ with $\alpha=e,R$. The separation between two points is characterized by the geodesic angle
\begin{equation}
\cos\gamma = \cos\theta_a\cos\theta_b + \sin\theta_a\sin\theta_b \cos(\phi_a-\phi_b),
\label{eq:supp-geodesic-angle}
\end{equation}
with arc length
\begin{equation}
r=R\gamma, \qquad R=\sqrt{Q}\ell_B .
\label{eq:supp-geodesic-distance}
\end{equation}
For the present $Q=3$ sphere, the region $r>2\ell_B$ corresponds to $\gamma/\pi>2/(\sqrt{3}\pi)\simeq0.368$, which defines the finite-size long-distance window used in the main text.

To identify the contribution of symmetry-sector leakage to the measured coherence, we further project the reconstructed state onto the target angular-momentum sector,
\begin{equation}
\rho^{(N_e,L_z=0)} = \frac{ P_{N_e,L_z=0} \rho_{\rm exp} P_{N_e,L_z=0} }{ \mathrm{Tr} ( P_{N_e,L_z=0} \rho_{\rm exp} ) }.
\label{eq:supp-read-lz-projection}
\end{equation}
The corresponding Read coherence provides a symmetry-resolved reference in which contributions outside the target Laughlin sector are removed.

Beyond the spatial coherence profile, we characterize the global organization of the correlation kernels through their natural modes. In the continuum limit, the eigenmodes of the unnormalized kernel satisfy
\begin{equation}
\int d\Omega_b\, G^\alpha(\Omega_a,\Omega_b) v_n^\alpha(\Omega_b) = \lambda_n^\alpha v_n^\alpha(\Omega_a), \qquad \alpha=e,R .
\label{eq:supp-natural-mode-continuum}
\end{equation}
For numerical evaluation, the sphere is discretized using an equal-area Fibonacci sampling with points $\Omega_a$. The continuous integral is approximated by a weighted sum,
\begin{equation}
\int d\Omega\,f(\Omega) \approx \sum_a w_a f(\Omega_a),
\end{equation}
where the equal-area quadrature weights are
\begin{equation}
w_a=\frac{4\pi}{N_\Omega}.
\end{equation}
To preserve the Hermitian structure of the integral eigenvalue problem, we diagonalize the weighted kernel
\begin{equation}
\widetilde G^\alpha_{ab} = \sqrt{w_a} G^\alpha(\Omega_a,\Omega_b) \sqrt{w_b},
\label{eq:supp-weighted-kernel}
\end{equation}
according to
\begin{equation}
\widetilde G^\alpha v_n^\alpha = \lambda_n^\alpha v_n^\alpha .
\label{eq:supp-natural-mode-eigenproblem}
\end{equation}
The normalized spectral fractions are defined as
\begin{equation}
f_n^\alpha = \frac{\lambda_n^\alpha}{\sum_m\lambda_m^\alpha}.
\label{eq:supp-mode-fraction}
\end{equation}
In Fig.~\ref{fig:supp-read-correlator}(b), we use $N_\Omega=48$ Fibonacci points with $w_a=4\pi/N_\Omega$. Both the electronic and Read spectra are evaluated from the same reconstructed $\rho^{(N_e)}$.
This natural-mode analysis characterizes off-diagonal coherence~\cite{Yang1962RMPODLRO}. In the present finite-size system, the leading Read fraction quantifies the redistribution of coherence into the vortex-dressed channel and should be interpreted as a finite-size coherence signature rather than evidence for thermodynamic condensation.

\section{Geometric manipulation and reconstruction of Laughlin quasiholes}
\label{sec:supp-quasihole-reconstruction}

The one-flux quasihole excitation provides a finite-dimensional manifold for geometric manipulation of the Laughlin state. The fermionic $\nu=1/3$ Laughlin state on the sphere occurs at $N_\phi=3(N_e-1)$,
and adding one flux quantum generates the single-quasihole sector, $N_\phi=3(N_e-1)+1$.
A quasihole localized at $\Omega_h=(\theta_h,\phi_h)$ is described by
\begin{equation}
\Psi_{\rm qh}(\Omega_h) = \mathcal N_{\rm qh} \prod_{i=1}^{N_e} (u_i v_h-v_i u_h) \Psi_{1/3},
\label{eq:supp-localized-quasihole-wavefunction}
\end{equation}
where $(u_h,v_h)$ denotes the spinor coordinate of the quasihole position.
For the north-pole quasihole, $(u_h,v_h)=(1,0)$, and the quasihole factor reduces, up to an overall phase, to $\prod_i v_i$.
The polar quasihole inherits the squeezed Fock-space skeleton of the seven-orbital Laughlin state.  The north-pole quasihole factor $\prod_i v_i$ embeds the Laughlin root pattern and its inward-squeezed descendants into the one-flux Hilbert space, leaving a five-component state in the fixed $(N,K)=(3,9)$ sector:
  \begin{equation}
  \begin{aligned}
  \ket{\Psi^{\rm qh}_{\rm pole}}
  ={}
  \sqrt{\frac{1}{10}}\ket{10010010}
  +\sqrt{\frac{6}{35}}\ket{01010100}
  -\sqrt{\frac{3}{14}}\ket{00111000}
  -\sqrt{\frac{3}{10}}\ket{10001100}
  -\sqrt{\frac{3}{14}}\ket{01100010}.
  \end{aligned}
  \label{eq:app-eight-orbital-polar-quasihole-state}
  \end{equation}
Equivalently, omitting the unoccupied $j=7$ orbital from the representation gives the seven-qubit configurations $\{1001001,0101010,0011100,1000110,0110001\}$. The eight-orbital space is retained for the subsequent quasihole rotation.
The polar quasihole is prepared with the same sparse digital strategy as the seven-orbital Laughlin state in Sec.~\ref{app:sparse}: the CNOT unfolding structure is unchanged, while the amplitude-loading rotations are updated to the polar-quasihole coefficients.
For the circuit used in the experiment, the $R_y$ rotations in the amplitude-loading block are assigned
\begin{equation}
    \begin{aligned}
  \theta_0=\arccos\frac15, \quad
  \theta_1=-\arccos\!\left(-\sqrt{\frac5{14}}\right),\quad
  \theta_2=\arccos\frac23,\quad
  \theta_3=-\frac{5\pi}{6}.
  \end{aligned}
\end{equation}
With these angles, the sparse circuit prepares Eq.~\eqref{eq:app-eight-orbital-polar-quasihole-state} exactly, up to an irrelevant global phase.

Because Eq.~\eqref{eq:supp-localized-quasihole-wavefunction} is homogeneous of degree $N_e$ in $(u_h,v_h)$, the single-quasihole zero modes transform as a spin-$L=N_e/2$ multiplet. For $N_e=3$, the polar state prepared above is the $M=-3/2$ member of this four-dimensional multiplet. A quasihole localized at $(u_h,v_h)$ is therefore
\begin{equation}
\begin{aligned}
|\Psi_{\rm qh}(u_h,v_h)\rangle ={}& u_h^3|\Psi_{-3/2}^{\rm qh}\rangle -\sqrt{3}u_h^2v_h|\Psi_{-1/2}^{\rm qh}\rangle
+ \sqrt{3}u_hv_h^2|\Psi_{1/2}^{\rm qh}\rangle -v_h^3|\Psi_{3/2}^{\rm qh}\rangle .
\end{aligned}
\label{eq:supp-three-electron-quasihole-coherent-state}
\end{equation}
The same state is obtained, up to an overall phase, by rotating the polar state:
\begin{equation}
|\Psi_{\rm qh}(\theta_h,\phi_h)\rangle = e^{-i\phi_hL_z} e^{-i\theta_hL_y} |\Psi_{\rm qh}^{\rm pole}\rangle ,
\label{eq:supp-rotated-quasihole-state}
\end{equation}
Using the SU(2) identity
$e^{-i\theta L_y}=e^{-i\frac{\pi}{2}L_z}
e^{-i\theta L_x} e^{i\frac{\pi}{2}L_z}
$ and the fact that the polar state is an eigenstate of $L_z$, the rightmost rotation contributes only a global phase. The implemented displacement is thus
\begin{equation}
U_{\rm move}(\theta,\phi)
=
e^{-i(\phi+\pi/2)L_z}
e^{-i\theta L_x}.
\end{equation}

In the orbital basis, the corresponding rotations are implemented through the second-quantized angular momentum operators. The azimuthal rotation is generated by
\begin{equation}
L_z = \sum_{j=0}^{N_\phi} (j-S)c_j^\dagger c_j ,
\label{eq:supp-Lz}
\end{equation}
where $S=N_\phi/2$ is the single-particle angular momentum of the LLL orbitals. The transverse rotation is generated by
\begin{equation}
L_x = \sum_{j=0}^{N_\phi-1} \frac{a_j}{2} (c_{j+1}^{\dagger}c_j+c_j^\dagger c_{j+1}),
\label{eq:supp-Lx}
\end{equation}
where $a_j$ are the corresponding spherical LLL angular-momentum matrix elements. For $N_\phi=7$, the orbital coupling coefficients are
\begin{equation}
(a_0,\ldots,a_6) = (\sqrt7,2\sqrt3,\sqrt{15},4,\sqrt{15},2\sqrt3,\sqrt7).
\end{equation}
Under the Jordan--Wigner transformation,
\begin{equation}
c_{j+1}^{\dagger}c_j+c_j^\dagger c_{j+1} = \frac12 (X_jX_{j+1}+Y_jY_{j+1}),
\label{eq:supp-XY-exchange}
\end{equation}
each term in $L_x$ becomes a nearest-neighbor exchange. With the hardware convention
\begin{equation}
XY(\theta)=\exp\!\left[-\frac{i\theta}{2}
\left(X_jX_{j+1}+Y_jY_{j+1}\right)\right],
\label{eq:app-xy-exchange-definition}
\end{equation}
the bond evolution generated by Eq.~\eqref{eq:supp-Lx} is
\begin{equation}
G_{j,j+1}(\gamma)
=\exp\!\left[-\frac{i\gamma}{4}
\left(X_jX_{j+1}+Y_jY_{j+1}\right)\right]
=XY(\gamma/2).
\label{eq:app-orbital-exchange}
\end{equation}
The exchange acts within sectors of fixed occupation on the orbital pair,
$[G_{j,j+1}(\gamma),n_j+n_{j+1}]=0$.
The exchange rotation is compiled into two CZ gates and local single-qubit rotations:
\begin{equation}
\resizebox{0.6\columnwidth}{!}{%
\begin{tikzpicture}[
  x=1cm,y=1cm,
  gate/.style={draw,fill=white,minimum width=1.35cm,minimum height=0.55cm,inner sep=1pt},
  dot/.style={circle,fill=black,inner sep=2pt}
]
\draw (-0.1,0.8) node[left] {$q_j$} -- (11.7,0.8);
\draw (-0.1,0) node[left] {$q_{j+1}$} -- (11.7,0);
\node[anchor=east,font=\large] at (-0.75,0.4) {$XY(\theta)\doteq$};
\node[gate] at (0.8,0.8) {$R_x(\pi/2)$};
\node[gate] at (0.8,0) {$R_y(\pi/2)$};
\node[gate] at (2.2,0) {$R_z(\pi)$};
\draw (3.25,0) -- (3.25,0.8);
\node[dot] at (3.25,0) {};
\node[dot] at (3.25,0.8) {};
\node[below] at (3.25,-0.18) {CZ};
\node[gate,minimum width=1.55cm] at (4.55,0.8) {$R_z(-\pi/2)$};
\node[gate,minimum width=1.75cm] at (6.45,0.8) {$R_y(\theta+\pi)$};
\node[gate,minimum width=1.55cm] at (6.45,0) {$R_y(-\theta)$};
\draw (7.65,0) -- (7.65,0.8);
\node[dot] at (7.65,0) {};
\node[dot] at (7.65,0.8) {};
\node[below] at (7.65,-0.18) {CZ};
\node[gate,minimum width=1.55cm] at (8.95,0.8) {$R_y(-\pi/2)$};
\node[gate,minimum width=1.55cm] at (8.95,0) {$R_y(-\pi/2)$};
\node[gate] at (10.75,0.8) {$R_z(\pi/2)$};
\end{tikzpicture}%
}
\label{eq:app-exchange-2cz-decomposition}
\end{equation}
Here equality holds up to a global phase. To compile the collective rotation $e^{-i\theta L_x}$, we group the exchange gates on disjoint bonds into even and odd layers. For $N_\phi=7$, their explicit forms are
\begin{subequations}
\begin{align}
E_x(\delta)
&=G_{01}(\sqrt7\,\delta)\,
  G_{23}(\sqrt{15}\,\delta)\,
  G_{45}(\sqrt{15}\,\delta)\,
  G_{67}(\sqrt7\,\delta),\\
O_x(\delta)
&=G_{12}(2\sqrt3\,\delta)\,
  G_{34}(4\delta)\,
  G_{56}(2\sqrt3\,\delta).
\end{align}
\label{eq:app-even-odd-exchange-layers}
\end{subequations}
A symmetric second-order Suzuki--Trotter decomposition, with $\delta=\theta/r$, then gives
\begin{equation}
U_x^{(r)}(\theta) = \left[ E_x\left(\frac{\delta}{2}\right) O_x\left({\delta}\right) E_x\left(\frac{\delta}{2}\right) \right]^r \simeq e^{-i\theta L_x},
\label{eq:symmetric-trotter}
\end{equation}
A single Trotter step is implemented by the circuit shown below.
\begin{center}
\definecolor{gateyellow}{RGB}{250,226,155}
\definecolor{gateyellowedge}{RGB}{190,145,55}
\definecolor{gateblue}{RGB}{174,204,235}
\definecolor{gateblueedge}{RGB}{66,110,160}
\begin{tikzpicture}[
  x=1cm,y=0.68cm,
  wire/.style={draw=black!65,line width=0.45pt},
  even/.style={draw=gateblueedge,fill=gateblue,rounded corners=2pt,
    minimum width=1.85cm,minimum height=0.82cm,inner sep=1pt,
    font=\scriptsize},
  odd/.style={draw=gateyellowedge,fill=gateyellow,rounded corners=2pt,
    minimum width=1.75cm,minimum height=0.82cm,inner sep=1pt,
    font=\scriptsize},
  layer/.style={font=\small},
  brace/.style={decorate,decoration={brace,mirror,amplitude=4pt}}
]
\foreach \j in {0,...,7} {
  \pgfmathsetmacro{\yy}{7-\j}
  \draw[wire] (0,\yy) -- (7.0,\yy);
}
\node[even] at (1.25,6.5) {$G_{01}(\sqrt7\,\delta/2)$};
\node[even] at (1.25,4.5) {$G_{23}(\sqrt{15}\,\delta/2)$};
\node[even] at (1.25,2.5) {$G_{45}(\sqrt{15}\,\delta/2)$};
\node[even] at (1.25,0.5) {$G_{67}(\sqrt7\,\delta/2)$};
\node[odd] at (3.50,5.5) {$G_{12}(2\sqrt3\,\delta)$};
\node[odd] at (3.50,3.5) {$G_{34}(4\delta)$};
\node[odd] at (3.50,1.5) {$G_{56}(2\sqrt3\,\delta)$};
\node[even] at (5.75,6.5) {$G_{01}(\sqrt7\,\delta/2)$};
\node[even] at (5.75,4.5) {$G_{23}(\sqrt{15}\,\delta/2)$};
\node[even] at (5.75,2.5) {$G_{45}(\sqrt{15}\,\delta/2)$};
\node[even] at (5.75,0.5) {$G_{67}(\sqrt7\,\delta/2)$};
\node[layer] at (1.25,-0.55) {$E_x(\delta/2)$};
\node[layer] at (3.50,-0.55) {$O_x(\delta)$};
\node[layer] at (5.75,-0.55) {$E_x(\delta/2)$};
\end{tikzpicture}
\end{center}
Its coherent Trotter error scales as $\mathcal O(\theta^3/r^2)$. The experiment uses a single step, $r=1$, for the displacement $\theta=\pi/6$.

The azimuthal part is diagonal in the orbital basis. Writing Eq.~\eqref{eq:supp-Lz} as
\begin{equation}
L_z=\sum_{j=0}^{7}m_j n_j,
\qquad
m_j=-\frac72+j,
\label{eq:app-Lz-orbital-diagonal}
\end{equation}
and using $n_j=(I-Z_j)/2$, its implementation is, up to a global phase,
\begin{equation}
e^{-i\lambda L_z}
\doteq
\prod_{j=0}^{7}R_z(-m_j\lambda),
\qquad
\lambda=\phi+\frac{\pi}{2},
\label{eq:app-azimuthal-virtual-z}
\end{equation}
where $\doteq$ denotes equality up to a global phase. These virtual-$Z$ frame updates are applied in parallel and add neither CZ gates nor physical pulse depth.

Using the LLL-projected field operator in Eq.~\eqref{eq:app-projected-field}, the quasihole density operator is
\begin{equation}
\hat n(\Omega) = \sum_{j,k=0}^{N_\phi} \phi_j^*(\Omega)\phi_k(\Omega)c_j^\dagger c_k .
\label{eq:app-projected-density-operator-rep}
\end{equation}
Taking its expectation value gives the real-space density
\begin{equation}
\rho(\Omega)
=\langle\hat n(\Omega)\rangle
=\sum_{j,k}\Gamma_{jk}\phi_j^*(\Omega)\phi_k(\Omega),
\qquad
\Gamma_{jk}\equiv\langle c_j^\dagger c_k\rangle.
\label{eq:supp-real-space-density}
\end{equation}
For the polar quasihole, the axial symmetry about the $z$ axis leads to a density profile determined solely by the orbital populations,
\begin{equation}
\rho(\theta) = \sum_j \langle n_j\rangle |\phi_j(\theta)|^2 .
\label{eq:supp-polar-density}
\end{equation}
For a displaced quasihole, however, the azimuthal position is encoded in the off-diagonal orbital coherences. Under a rotation around the $z$ axis,
\begin{equation}
\Gamma_{j,j+q} \rightarrow e^{-iq\phi_h} \Gamma_{j,j+q},
\label{eq:supp-coherence-phase-winding}
\end{equation}
and the associated phase winding determines the angular displacement of the reconstructed density profile. Therefore, resolving a quasihole away from the pole requires the coherent orbital correlations contained in the one-body density matrix.

\begin{figure}[!htbp] \centering \includegraphics[width=0.98\columnwidth]{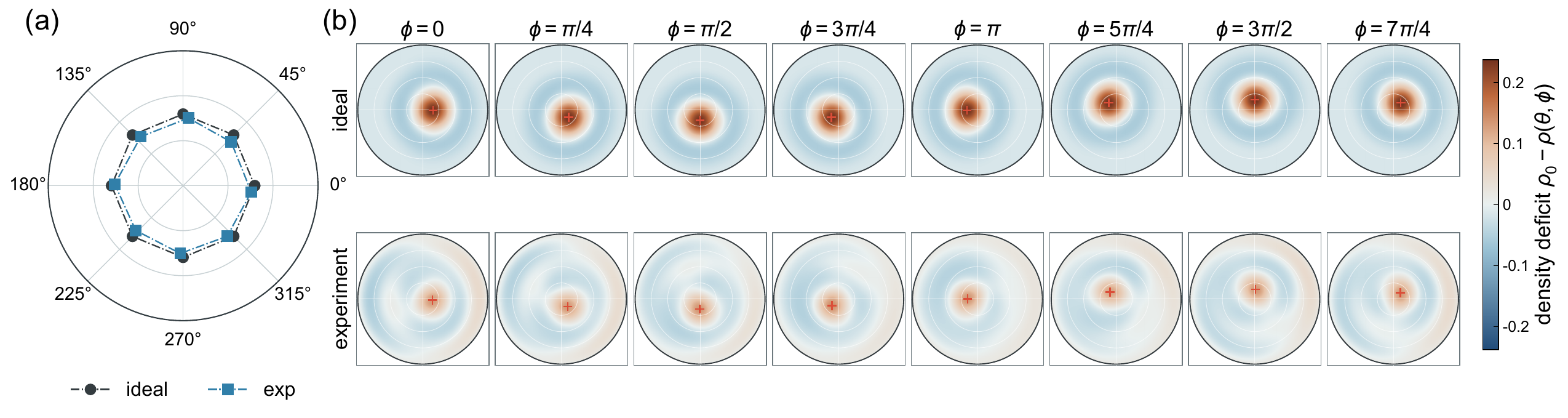} 
  \caption[OBDM-based reconstruction of the quasihole density deficit]{
  Quasihole position and density deficit reconstructed from the measured one-body density matrix (OBDM). (a) Programmed quasihole longitudes and longitudes inferred from the reconstructed density deficits. (b) Ideal (upper row) and experimental (lower row) maps of $\rho_0-\rho(\theta,\phi)$ for quasiholes displaced to the fixed polar angle $\theta_h=\pi/6$, with $\rho_0=N_e/(4\pi)$.
  }
\label{fig:app-obdm-density-deficit-scan} 
\end{figure}

\begin{table*}[!htbp]
\centering
\caption{The 64 product-Pauli bases used to reconstruct the off-diagonal OBDM elements; diagonal occupations are obtained from a separate all-$Z$ measurement.}
\label{tab:app-obdm-optimal-settings}
\renewcommand{\arraystretch}{1.0}
\setlength{\tabcolsep}{3pt}
\scriptsize
\begin{tabular}{ccccc@{\hspace{10pt}}ccccc}
\toprule
$\mathcal P$ & $XX\cdots X$ & $YY\cdots Y$ & $XYXY\cdots$ & $YXYX\cdots$
& $\mathcal P$ & $XX\cdots X$ & $YY\cdots Y$ & $XYXY\cdots$ & $YXYX\cdots$ \\
\midrule
$(0,4,5)$ & $\mathtt{XZZZXXZZ}$ & $\mathtt{YZZZYYZZ}$ & $\mathtt{XZZZYXZZ}$ & $\mathtt{YZZZXYZZ}$ & $(0,2,4,7)$ & $\mathtt{XZXZXZZX}$ & $\mathtt{YZYZYZZY}$ & $\mathtt{XZYZXZZY}$ & $\mathtt{YZXZYZZX}$ \\
$(0,5,6)$ & $\mathtt{XZZZZXXZ}$ & $\mathtt{YZZZZYYZ}$ & $\mathtt{XZZZZYXZ}$ & $\mathtt{YZZZZXYZ}$ & $(1,2,5)$ & $\mathtt{ZXXZZXZZ}$ & $\mathtt{ZYYZZYZZ}$ & $\mathtt{ZXYZZXZZ}$ & $\mathtt{ZYXZZYZZ}$ \\
$(0,6,7)$ & $\mathtt{XZZZZZXX}$ & $\mathtt{YZZZZZYY}$ & $\mathtt{XZZZZZYX}$ & $\mathtt{YZZZZZXY}$ & $(2,6)$ & $\mathtt{ZZXZZZXZ}$ & $\mathtt{ZZYZZZYZ}$ & $\mathtt{ZZXZZZYZ}$ & $\mathtt{ZZYZZZXZ}$ \\
$(0,7)$ & $\mathtt{XZZZZZZX}$ & $\mathtt{YZZZZZZY}$ & $\mathtt{XZZZZZZY}$ & $\mathtt{YZZZZZZX}$ & $(2,7)$ & $\mathtt{ZZXZZZZX}$ & $\mathtt{ZZYZZZZY}$ & $\mathtt{ZZXZZZZY}$ & $\mathtt{ZZYZZZZX}$ \\
$(0,1,4,6)$ & $\mathtt{XXZZXZXZ}$ & $\mathtt{YYZZYZYZ}$ & $\mathtt{XYZZXZYZ}$ & $\mathtt{YXZZYZXZ}$ & $(0,3,4)$ & $\mathtt{XZZXXZZZ}$ & $\mathtt{YZZYYZZZ}$ & $\mathtt{XZZYXZZZ}$ & $\mathtt{YZZXYZZZ}$ \\
$(1,5,7)$ & $\mathtt{ZXZZZXZX}$ & $\mathtt{ZYZZZYZY}$ & $\mathtt{ZXZZZYZX}$ & $\mathtt{ZYZZZXZY}$ & $(1,3,5)$ & $\mathtt{ZXZXZXZZ}$ & $\mathtt{ZYZYZYZZ}$ & $\mathtt{ZXZYZXZZ}$ & $\mathtt{ZYZXZYZZ}$ \\
$(1,6)$ & $\mathtt{ZXZZZZXZ}$ & $\mathtt{ZYZZZZYZ}$ & $\mathtt{ZXZZZZYZ}$ & $\mathtt{ZYZZZZXZ}$ & $(2,3,6)$ & $\mathtt{ZZXXZZXZ}$ & $\mathtt{ZZYYZZYZ}$ & $\mathtt{ZZXYZZXZ}$ & $\mathtt{ZZYXZZYZ}$ \\
$(1,7)$ & $\mathtt{ZXZZZZZX}$ & $\mathtt{ZYZZZZZY}$ & $\mathtt{ZXZZZZZY}$ & $\mathtt{ZYZZZZZX}$ & $(3,7)$ & $\mathtt{ZZZXZZZX}$ & $\mathtt{ZZZYZZZY}$ & $\mathtt{ZZZXZZZY}$ & $\mathtt{ZZZYZZZX}$ \\
\bottomrule
\end{tabular}
\end{table*}

\section{Direct reconstruction of quasihole density from one-body correlations}
\label{app:quasihole-real-space-reconstruction}

Each spherical LLL orbital is extended around the azimuthal direction, so a localized quasihole cannot be inferred from orbital populations alone. Its position is encoded in the interference between orbitals, captured by the one-body density matrix $\Gamma_{ij}=\langle c_i^\dagger c_j\rangle$. The diagonal elements describe the azimuthally averaged population profile, whereas the off-diagonal elements carry the phase coherence that localizes the density depletion. Equation~\eqref{eq:supp-real-space-density} then converts $\Gamma$ into the real-space density without requiring many-body state tomography.

The orbital populations follow directly from computational-basis measurements,
\begin{equation}
\Gamma_{ii} = \langle n_i \rangle = \frac{1 - \langle Z_i \rangle}{2}.
\label{eq:app-diagonal-gamma-z}
\end{equation}
The orbital coherences require phase-sensitive measurements. Under the Jordan--Wigner mapping,
\begin{equation}
c_j = \left( \prod_{k=0}^{j-1} Z_k \right) \frac{X_j + iY_j}{2},
\label{eq:app-jordan-wigner-convention}
\end{equation}
which gives, for $i<j$,
\begin{equation}
\begin{aligned}
c_i^\dagger c_j
= \frac{1}{4} \Big( & X_i P_{ij} X_j + Y_i P_{ij} Y_j 
 + i X_i P_{ij} Y_j - i Y_i P_{ij} X_j \Big),
\end{aligned}
\label{eq:app-fermionic-bilinear-pauli}
\end{equation}
where $
P_{ij} = \prod_{\ell = i+1}^{j-1} Z_\ell
$
is the fermion-parity string between orbitals $i$ and $j$. The string preserves the fermionic sign accumulated between the two orbitals. Defining
\begin{equation}
C_{ij}^{\mu\nu} = \left\langle \sigma_i^\mu P_{ij} \sigma_j^\nu \right\rangle, \qquad \mu,\nu \in \{X,Y\},
\label{eq:app-pauli-correlator-definition}
\end{equation}
the orbital coherence is reconstructed as
\begin{equation}
\Gamma_{ij} = \frac{1}{4} \left( C_{ij}^{XX} + C_{ij}^{YY} \right) + \frac{i}{4} \left( C_{ij}^{XY} - C_{ij}^{YX} \right).
\label{eq:app-gamma-from-pauli-correlators}
\end{equation}
Adjacent orbitals have $P_{i,i+1}=I$; nonadjacent coherences retain the intervening parity information through the $Z$ string.

Each off-diagonal element $\Gamma_{ij}$ with $i<j$ is obtained from four product-Pauli settings, corresponding to the endpoint combinations $XX$, $YY$, $XY$, and $YX$ in Eq.~\eqref{eq:app-gamma-from-pauli-correlators}. The required $X$ and $Y$ measurements are implemented by local basis rotations followed by computational-basis readout. Hermiticity fixes the opposite matrix element, $\Gamma_{ji}=\Gamma_{ij}^*$, while an all-$Z$ setting gives the diagonal occupations. Measuring all orbital pairs independently would therefore require
\begin{equation}
N_{\mathrm{set}}^{\mathrm{pairwise}} = 4 \binom{8}{2} + 1 = 113,
\label{eq:app-obdm-setting-pairwise}
\end{equation}
Pauli strings that share a common tensor-product eigenbasis can be measured simultaneously in a single setting~\cite{Huggins2021npjQuantumMeasurement,Hamamura2020npjEntangledMeasurements}. Grouping the Jordan--Wigner correlators by this criterion reduces the required number of settings below the pairwise count, so $N_{\mathrm{set}}^{\mathrm{pairwise}}=113$ is only an upper bound.

The Jordan--Wigner strings constrain how far the grouping can be taken. 
Within local product measurements over \(\{X,Y,Z\}^{\otimes8}\), independently measuring all four Jordan--Wigner endpoint
correlators for every orbital pair requires at least \(N_{\mathrm{set,off}}^{\min}=64\) settings.
To establish the lower bound, divide the orbital chain across its middle,
\begin{equation}
\{0,1,2,3\} \mid \{4,5,6,7\}.
\label{eq:app-obdm-middle-cut}
\end{equation}
The cut is crossed by $4\times4=16$ orbital pairs, each requiring the four endpoint combinations $XX$, $YY$, $XY$, and $YX$. A Jordan--Wigner correlator is accessible only when its endpoints are consecutive among the non-$Z$ positions of the global basis; otherwise an intervening $X$ or $Y$ would break the required parity string. At most one such consecutive pair can cross the middle cut in a given setting. The 64 cross-cut correlators therefore require $N_{\mathrm{set,off}}\ge64$.
The lower bound is attained by organizing the orbital pairs into ordered paths
\begin{equation}
\mathcal P = (p_1, p_2, \ldots, p_m), \qquad p_1 < p_2 < \cdots < p_m,
\end{equation}
and assigning four settings to each path: uniform $X$, uniform $Y$, and two alternating $X/Y$ patterns. Orbitals outside the path are measured in $Z$. Every consecutive edge $(p_r,p_{r+1})$ then supplies all four endpoint correlators.
The sixteen paths in Table~\ref{tab:app-obdm-optimal-settings} partition all $\binom{8}{2}=28$ orbital pairs into consecutive path edges. Four settings per path reconstruct all off-diagonal elements, requiring 64 settings. A
separate all-$Z$ setting determines the diagonal occupations. The implemented OBDM measurement therefore uses 65 settings in total.
To ensure that the reconstructed correlations define a valid three-fermion one-body state, we impose
\begin{equation}
\Gamma = \Gamma^\dagger, \qquad 0 \preceq \Gamma \preceq I, \qquad \operatorname{Tr} \Gamma = N_e.
\label{eq:app-obdm-physical-constraints}
\end{equation}
The constrained $\Gamma^{\mathrm{exp}}$ is converted to real space through Eq.~\eqref{eq:supp-real-space-density}. Figure~\ref{fig:app-obdm-density-deficit-scan} shows the corresponding deficit $\rho_0-\rho(\theta,\phi)$, with the localized depletion following the programmed quasihole position along the longitude scan.

\begin{figure}[t]
  \centering
  \includegraphics[width=0.85\columnwidth]
  {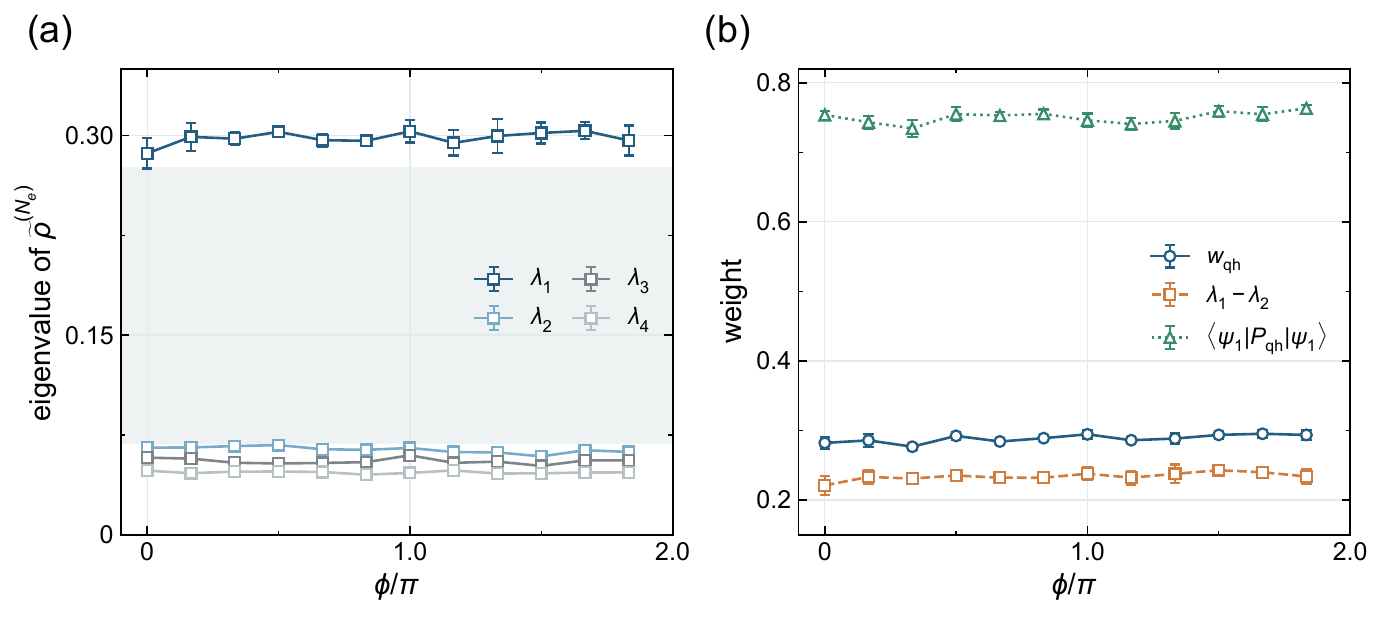}
  \caption{
  Spectral and quasihole-manifold diagnostics of the reconstructed states.
  (a) Leading eigenvalue branches of $\widetilde\rho^{(N_e)}(\phi)$ along the 12-point longitude scan. The shaded region indicates the spectral gap between $\lambda_1$ and $\lambda_2$.
  (b) Total quasihole-manifold weight $w_{\rm qh}$, spectral gap $\lambda_1-\lambda_2$, and quasihole-manifold weight
  $\langle\psi_1|P_{\rm qh}|\psi_1\rangle$ of the leading eigenvector. Symbols show mean values, and error bars denote standard errors of the mean at each sampled longitude.
  }
  \label{fig:app-moving-qh-branch-diagnostics}
  \end{figure}

\section{Tomographic reconstruction of quasihole motion}
\label{app:quasihole-motion-tomography}

The quasihole mover preserves the total fermion number,
\begin{equation}
[U_{\mathrm{move}}(\varphi),\hat N]=0.
\label{eq:app-moving-qh-number-conservation}
\end{equation}
The ideal quasihole mover is block diagonal in particle number.
We reconstruct the moving quasihole using a tomographic
estimator restricted to the prescribed fixed-\(N_e\)
Hilbert space,
\begin{equation}
\mathcal H^{(N_e)}
=
\operatorname{span}\!\left\{\ket{b}:|b|=N_e\right\},
\quad
d_{N_e}=56,
\label{eq:app-fixed-n3-sector}
\end{equation}
where $N_e=3$. The reconstructed state obeys
\begin{equation}
\widetilde\rho^{(N_e)}\succeq0,
\qquad
\operatorname{Tr}\widetilde\rho^{(N_e)}=1.
\label{eq:app-sector-density-matrix}
\end{equation}
We adopt a greedy rank-increasing procedure to select 91 product-Pauli
settings whose restricted measurement operators span the complete
$d_{N_e}^2$-dimensional operator space on $\mathcal H^{(N_e)}$.
For a product-Pauli setting
\begin{equation}
s=s_0s_1\cdots s_7,
\qquad s_j\in\{X,Y,Z\},
\label{eq:app-pauli-setting}
\end{equation}
let $U_s$ denote the corresponding premeasurement rotations. The outcome
$b$ is represented in the fixed-number sector by
\begin{equation}
E_{s,b}^{(N_e)}
=
P^{(N_e)}U_s^\dagger\ket{b}\bra{b}U_sP^{(N_e)},
\label{eq:app-projected-povm-element}
\end{equation}
where $P^{(N_e)}$ projects onto $\mathcal H^{(N_e)}$. Its ideal probability is
\begin{equation}
q_{s,b}\!\left(\widetilde\rho^{(N_e)}\right)
=
\operatorname{Tr}\!\left[
\widetilde\rho^{(N_e)}E_{s,b}^{(N_e)}
\right].
\label{eq:app-ideal-sector-probability}
\end{equation}
The Laughlin quasihole manifold studied here is defined entirely within the
fixed-particle-number space $\mathcal H^{(N_e)}$, which is preserved by the
ideal quasihole mover. 
The reconstruction is therefore targeted at this
physically specified subspace. As the $X/Y$ rotations do not commute with
particle number, shot-wise sector selection is not available.
Here, the
complete readout-corrected distributions are fitted jointly using measurement
operators restricted to $\mathcal H^{(N_e)}$. The resulting positive,
unit-trace matrix $\widetilde\rho^{(N_e)}$ is the maximum-likelihood estimate
within the target sector, while the independently measured all-$Z$ statistics
provide its experimental population $w^{(N_e)}$.
Following the readout correction described in
Sec.~\ref{app:full-quantum-state-tomography}, the corrected distributions $\bar q_{s,b}$ from all settings are fitted jointly to a physical state in
$\mathcal H^{(N_e)}$.
With $R_s$ repetitions for setting $s$, the maximum-likelihood estimate is
\begin{equation}
\widetilde\rho^{(N_e)}
=
\arg\max_{\sigma\succeq0,\,\operatorname{Tr}\sigma=1}
\sum_{s\in\mathcal S}R_s\sum_b\bar q_{s,b}
\log\!\left[\operatorname{Tr}(\sigma E_{s,b}^{(N_e)})\right].
\label{eq:app-sector-effective-mle}
\end{equation}
We exploit the particle-number structure of the target quasihole states by restricting the tomographic estimator to the fixed-$N_e$ Hilbert space $\mathcal H^{(N_e)}$.
This symmetry constraint substantially reduces the number of free parameters while retaining a general positive, unit-trace density matrix within the target sector.
The resulting reconstruction therefore imposes no
additional assumptions on the state's rank or
entanglement structure~\cite{Tan2026Critical}.
Experimentally, the maximum-likelihood reconstruction yields $\widetilde\rho^{(N_e)}(\varphi_k)$ at each programmed
longitude. The all-$Z$ data independently give the sector population,
$w^{(N_e)}=0.461\pm0.003$. The corresponding one-body
density matrix,
\begin{equation}
\widetilde\Gamma_{jk}
=
\operatorname{Tr}\!\left(
\widetilde\rho^{(N_e)}c_j^\dagger c_k
\right),
\label{eq:app-reconstructed-one-body-density-matrix}
\end{equation}
gives the real-space density through
Eq.~\eqref{eq:supp-real-space-density}.

The reconstructed subspace states are mixed, but their leading eigenvalue remains nondegenerate and spectrally separated throughout the measured loop. The spectral decomposition is
\begin{equation}
\widetilde\rho^{(N_e)}(\phi_k)=
\sum_{\alpha}
\lambda_\alpha(\phi_k)
\ket{\psi_\alpha(\phi_k)}
\bra{\psi_\alpha(\phi_k)},
\qquad
\lambda_1\geq\lambda_2\geq\cdots .
\label{eq:app-moving-qh-spectral-decomposition}
\end{equation}
As shown in Fig.~\ref{fig:app-moving-qh-branch-diagnostics}(a), the leading eigenvalue remains separated from the rest of the spectrum, ${\lambda}_1=0.298\pm0.003$ and
  ${\lambda_1}-{\lambda_2}=0.234\pm0.004$. 
The associated principal eigenvector $\ket{\psi_k}$ therefore defines a well-resolved dominant quasihole branch across the sampled longitudes. 
For $N_e=3$ and $N_\phi=7$, the exact quasihole manifold forms the four-dimensional $L=3/2$ multiplet $\{\ket{\chi_\mu}\}_{\mu=0}^{3}$, with projector $P_{\rm qh}=\sum_{\mu=0}^{3}|\chi_{\mu}\rangle\langle\chi_{\mu}|$. 
We characterize the support of the reconstructed state and its leading eigenvector in this manifold through
\begin{equation}
\begin{aligned}
  w_{\rm qh}(\phi_k)
  &=
  \operatorname{Tr}\!\left[
  P_{\rm qh}\widetilde\rho^{(N_e)}(\phi_k)
  \right],\quad
  p_{\rm qh}^{(1)}(\phi_k)
  &=
  \bra{\psi_1(\phi_k)}
  P_{\rm qh}
  \ket{\psi_1(\phi_k)}.
  \label{eq:app-leading-qh-weight}
  \end{aligned}
\end{equation}
While $w_{\rm qh}$ measures the quasihole-manifold support of the state, $p_{\rm qh}^{(1)}$ isolates the quasihole character of its leading eigenvector and remains near $0.75$ throughout the scan [Fig.~\ref{fig:app-moving-qh-branch-diagnostics}(b)].
The persistent spectral gap and quasihole-manifold overlap identify
$\ket{\psi_1(\phi_k)}$ as a well-resolved quasihole branch.

To characterize the reconstructed principal-eigenstate branch and compare it with the ideal pure-state quasihole trajectory, we evaluate the gauge-invariant discrete overlap product (Wilson loop)~\cite{SamuelBhandari1988PRLGeometricPhase},
\begin{equation}
|\gamma_{\rm eff}|
=\left|\arg\prod_{k=0}^{N_{\rm pt}-1}
\langle\psi_k|\psi_{k+1}\rangle\right|,
\qquad |\psi_{N_{\rm pt}}\rangle=|\psi_0\rangle,
\label{eq:app-wilson-loop-principal-vector}
\end{equation}
where we denote the leading eigenvector at $\phi_k$ by $\ket{\psi_k}\equiv\ket{\psi_1(\phi_k)}$. Here $N_{\rm pt}$ is the number of sampled points along the closed path.
The reported quantity is the Wilson-loop phase magnitude of the tracked dominant eigenstates $\ket{\psi_k}$ extracted from the reconstructed density matrices. The argument is taken on its principal branch, and the plotted quantity is the dimensionless magnitude $|\gamma_{\rm eff}|/\pi$. Because the reconstructed states are mixed, we use this branch-based quantity as an effective geometric-phase diagnostic of the reconstructed trajectory.
For the eight- and twelve-point paths, this gives $|\gamma_{\rm eff}|/\pi=0.353\pm0.003$ and $0.381\pm0.014$, respectively, with uncertainties denoting the standard error of the mean across independent scans. The noiseless Trotter circuits at the same sample points give $0.371$ and $0.381$; their dense-sampling limit is $0.382$. The continuous ideal spin-$3/2$ loop at $\theta_h=\pi/6$ gives the reference phase magnitude $|\gamma_{\rm id}|/\pi\simeq0.402$. These are the experimental and theoretical quantities compared in Fig.~4(c) of the main text.

%